\documentclass[preprint]{aastex}

\usepackage{graphicx}
\usepackage{dcolumn}
\usepackage{bm}
\usepackage{amsfonts}
\usepackage{color}
\usepackage{amsmath}

\begin{document}

\shorttitle{A NEW GR NEUTRINO RADIATION-HYDRO CODE}
\shortauthors{Kuroda, Takiwaki \& Kotake}

\title{A New Multi-Energy Neutrino Radiation-Hydrodynamics Code in Full General Relativity and Its Application to Gravitational Collapse of Massive Stars}

\author{Takami Kuroda\altaffilmark{1}, Tomoya Takiwaki\altaffilmark{2} and Kei Kotake\altaffilmark{3,4}}

\affil{$^1$Department of Physics, University of Basel,
Klingelbergstrasse 82, 4056 Basel, Switzerland}
\affil{$^2$Astrophysical Big Bang Laboratory, RIKEN, Saitama, 
351-0198, Japan}
\affil{$^3$Department of Applied Physics, Fukuoka University,
 8-19-1, Jonan, Nanakuma, Fukuoka, 814-0180, Japan}
\affil{$^4$Division of Theoretical Astronomy, National Astronomical Observatory of Japan, 2-21-1, Osawa, Mitaka, Tokyo, 181-8588, Japan}

\begin{abstract}
We present a new multi-dimensional radiation-hydrodynamics code 
for massive stellar core-collapse in full general relativity (GR).
Employing an M1 analytical closure scheme, 
 we solve spectral neutrino transport of the radiation energy and momentum 
based on a truncated moment formalism.
Regarding neutrino opacities, we take into account a baseline
 set in state-of-the-art simulations, in which inelastic neutrino-electron scattering, 
thermal neutrino production via pair annihilation and 
nucleon-nucleon bremsstrahlung are included. 
While the Einstein field equations and the spatial advection terms in the radiation-hydrodynamics equations
are evolved explicitly, the 
source terms due to neutrino-matter interactions
and energy shift in the radiation moment equations are integrated 
implicitly by an iteration method.
To verify our code, we first perform a series of 
 standard radiation tests with analytical solutions that 
include the check of 
gravitational redshift and Doppler shift. A good agreement in these
 tests supports the reliability of the GR multi-energy neutrino transport scheme.
We then conduct several test simulations 
 of core-collapse, bounce, and shock-stall of 
 a 15M$_\odot$ star in the Cartesian coordinates and make a detailed 
comparison with published results. Our code performs 
quite well to reproduce the results of full-Boltzmann 
neutrino transport especially before bounce. In the postbounce phase,
our code basically performs well, 
however, there are several differences that are most 
likely to come 
 from the insufficient spatial resolution in our current 3D-GR models.
  For clarifying the resolution dependence and extending 
the code comparison in the late postbounce phase, we discuss that
next-generation Exaflops-class supercomputers are at least needed.
\end{abstract}

\maketitle

\section{Introduction}
\label{sec:Introduction}
 Neutrino transport is an essential ingredient for
 numerical simulations to clarify the theory of 
 core-collapse of massive stars and the formation mechanisms of compact 
 objects (see, e.g., \citet{thierry15,Mezzacappa2014,Burrows13,Janka12,Kotake12} for
 recent reviews). 
 In the collapsing iron core, neutrinos play crucial roles 
in every phase; deleptonization in the core 
determines the time of bounce and the mass of 
 the proto-neutron star (PNS) (e.g., \citet{langanke03}); the gigantic 
internal energy tapped in the PNS is almost completely carried away by 
neutrinos, a tiny fraction of which contributes to the heating of the 
postshock material, leading to the onset of core-collapse supernovae (CCSNe) 
 in the context of the neutrino-heating mechanism \citep{Wilson85,Bethe85}.
Since these SN neutrinos are generally not in both thermal and chemical
 equilibrium, the distribution of neutrinos in the phase space 
should be computationally determined. This is a
 seven dimensional (7D) problem; (3D+1D in space-time and 3D 
in momentum space), the reason why CCSN 
simulations are considered as one of the most challenging
 subjects in computational astrophysics.

Primarily due to neutrino-driven convection (e.g., \citet{bethe,
herant,burr95,Janka96,mujan}) and the 
standing-accretion-shock instability (SASI, e.g., 
\citet{Blondin03,Foglizzo06,Foglizzo07,Ohnishi06,Blondin07_nat,Iwakami08,Iwakami09,rodrigo09,jerome12,Hanke12,thierry12,couch13a,Fernandez14}),  
the postbounce cores are essentially of 
 multi-dimensional (multi-D) nature.\footnote{Recently multi-dimensionalities
 in the precollapse core (e.g., \citet{meakin11}) are
 also attracting much attention (e.g., \cite{couch_ott,bernhard14}).} 
Due to the high compactness of the PNS, these 
 multi-D fluid motions are all under the influence of 
the general relativistic (GR) gravity, the consideration of which
 used to be standard in the pioneering era of CCSN simulations 
(e.g., \cite{May,schwartz}). In rapidly rotating supernova cores
 (e.g., \citet{Woosley06}),
 magnetohydrodynamics (MHD) instabilities naturally
 make the core dynamics intrinsically non-spherical (e.g.,
 \citet{arde00,kota04b,kotake06,ober06a,martin14,burr07,masada12,masada14,sawai13,Nakamura3D14,iwakami14}). All in all, in order to 
 have the final word on the CCSN mechanisms quantitatively, one needs
 to deal with the 7D Boltzmann neutrino 
transport simulations in full GR MHD. Unfortunately, however, this still remains
 to be 
 computationally very demanding even using exa-scale computing platforms 
 in the next decade(s) to come (see discussions in 
\citet{Kotake12_ptep})\footnote{We here mean the feasibility of 7D-GR Boltzmann 
neutrino transport simulation following $\sim $1 s after bounce 
with sufficient numerical resolutions in both space and momentum space.}.


Since the late 1990s, the ultimate spherically-symmetric (1D) simulations where the 
GR Boltzmann transport is coupled to 1D-GR hydrodynamics have 
 been made feasible by \citet{Liebendorfer01} and \cite{Sumiyoshi05}
 (see \citet{Mezzacappa89,tony93a,tony93b,tony93c,Yamada97,Yamada99,Bruenn01,Liebendorfer01,Liebendorfer04} for the code developments).
 Unfortunately, however,
these full-fledged 1D simulations fail to produce 
explosions except for super-AGB stars at the low-mass end
\citep{kitaura}. In the context of the full Boltzmann calculations, 
\citet{Livne04} were the first to perform two-dimensional (2D) 
multi-angle (i.e., assuming axisymmetry in both space and momentum space)
neutrino hydrodynamics simulations using the discrete ordinate ($S_n$) method. 
Then it was demonstrated by \citet{Ott08} that the multi-angle transport 
is really important
especially when the neutrino radiation field deviates significantly 
from spherical symmetry such as in the rapidly rotating cores 
(see also \citet{Brandt}). Going beyond the assumption of 
 axisymmetry in the multi-angle transport,
 \cite{Sumiyoshi12} were the first to develop the 
fully multi-angle Boltzmann transport code and then apply it
for static backgrounds \citep{sumi14}.
More recently, \cite{Nagakura14} extended the code to include special relativistic (SR) corrections and showed the ability of the code by performing
 1D core-collapse simulation of a 15$M_\odot$ model.
Albeit not yet implemented in hydrodynamics simulations, several 
novel formulations and schemes of the full Boltzmann equation have recently
 been reported in \citet{cardall12,cardall13,shibata14}, and \citet{peres}.

At present, direct discretization of the Boltzmann transport equation fully 
into the neutrino angle and energy (such as by the $S_n$ method mentioned 
above) is still computationally very expensive. An approximation often made in 
previous works is to remove the full angular dependence 
of the Boltzmann equation by expanding the neutrino distribution function as a 
series of moments. The simplest version, in which one closes the moment
 expansion after the zeroth angular moment, is multi-group 
flux limited diffusion (MGFLD) scheme 
(e.g., \citet{Bruenn85,Livne04,kotake06fld,Swesty09,zhang13,Bruenn13}).
In FLD schemes, the basic equation is a diffusion equation for 
 the neutrino energy density. In solving it, 
an appropriate flux-limiter should be employed 
(e.g., \citet{minerbo78,pomraning81,levermore84,janka92})
 to ensure the causality of the 
radiation field in the transparent regions where the diffusion 
approximation breaks down.
The isotropic diffusion source approximation (IDSA) scheme 
developed by \citet{Liebendorfer09} is basically positioned 
at a similar approximation level compared to the MGFLD scheme.
  In the IDSA, the neutrino distribution function is 
divided into two components, 
the one which is trapped with matter and has isotropic distribution 
function and the other in the free-streaming limit, each of which 
 is solved independently, while satisfying the 1D Boltzmann equation 
as a whole.
Due to the high computational efficiency, the IDSA has been extensively
 employed in both 2D \citep{Suwa10,Suwa11,suwa13,suwa14,Nakamura2D14} and 
3D simulations \citep{Takiwaki12,Takiwaki14}. 
One can also truncate the angular moment at the second order and transport 
the zeroth and first order angular moment.
In this case, higher or equal to the second order moment needs to be 
determined independently to close the set of the transport equations.
In the M1 moment scheme (e.g., \citet{Pons00,Shibata11}), one applies an 
analytic formula for the closure relation (see examples applied in 
post-Newtonian MHD simulations \citep{martin14} and GR simulations
 in 1D \citep{O'Connor13,O'Connor14} and in 3D \citep{KurodaT12,KurodaT14}).
In contrast, one can self-consistently determine the closure relation
 by the variable Eddington factor (VEF) method (e.g., \citet{BMuller10} and
 see references therein).
 In these cases, a model Boltzmann equation is integrated to iteratively obtain the solution 
up to the higher moments (i.e., the Eddington tensor) until the system is 
converged. Currently, the state of the art in
 multi-D simulations of CCSNe is defined by multi-group (spectral) neutrino
 hydrodynamics simulations. More severe approximations
 include gray transport \citep{fryer1999,scheck06} 
or the light-bulb and leakage 
schemes 
(e.g., \citet{Janka96,ruffert96,ross03,Kotake03,Murphy08,oconnor,perego}), 
which have been often employed in many recent studies of 
multi-D instabilities and the MHD mechanism of CCSNe.

In addition to the multi-D and multi-angle/truncated neutrino 
transport, the accurate treatment of GR is highly ranked among the 
to-do lists towards the ultimate CCSN simulations. In most of previous multi-D models with multi-group neutrino 
transport, GR effects are attempted to be modelled by using a 
modified gravitational potential that takes into account a 1D GR
 effect by replacing the monopole term of Newtonian gravity with the TOV 
potential \citep{Buras06a,Buras06b,Marek09,Bruenn09,Hanke13}.
While there are a number of GR core-collapse 
simulations in 2D (e.g., \citet{Dimmelmeier02,shibaseki,BMuller12a})
 and in 3D (e.g., \citet{shibata05,Ott12a,KurodaT12,KurodaT14,moesta14}), 
many of them especially in 3D have been made,
 for the sake of computational cost, to employ
 a simplified microphysics such as by the $Y_e$ parametrization scheme 
\citep{Liebendorfer05} or by the neutrino leakage scheme
 \citep{Sekiguchi10}.
In our previous study \citep{KurodaT12}, 
we performed 3D GR/SR hydrodynamics simulations of a $15 M_{\odot}$ star 
with the gray M1 scheme. We demonstrated that due to deeper gravitational well
 of GR the neutrino luminosity
 and the average neutrino energy in the postbounce phase increase
 when switching from SR to GR hydrodynamics. Since the neutrino heating 
rates in the postshock regions are sensitively affected by
 the emergent neutrino spectra, the GR effect whether it will or will not
 help the onset of neutrino-driven explosions needs to be investigated 
 by multi-D GR simulations with more sophisticated neutrino transport scheme.\footnote{It is worth mentioning that 2D GR models with the VEF method tend to
 explode more easily than the corresponding 2D Newtonian models with
 and without the GR correction
\citep[e.g.][]{BMuller10,BMuller12a,BMuller14}. This may support the 
 speculation that GR is helpful for the working of the neutrino mechanism
 in multi-D simulations.}

 In this paper, we present a new 3D-GR radiation hydrodynamics code that is meant 
 to apply for stellar core-collapse simulations. The
 spacetime treatment is based on the Arnowitt-Deser-Misner 3+1 formalism and
 we employ the Baumgarte-Shapiro-Shibata-Nakamura (BSSN) formalism 
\citep{Shibata95,Baumgarte99} to evolve the metric variables.
The full GR radiation-hydrodynamics equations 
 are evolved in a conservative form, in which 
we solve the energy-dependent set of radiation moments up to the first order
with the M1 moment scheme.
 This part is based on the partial implementation of the Thorne's moment 
formalism \citep{Thorne81}, which is extended by \citet{Shibata11} in 
a more suitable manner applicable to the neutrino transport problem.
Regarding the neutrino-matter interaction terms, we employ
a baseline set of weak interactions as given in 
\citet{Bruenn85} and \citet{Rampp02} where nucleon-nucleon bremsstrahlung is additionally 
taken into account.
Our newly developed code is designed to evolve the Einstein field
 equation together with the GR radiation hydrodynamic equations
 in a self-consistent manner while satisfying the Hamiltonian and momentum 
constraints. A nested structure embedded in the 3D Cartesian computational domain enables 
us to follow the dynamics starting from the onset of gravitational collapse 
of a 15 $M_{\odot}$ star \citep{WW95}, through bounce, 
up to about $\sim$50 ms postbounce.
Since, it is still computationally too expensive to follow long-term evolution in
 full 3D until the neutrino-driven explosion takes place
(e.g., at the earliest $\sim 200$ ms after bounce \citep{Bruenn09,Marek09} or $\sim 500$ ms in 2D GR calculation
\citep{BMuller14}), we mainly focus on detailed comparisons between our pseudo-1D neutrino profiles computed in the 3D Cartesian coordinates
and previous 1D results to check the validity of our new code
 in the early postbounce phase.

This paper is organized as follows: In section \ref{sec:Formalism}, after we 
shortly introduce the formulation of the GR transport scheme, we describe
 the governing equations of hydrodynamics and neutrino transport in detail.
Some practical implementation schemes how to satisfy important 
conservative quantities such as lepton number, energy, and momentum
 are given in section \ref{sec:Numerical Method}.
  The main results and detailed comparisons with previous studies are presented
  in section \ref{sec:Core Collapse of a 15M star}.
Note that geometrized unit system is used in sections \ref{sec:Formalism} and \ref{sec:Numerical Method},
i.e. the speed of light, the gravitational constant and the Planck constant are set to unity: $c= G = h=1$,
and cgs unit is used in section \ref{sec:Core Collapse of a 15M star}.
Greek indices run from 0 to 3 and Latin indices do from 1 to 3.

\section{Formalism}
\label{sec:Formalism}
This section starts with a brief summary of the 
basic equations and the numerical schemes of GR radiation hydrodynamics.

Following our previous work \citep{KurodaT12}, our code 
  consists of the following three parts, where
 the evolution equations of metric, hydrodynamics, 
and neutrino radiation are solved, respectively.
Each of them is solved in an operator-splitting manner, but the system evolves self-consistently
as a whole satisfying the Hamiltonian and momentum constraints.
Regarding the metric evolution, the spatial metric $\gamma_{ij}$ 
(in the standard (3+1) form: $ds^2=-\alpha^2dt^2+\gamma_{ij}(dx^i+\beta^idt)(dx^j+\beta^jdt),$ with $\alpha$ and $\beta^i$ being the lapse and shift,
respectively)
and its extrinsic curvature $K_{ij}$ are evolved using the BSSN 
formulation \citep{Shibata95,Baumgarte99} (see 
\cite{KurodaT12,KurodaT14} for more details).

\subsection{Radiation Hydrodynamics}
\label{sec:Radiation Hydrodynamics}
Major differences compared to our previous code 
\citep{KurodaT12} are, on the one hand we evolved energy integrated 
(``{\it gray}'') neutrino radiation field with an approximate
description of neutrino-matter interaction based on the neutrino 
leakage scheme, 
on the other hand we now solve the spectral neutrino transport 
 where the source terms are treated self-consistently following 
 a standard procedure of the M1 closure scheme \citep{Shibata11}.
For convenience, we briefly summarize 
 the basic equations of our newly developed code in the following
 (see \citet{Shibata11} and \citet{cardall13} for the complete derivation).

The total stress-energy tensor $T^{\alpha\beta}_{\rm (total)}$ is expressed as
\begin{equation} T_{\rm (total)}^{\alpha\beta} = 
T_{\rm (fluid)}^{\alpha\beta} +\int d\varepsilon \sum_{\nu\in\nu_e,\bar\nu_e,\nu_x}T_{(\nu,\varepsilon)}^{\alpha\beta},
\label{TotalSETensor}
\end{equation}
where $T_{\rm (fluid)}^{\alpha\beta}$ and $T_{(\nu,\varepsilon)}^{\alpha\beta}$ is 
the stress-energy tensor of fluid and energy-dependent neutrino radiation 
field, respectively.
Note in the above equation, summation is taken for 
all species of neutrinos ($\nu_e,\bar\nu_e,\nu_x$) with $\nu_x$ 
representing heavy-lepton neutrinos (i.e. 
$\nu_{\mu}, \nu_{\tau}$ and their anti-particles), 
and $\varepsilon$ represents neutrino energy measured in the 
comoving frame with the fluid.
For simplicity, the neutrino flavor index $\nu$ is omitted below.

With introducing radiation energy ($E_{(\varepsilon)}$),
radiation flux ($F^{\mu}_{(\varepsilon)}$) and radiation pressure ($P^{\mu\nu}_{(\varepsilon)}$),
measured by an Eulerian observer or
($J_{(\varepsilon)}$, $H^{\mu}_{(\varepsilon)}$ and $L^{\mu\nu}_{(\varepsilon)}$)
measured in a comoving frame, $T_{(\varepsilon)}^{\mu\nu}$ can be written
 in covariant form as
{\setlength\arraycolsep{2pt}
\begin{eqnarray}
T_{(\varepsilon)}^{\mu\nu}&=& E_{(\varepsilon)} n^\mu n^\nu+ F_{(\varepsilon)}^\mu n^\nu+
F_{(\varepsilon)}^\nu n^\mu +
P_{(\varepsilon)}^{\mu\nu},
\label{t_lab}\\
&=& J_{(\varepsilon)} u^\mu u^\nu+ H_{(\varepsilon)}^\mu u^\nu+
H_{(\varepsilon)}^\nu u^\mu +
L_{(\varepsilon)}^{\mu\nu}.
\label{t_com}
\end{eqnarray}}
In the above equations, $n^{\mu}=(1/\alpha,-\beta^k/\alpha)$ is a unit vector orthogonal to the spacelike hypersurface
and $u^\mu$ is the four velocity of fluid.
In the truncated moment formalism \citep{Thorne81,Shibata11},
one evolves radiation energy ($E_{(\varepsilon)}$) and radiation flux ($F^{\alpha}_{(\varepsilon)}$)
in a conservative form and $P^{\mu\nu}_{(\varepsilon)}$ is determined by an analytic closure relation (e.g., Eq.(\ref{neu_p})).
The evolution equations for $E_{(\varepsilon)}$ and $F^{\alpha}_{(\varepsilon)}$ 
are given by
\begin{eqnarray}
\partial_t \sqrt{\gamma}E_{(\varepsilon)}+\partial_i \sqrt{\gamma}(\alpha F_{(\varepsilon)}^i-\beta^i E_{(\varepsilon)})
+\sqrt{\gamma}\alpha \partial_\varepsilon \bigl(\varepsilon \tilde M^\mu_{(\varepsilon)} n_\mu\bigr)  =\nonumber \\
\sqrt{\gamma}(\alpha P^{ij}_{(\varepsilon)}K_{ij}-F_{(\varepsilon)}^i\partial_i \alpha-\alpha S_{(\varepsilon)}^\mu n_\mu),
\label{eq:rad1}
\end{eqnarray}
and 
\begin{eqnarray}
\partial_t \sqrt{\gamma}{F_{(\varepsilon)}}_i+\partial_j \sqrt{\gamma}(\alpha 
{P_{(\varepsilon)}}_i^j-\beta^j {F_{(\varepsilon)}}_i)
-\sqrt{\gamma}\alpha \partial_\varepsilon\bigl(\varepsilon \tilde M^\mu_{(\varepsilon)} \gamma_{i\mu}\bigr)=\nonumber \\
\sqrt{\gamma}[-E_{(\varepsilon)}\partial_i\alpha +{F_{(\varepsilon)}}_j\partial_i \beta^j+(\alpha/2) 
P_{(\varepsilon)}^{jk}\partial_i \gamma_{jk}+\alpha S^\mu_{(\varepsilon)} \gamma_{i\mu}],
\label{eq:rad2}
\end{eqnarray}
respectively.
Here $\gamma$ is the determinant of the three metric $\gamma\equiv{\rm det}(\gamma_{ij})$ and $S^{\mu}_{(\varepsilon)}$ is the source term for neutrino matter interactions
(see appendix \ref{sec:Neutrino Matter Interaction Terms} for the currently implemented processes).
$\tilde M^\mu_{(\varepsilon)}$ is defined by $\tilde M^\mu_{(\varepsilon)}\equiv M^{\mu\alpha\beta}_{(\varepsilon)}
\nabla_\beta u_\alpha$ where $M^{\mu\alpha\beta}_{(\varepsilon)}$ denotes the third 
rank moment of neutrino distribution function
(\citet{Shibata11} for the explicit expression).

By adopting the M1 closure scheme, the radiation pressure can be expressed as
\begin{eqnarray}
P_{(\varepsilon)}^{ij}=\frac{3\chi_{(\varepsilon)}-1}{2}P^{ij}_{\rm thin(\varepsilon) }
+\frac{3(1-\chi_{(\varepsilon)})}{2}P^{ij}_{\rm thick(\varepsilon)},
\label{neu_p}
\end{eqnarray}
where $\chi_{(\varepsilon)}$ represents the variable Eddington factor,
$P^{ij}_{\rm thin(\varepsilon)}$ and $P^{ij}_{\rm thick(\varepsilon)}$ corresponds 
to the radiation pressure in the optically thin and thick limit, respectively.
They are written in terms
of $J_{(\varepsilon)}$ and $H^\mu_{(\varepsilon)}$ \citep{Shibata11}.
Following \citet{minerbo78,Cernohorsky94} and \citet{Obergaulinger11}, 
we take the variable Eddington factor $\chi_{(\varepsilon)}$ as
\begin{eqnarray}
\label{eq:Closure_Chi}
\chi_{(\varepsilon)}&=&\frac{5+6\bar{F}^2_{(\varepsilon)}-2\bar{F}^3_{(\varepsilon)}+6\bar{F}^4_{(\varepsilon)}}{15},
\end{eqnarray}
where
\begin{eqnarray}
\label{eq:F^2}
\bar{F}^2_{(\varepsilon)}&\equiv&\frac{h_{\mu\nu}H^\mu_{(\varepsilon)} H^\nu_{(\varepsilon)}}{J^2_{(\varepsilon)}}.
\end{eqnarray}
In Eq.(\ref{eq:F^2}), $h_{\mu\nu}\equiv g_{\mu\nu}+u_\mu u_\nu$ is the projection operator.
As we will discuss later, by the definition of $\bar{F}_{(\varepsilon)}$ in
 Eq.(\ref{eq:F^2}), one can appropriately reproduce several important 
 neutrino behaviours, for example, neutrino trapping 
in the rapidly collapsing opaque core.
We iteratively solve the simultaneous equations (\ref{eq:Closure_Chi}-\ref{eq:F^2}) to find
 the converged solution of $\chi_{(\varepsilon)}$.

The hydrodynamic equations are written in a conservative form as,
{\setlength\arraycolsep{2pt}
\begin{eqnarray}
\label{eq:GRmass}
\partial_t \rho_{\ast}&+&\partial_i(\rho_\ast v^i)=0,\\
\label{eq:GRmomentum}
\partial_t \sqrt{\gamma} S_i&+&\partial_j \sqrt{\gamma}( S_i v^j+\alpha P\delta_i^j)=\nonumber \\
&&-\sqrt{\gamma}\biggl[ S_0\partial_i \alpha- S_k\partial_i \beta^k-2\alpha S_k^k\partial_i \phi 
+\alpha e^{-4\phi} ({S}_{jk}-P \gamma_{jk}) \partial_i 
\tilde{\gamma}^{jk}/2+\alpha \int d\varepsilon S_{(\varepsilon)}^\mu \gamma_{i\mu} \biggr],\nonumber\\ \\
\label{eq:GRenergy}
\partial_t \sqrt{\gamma} \tau&+&\partial_i \sqrt{\gamma}(\tau v^i+P(v^i+\beta^i))=\nonumber \\
&&\sqrt{\gamma}\biggl[ \alpha K S_k^k /3+\alpha e^{-4\phi} ({S}_{ij}-P \gamma_{ij})\tilde{A^{ij}}  
- S_iD^i\alpha+\alpha \int d\varepsilon S_{(\varepsilon)}^\mu n_\mu \biggr],\\
\label{eq:GRYe}
\partial_t (\rho_\ast Y_e)&+&\partial_i (\rho_\ast Y_e v^i)=\sqrt{\gamma}\alpha m_{\rm u}\int \frac{d\varepsilon}{\varepsilon}
(S_{(\nu_e,\varepsilon)}^\mu-S_{(\bar\nu_e,\varepsilon)}^\mu) u_\mu,
\end{eqnarray}}
where $\rho_\ast=\rho \sqrt{\gamma}W$, $S_i=\rho hW u_i $, $S_{ij}=\rho h u_i u_j+P\gamma_{ij}$, $S_k^k=\gamma^{ij}S_{ij}$,
$S_0=\rho h W^2-P$ and $\phi={\rm log}(\gamma)/12$.
$\rho$ is the rest mass density, $W$ is the Lorentz factor, $h= 1+e+P/\rho$ is the specific enthalpy,
$v^i=u^i/u^t$, $\tau= S_0-\rho W$, $Y_e\equiv n_e/n_b$ is the electron fraction ($n_X$ is the number density of $X$),
$e$ and $P$ are the specific internal energy and pressure of matter, respectively and $m_{\rm u}$ is the atomic mass unit.
$P(\rho,s,Y_e)$ and $e(\rho,s,Y_e)$ are given by an equation of state (EOS) 
with $s$ denoting the entropy per baryon. We employ an EOS by 
Lattimer \& Swesty (1991) (LS220, see section section \ref{sec:Numerical Setups} for more detail).
In the right hand side of Eq.(\ref{eq:GRenergy}), $D^i$ represents the covariant derivative with respect to the three metric $\gamma_{ij}$.

\subsection{Conservation of Energy and Lepton Number}
\label{sec:Conservation of Energy and Lepton Number}
As explained in Section \ref{sec:Radiation Hydrodynamics}, 
 the formalism of our code that treats the radiation-hydrodynamics 
 equations in a conservative form is suitable
 to satisfy the energy conservation of the total 
system (neutrinos and matters, see also 
\citet{KurodaT10,KurodaT12} for more details). 
Let us first show how the energy conservation law is obtained in our code.

To focus only on the energy exchange between the matter and neutrino radiation 
field, we omit the gravitational source term in the following discussion.
Then, the equations of energy conservation of matter and neutrinos
(e.g., Eqns. (\ref{eq:rad1}) and (\ref{eq:GRenergy})) become
{\setlength\arraycolsep{2pt}
\begin{eqnarray}
\partial_t \sqrt{\gamma} \tau+\partial_i \sqrt{\gamma} (\tau v^i+P(v^i+\beta^i))&\ \ \ \ \ \ \ \ \ \ \ \ \ \ \ \ \ \ \ \ \ \ \ &
 =\int d\varepsilon \sqrt{\gamma}\alpha S_{(\varepsilon)}^\mu n_\mu,\\
\partial_t \sqrt{\gamma}E_{(\varepsilon)}+\partial_i \sqrt{\gamma}(\alpha F_{(\varepsilon)}^i-\beta^i E_{(\varepsilon)})
&+\sqrt{\gamma}\alpha \partial_\varepsilon \bigl(\varepsilon \tilde M^\mu_{(\varepsilon)} n_\mu\bigr)&  =-\sqrt{\gamma}\alpha S_{(\varepsilon)}^\mu n_\mu.
\end{eqnarray}}
From the above two equations, one can readily see that 
the total energy (sum of matter and neutrinos) 
contained in the computational domain,
$E_{\rm \nu m}\equiv\int dx^3 \sqrt{\gamma}(\tau+\int d\varepsilon E_{(\varepsilon)})$,
is conserved in our basic equations as long as there is no net energy flux
through the numerical and momentum space boundaries
(i.e. $\int d\varepsilon \left[\partial_\varepsilon \bigl(\varepsilon \tilde M^\mu_{(\varepsilon)} n_\mu\bigr)\right]=0$).

The lepton number conservation needs to be satisfied with good accuracy
 because it determines the PNS mass and the postbounce supernova dynamics.
We here explain how we treat it in our code.
As for the electron and neutrino number conservation, the 
basic equations are given by
{\setlength\arraycolsep{2pt}
\begin{eqnarray}
\label{eq:LeptonYe}
\partial_t \biggl(\frac{\rho_\ast Y_e}{ m_{\rm u}}\biggr)&+&\partial_i \biggl(\frac{\rho_\ast Y_e v^i}{m_{\rm u}}\biggr)
\ \ \ \ \ \ \ \ \ \ \ \ \ \ \ \ \ \ \ \ \ \ \ \ \ \ \ \ =\sqrt{\gamma}\alpha \int \frac{d\varepsilon}{\varepsilon}
(S_{(\nu_e,\varepsilon)}^\mu-S_{(\bar\nu_e,\varepsilon)}^\mu) u_\mu, \\
\label{eq:LeptonNeutrino}
\partial_t ( q_{(\varepsilon)}^0\sqrt{\gamma}\alpha)& +&\partial_i ( q_{(\varepsilon)}^i \sqrt{\gamma}\alpha)
 -\sqrt{\gamma}\alpha\partial_\varepsilon \left( \varepsilon q_{(\varepsilon)}^{\alpha \beta} \right)
 \nabla_\beta u_\alpha=-\frac{\sqrt{\gamma}\alpha}{\varepsilon}S_{(\varepsilon)}^\mu u_\mu,
\end{eqnarray}}
where
{\setlength\arraycolsep{2pt}
\begin{eqnarray}
\label{eq:Q^A}
q_{(\varepsilon)}^\alpha &\equiv&-\varepsilon^{-1} T_{(\varepsilon)}^{\alpha\beta}u_\beta=
  ({\mathcal J}_{(\varepsilon)} u^\alpha+{\mathcal H}_{(\varepsilon)}^\alpha), \\
\label{eq:Q^AB}
q_{(\varepsilon)}^{\alpha \beta} &\equiv&\varepsilon^{-1} T_{(\varepsilon)}^{\gamma\beta}h^\alpha_\gamma=
({\mathcal H}_{(\varepsilon)}^\alpha u^\beta+{\mathcal L}_{(\varepsilon)}^{\alpha \beta}),
\end{eqnarray}}
with $ (\mathcal J, \mathcal H^\alpha,\mathcal L^{\alpha\beta})\equiv \varepsilon^{-1}(J,H^\alpha,L^{\alpha\beta})$.
The conservation equation for neutrino number (\ref{eq:LeptonNeutrino}) corresponds to Eq.(3.22) (divided by $\varepsilon$)
 in \citet{Shibata11}.
Since the neutrino number density measured by an Eulerian observer is expressed as
\begin{eqnarray}
\label{eq:Nnu_Eulerian}
n_{\nu,{\rm Euler}}=\int d\varepsilon q_{(\nu,\varepsilon)}^0 \sqrt{\gamma}\alpha,
\end{eqnarray}
the equation of total lepton number conservation becomes
\begin{eqnarray}
\label{eq:GRYl}
\partial_t (\rho_\ast Y_l)+
\partial_i \biggl(\rho_\ast Y_e v^i+ m_{\rm u}\sqrt{\gamma}\alpha \int d\varepsilon \Bigl[q_{(\nu_e,\varepsilon)}^i-q_{(\bar\nu_e,\varepsilon)}^i\Bigr] \biggr)=0.
\end{eqnarray}
Here the total lepton fraction $Y_l$ is defined by
{\setlength\arraycolsep{2pt}
\begin{eqnarray}
\label{eq:LeptonFraction}
Y_l&=&Y_e+Y_{\nu_e}-Y_{\bar\nu_e}\nonumber \\
&=&Y_e + \frac{m_{\rm u}}{\rho u^t}\int d\varepsilon (q_{(\nu_e,\varepsilon)}^0-q_{(\bar\nu_e,\varepsilon)}^0).
\end{eqnarray}}
From Eq.(\ref{eq:GRYl}), one can readily see that 
the total lepton number is conserved irrespective of the included neutrino matter interaction processes
in case that there is no net flux through the numerical and energy space boundaries.
The distribution of $Y_l$ into the each component
 (e.g., $Y_e, Y_{\nu_e}, Y_{{\bar \nu}_e}$) is determined by the details of the 
 implemented microphysics, which should be checked carefully and will be reported 
in Sec. \ref{sec:Core Collapse of a 15M star}.

\subsubsection{Neutrino Number Transport in the Diffusion Limit}
\label{sec:Neutrino Number Transport in Diffusion Limit}
In the collapsing iron core, it is well known that 
the central core becomes opaque to neutrinos
due mainly to scattering off heavy nuclei when the central density exceeds $\sim10^{11-12}$ g cm$^{-3}$ \citep{Sato75}
and neutrinos are {\it trapped} with matter afterward.
In the diffusion limit at large neutrino opacities, 
 the trapped
 neutrinos move with the matter velocity $v_i$
for an Eulerian observer. Thus their advection equation can be 
described with the same form of $Y_e$ as
\begin{eqnarray}
\label{eq:NnuSlowMotionlimit}
\partial_t \rho_\ast Y_\nu+\partial_i \rho_\ast Y_\nu v^i= \,{(\rm Source}\,{\rm terms)}.
\end{eqnarray}
Because the source terms in the above equation amount equal to the negative value
of the source terms in the electron number conservation equation,
the core lepton number is conserved 
in a good accuracy until it gradually decreases by diffusion 
 in the PNS cooling phase.
Since the central core mass depends on the core lepton number,
the CCSN simulation should be able to capture this important phenomena 
appropriately.
 
In our formalism, however, this is not a trivial problem because we solve the energy-momentum conservation equations
(\ref{eq:rad1})-(\ref{eq:rad2}) and not the neutrino number conservation equation (\ref{eq:LeptonNeutrino}).
In this section we check whether our basic equations can
reproduce the neutrino diffusion limit adequately, i.e., they reach asymptotically to Eq.(\ref{eq:NnuSlowMotionlimit}).

For simplicity, we assume in the following that the spacetime is 
 flat and the typical velocity of the matter field ($v$) is much smaller 
than the speed of light (slow motion limit; neglecting terms higher 
 than the second order with respect to $(v/c)$).

Let us first check whether Eq.(\ref{eq:LeptonNeutrino})
 can satisfy the local neutrino number conservation 
in the trapped region.
In this limit, neutrino number density at each energy bin $q^0$ and its flux $q^i$ approach
{\setlength\arraycolsep{2pt}
\begin{eqnarray}
\label{eq:SlowMotionlimitCom0}
 q^0&=&\mathcal Ju^t+\mathcal H^0\sim \mathcal J+\frac{\mathcal H_i v^i}{W}\sim \mathcal J,\\
\label{eq:SlowMotionlimitCom1}
q^i&=&\mathcal Ju^i+\mathcal H^i\sim \mathcal Jv^i+\mathcal H^i\sim q^0(v^i+\mathcal H^i/\mathcal J).
\end{eqnarray}}
From these relations, it is obvious that the neutrino number density at each energy bin $q^0$ is transferred
with the matter velocity $v^i$ plus the diffusion velocity $\mathcal H^i/\mathcal J$ and the equation of the 
total lepton number (Eq. (\ref{eq:GRYl})) in the slow motion limit becomes
{\setlength\arraycolsep{2pt}
\begin{eqnarray}
\label{eq:GRYlDiffusionLimit}
\partial_t (\rho_\ast Y_l)+
\partial_i \biggl(\rho_\ast Y_e v^i+ m_{\rm u}\sqrt{\gamma}\alpha \int d\varepsilon \Bigl[q_{(\nu_e,\varepsilon)}^0-q_{(\bar\nu_e,\varepsilon)}^0\Bigr] v^i\biggr)&=&
\partial_t (\rho_\ast Y_l)+\partial_i (\rho_\ast Y_l v^i)\nonumber \\&=&0,
\end{eqnarray}}
 demonstrating that Eq.(\ref{eq:GRYl}) satisfies 
 the local lepton number conservation in the trapped region. 

Next, we take the diffusion limit of Eq.(\ref{eq:rad1}).
In this limit, $H^i/J$ should approach 0 (i.e., the 
radiation flux ($H^i$) in the comoving frame vanishes). From this,
 the following relation can be derived,
\begin{eqnarray}
\label{eq:SlowMotionlimitH}
H^i\sim-Ev^i+F^i-P^{ij}v_j\sim 0 \ \ \ \ \ \ \Longrightarrow \ \ \ \ \ F^i\sim\frac{4}{3}Ev^i,
\end{eqnarray}
where we take a simple closure relation $P^{ij}=\frac{E}{3}\delta^{ij}$ 
in the diffusion limit.
Inserting this into the left hand side of Eq.(\ref{eq:rad1}), 
 one can get
{\setlength\arraycolsep{2pt}
\begin{eqnarray}
\label{eq:SlowMotionlimit:eq:rad1_1}
\partial_t E+\partial_i F^i+\partial_\varepsilon ( - \varepsilon P^{ij}\partial_i v_j )&\sim&
\partial_t E+\partial_i \biggl(\frac{4}{3}Ev^i\biggr)-\frac{E}{3}\delta^{ij} \partial_i v_j -\varepsilon\left(\partial_\varepsilon P^{ij}\right)\partial_i v_j
 \\
\label{eq:SlowMotionlimit:eq:rad1_2}
&\sim&\partial_t E+\partial_i \biggl(\frac{4}{3}Ev^i\biggr)
-\partial_i \biggl(\frac{1}{3}Ev^i\biggr) -\varepsilon\left(\partial_\varepsilon P^{ij}\right)\partial_i v_j \\  
\label{eq:SlowMotionlimit:eq:rad1_3}
&\sim&\partial_t E+\partial_i (Ev^i) -\varepsilon\left(\partial_\varepsilon P^{ij}\right)\partial_i v_j.
\end{eqnarray}}
Moving from the right hand side of Eq.(\ref{eq:SlowMotionlimit:eq:rad1_1}) to that of Eq.(\ref{eq:SlowMotionlimit:eq:rad1_2}),
 we assumed that $E$ has almost no spatial gradient 
well below the neutrino spheres (at high opacities) in the prebounce core.
This is satisfied quite well as shown in the literature
(\cite{Bruenn85}), in which a nearly flat
$Y_{\nu_e}$ profile is shown in their standard models within the central core with mass $\sim$0.5-0.6M$_\odot$
after the central density exceeds $\sim5\times10^{13}$ g cm$^{-3}$.
Note that, in Eq. (\ref{eq:SlowMotionlimit:eq:rad1_2}), the third term $-\partial_i (Ev^i/3)$
which is originally a part of the advection terms in the energy space
$\partial_\varepsilon (-\varepsilon P^{ij}\partial_i v_j )$ balances
with a part of the spatial advection term $\partial_i (4Ev^i/3)$ and they lead to the second term in Eq. (\ref{eq:SlowMotionlimit:eq:rad1_3}).
From this, one can clearly see how the {\it apparent} advection speed of $E$ at each energy bin $\varepsilon$
approaches the matter velocity $v^i$ in the diffusion limit.
Then, assuming that the neutrino number density at each energy bin can be approximately
 expressed as
\begin{eqnarray}
\label{eq:SlowMotionlimitJ}
q^0\sim \mathcal J\sim \mathcal E-2\mathcal F^iv_i\sim \mathcal E,
\end{eqnarray}
 in the slow motion limit, when we divide Eq. (\ref{eq:SlowMotionlimit:eq:rad1_3}) by $\varepsilon$ and integrate it in energy space, 
it can be summarized as
{\setlength\arraycolsep{2pt}
\begin{eqnarray}
\label{eq:SlowMotionlimit2:eq:rad1}
\int d\varepsilon \left[\partial_t q^0+\partial_i (q^0v^i) -\left(\partial_\varepsilon P^{ij}\right)\partial_i v_j\right]=\partial_t n_\nu +\partial_i n_\nu v^i. 
\end{eqnarray}}
From this, one can clearly see that the neutrino number density $n_\nu(=\rho Y_\nu)$ 
is transported with the same matter velocity $v^i$ in the diffusion limit.

\section{Numerical Method}
\label{sec:Numerical Method}
In this section we describe how to evolve the radiation-hydrodynamics 
variables\footnote{We omit our numerical method to evolve 
 the space-time variables that is essentially the same as in
 \citet{KurodaT12}.}. As we explained in the previous section, we solve Eqs. (\ref{eq:rad1}), (\ref{eq:rad2}) and (\ref{eq:GRmass})-(\ref{eq:GRYe})
as our basic equations which are collectively expressed as
\begin{eqnarray}
\label{eq:ExpImpEq}
\partial_t {\bf Q}+{\bf S_{\rm adv,s}}+{\bf S_{\rm adv,e}}+{\bf S_{\rm grv}}+{\bf S_{\rm \nu m}}=0,
\end{eqnarray}
where $\bf Q$ denotes conservative variables
\begin{eqnarray}
{\bf Q}=\left[
\begin{array}{c}
\rho_\ast  \\
\sqrt{\gamma} S_i \\
\sqrt{\gamma} \tau \\
\rho_\ast Y_e \\
\sqrt{\gamma} E_{(\nu,\varepsilon)} \\
\sqrt{\gamma} {F_{(\nu,\varepsilon)}}_i \\
 \end{array}
\right].
\end{eqnarray}
In Eq.(\ref{eq:ExpImpEq}), $\bf S_{\rm adv,s}$, $\bf S_{\rm adv,e}$, $\bf S_{\rm grv}$ and $\bf S_{\rm \nu m}$ denote
advection term in space, advection term in momentum space, gravitational source and neutrino-matter interaction term,
respectively.
Throughout this paper other than Appendix \ref{sec:Comparison in 1D Spherical Coordinate},
we divide this equation into the following two parts which are expressed in the finite difference expression as,
\begin{eqnarray}
\label{eq:ExpEq}
\frac{{\bf Q}^\ast-{\bf Q}^n}{\Delta t}+{\bf S_{\rm adv,s}}^n+{\bf S_{\rm grv}}^n=0,
\end{eqnarray}
for the explicit part and
\begin{eqnarray}
\label{eq:ImpEq}
\frac{{\bf Q}^{n+1}-{\bf Q}^\ast}{\Delta t}+{\bf S_{\rm adv,e}}^{n+1}+{\bf S_{\rm \nu m}}^{n+1}=0,
\end{eqnarray}
for the implicit part, respectively.
In Eqs.(\ref{eq:ExpEq})-(\ref{eq:ImpEq}),
$\Delta t$ is the time step size between $n$-th and $n+1$-th time steps and the upper indices ``$n$'' represents $n$-th time step.
Variables with ``$\ast$'' denote the time updated variables 
 during an operator-splitting procedure.

 In Eq.(\ref{eq:ExpEq}), ${\bf S_{\rm adv,s}}^n$ and ${\bf S_{\rm grv}}^n$ 
 represent the terms with respect to advection in space and gravitational fields at $n$-th time step, both of which 
are added first in an explicit manner to obtain conservative variables at a 
middle time step ${\bf Q}^\ast$.
Next in Eq. (\ref{eq:ImpEq}), the rest of terms, advection in energy space 
(${\bf S_{\rm adv,e}}$) and neutrino-matter interaction terms (${\bf S_{\rm \nu m}}$) at $(n+1)$-th time step are added to ${\bf Q}^\ast$  in order 
 to find the converged solution of ${\bf Q}^{n+1}$ by an iterative method.
The reason why we separate source terms into the 
two parts, explicit and implicit ones, is that the typical time step size, $\Delta t\sim 4\times10^{-7}$s,
which is determined by the speed of light $c$,
typical minimum grid width in our calculation $\Delta x\sim500$m and the Courant-Friedrichs-Lewy number CFL, e.g., 0.25,
is sufficiently short for the advection term in space ${\bf S_{\rm adv,s}}$
and the gravitational source term ${\bf S_{\rm grv}}$
as well as for all the geometrical variables by an explicit update.
However it is too long to follow, e.g., the weak-interaction term,
which has significantly shorter time scale ($\lesssim10^{-9}$s).
We thus need to treat these terms in an implicit way through
 Eq.(\ref{eq:ImpEq}) to ensure a numerical convergence and stability.

Here, we should comment on the treatment of the advection terms in energy space ${\bf S_{\rm adv,e}}$.
As we mentioned above, we solve them implicitly in time as Eq.(\ref{eq:ImpEq}) in our main results.
It means that there is a time gap $\Delta t$ between moment of evaluation for advection terms
in real space ${\bf S_{\rm adv,s}}^n$ and that in energy space ${\bf S_{\rm adv,e}}^{n+1}$.
Although we can generally obtain consistent results with previous studies regardless of the time gap,
it is noted that the treatment of the energy advection either by the implicit or explicit scheme
leads to visible changes in the postbounce features.
We will discuss this point further in Appendix \ref{sec:Comparison in 1D Spherical Coordinate}.

In the following sections, we are going to describe how to
 evaluate the advection terms in space (Sec. \ref{sec:Advection in Space})
 and in energy space (Sec. \ref{sec:Advection in Energy Space}), and 
then move on to describe the implicit time update in Sec. \ref{sec:Implicit Time Update}.

\subsection{Advection in Space}
\label{sec:Advection in Space}
We employed a standard high-resolution-shock-capturing scheme and utilize the  
HLL (Harten-Lax-van Leer) scheme \citep{Harten83} to evaluate the numerical fluxes in space \citep{KurodaT10}.
As for the fastest and slowest characteristic wave speeds of the radiation field system
(Eqs.(\ref{eq:rad1}) and (\ref{eq:rad2})),
we again use the same definition as in \citet{KurodaT12} (see also \cite{Shibata11})
and connect $\lambda_{\rm rad,thin}$ and $\lambda_{\rm rad,thick}$ smoothly via the variable Eddington factor
$\chi$ as
\begin{eqnarray}
\label{eq:LambdaRad}
\lambda_{\rm rad}=\frac{3\chi-1}{2}\lambda_{\rm rad,thin}+\frac{3(1-\chi)}{2}
\lambda_{\rm rad,thick},
\end{eqnarray}
where $\lambda_{\rm rad,thin}$ and $\lambda_{\rm rad,thick}$ are the wave speed in the optically thin and thick limit,
respectively.

To enforce the numerical flux of the radiation field in the opaque
region asymptotically approach to the diffusion limit,
 we evaluate the energy flux ($F_{\rm hll}^0$) and the momentum flux 
($F_{\rm hll}^i$) as
\begin{eqnarray}
\label{eq:AuditEnergyFlux}
F_{\rm hll}^0=\frac{\tilde\lambda_+F_L^0-\tilde\lambda_-F_R^0+\epsilon\tilde\lambda_-\tilde\lambda_+(Q_R^0-Q_L^0)}
{\tilde\lambda_+-\tilde\lambda_-},
\end{eqnarray}
 and
\begin{eqnarray}
\label{eq:AuditMomentumFlux}
F_{\rm hll}^i=\frac{\epsilon^2(\tilde\lambda_+F_L^i-\tilde\lambda_-F_R^i)
+\epsilon\tilde\lambda_-\tilde\lambda_+(Q_R^i-Q_L^i)}{\tilde\lambda_+-\tilde\lambda_-}
+(1-\epsilon^2)\frac{F_L^i+F_R^i}{2},
\end{eqnarray}
 respectively \citep{Audit02,O'Connor13}.
Here, $Q_{L/R}^\alpha$ and $F_{L/R}^\alpha$ are the conservative variables and their corresponding fluxes, respectively,
with $L/R$ denoting the left/right states for the Riemann problem.
All the radiation-hydrodynamical variables are defined at 
the cell center. For those cell centered (primitive) variables, we use a monotonized central reconstruction,
which has second order accuracy in space,
and obtain left/right states at the cell surface \citep[see][for more detailed explanation]{KurodaT10}.
After the reconstruction, all the characteristic wave speeds of 
the matter and radiation fields are evaluated.

$\epsilon$ is a modification parameter to fit the numerical flux to 
 the diffusion limit, which we take 
 as \begin{eqnarray}
\label{eq:AuditEpsilon}
\epsilon={\rm min}\biggl(1,\frac{1}{\kappa \Delta x}\biggr),
\end{eqnarray}
where $\kappa$ is the total opacity and $\Delta x$ is the grid width
 \citep{Audit02,O'Connor13}.


\subsection{Advection in Energy Space}
\label{sec:Advection in Energy Space}
Regarding the advection term in energy space in all conservation equations 
(Eqns.(\ref{eq:rad1}), (\ref{eq:rad2}) and (\ref{eq:LeptonNeutrino})),
we define the advection fluxes at the interface of the energy bin as the same as in \citet{BMuller10}
so that all energy integrated advection term will vanish.
This can be achieved since both terms $\tilde M^{\alpha}$, appearing in energy and momentum conservation equations,
and $\varepsilon q^{\alpha\beta}$ in number conservation equation are expressed in terms of
linear combinations of radiation momenta, $J$, $H^\alpha$, $L^{\alpha\beta}...$ , 
and we therefore can define their cell surfaced values with an appropriate weighting function
to suppress violation of all conservations simultaneously.

For all orders of the radiation momenta $X\in{\{J, H^\alpha, L^{\alpha\beta}...\}}$, the following conditions need to be satisfied for number conservation,
\begin{eqnarray}
\label{eq:EnergyShiftNumber=0}
\int d\varepsilon\partial_\varepsilon X=0\ \ \  \Longrightarrow\ \ \  \sum_i \Delta\varepsilon_i \frac{X_{i+1/2}-X_{i-1/2}}{\Delta\varepsilon_i}=0,
\end{eqnarray}
and for energy and momentum conservations,
\begin{eqnarray}
\label{eq:EnergyShiftEnergy=0}
\int d\varepsilon\partial_\varepsilon \left(\varepsilon X\right)=\int d\varepsilon\left(\varepsilon\partial_\varepsilon X+X \right)=0
\ \ \  \Longrightarrow\ \ \  \sum_i \Delta\varepsilon_i \left[ 
\varepsilon_i \frac{X_{i+1/2}-X_{i-1/2}}{\Delta\varepsilon_i}+X_i
\right]=0, 
\end{eqnarray}
 respectively.
The RHS of the above equations represent the finite difference 
expressions with 
$i$ and $i+1/2$ denoting $i$-th energy bin and the interface
between $i$- and $(i+1)$-th energy bins, respectively.
$\Delta \varepsilon_i$ is energy grid width $\Delta \varepsilon_i=\varepsilon_{i+1/2}-\varepsilon_{i-1/2}$.
It is straightforwad to show that Eq.(\ref{eq:EnergyShiftNumber=0}) can be 
automatically satisfied for any cell-surfaced quantities as long as 
they vanish at the outer boundary in the energy space.
By introducing a definition 
$X_{i+1/2}\equiv X_i^{\rm L}+X_{i+1}^{\rm R}$, the RHS of Eq.(\ref{eq:EnergyShiftEnergy=0}) can be
 expressed as,
\begin{eqnarray}
\sum_i \left[ 
-(\varepsilon_{i+1}- \varepsilon_i)X_i^{\rm L}  - (\varepsilon_i- \varepsilon_{i-1})X_i^{\rm R}+\Delta\varepsilon_iX_i
\right]=0.
\end{eqnarray}
As in \citet{BMuller10}, we can get the solution of Eq.(42) as
\begin{eqnarray}
X^{\rm L}_{i}&\equiv& \frac{\Delta \varepsilon_i}{\varepsilon_{i+1}-\varepsilon_i } X_{i}\xi_i,\\
X^{\rm R}_{i}&\equiv& \frac{\Delta \varepsilon_i}{\varepsilon_i -\varepsilon_{i-1}} X_{i}(1-\xi_i),
\end{eqnarray}
where $\xi_i$ is a weighting function and is expressed as
\begin{eqnarray}
\xi_i&\equiv& \frac{f_{i+1/2}^\sigma}{f_{i-1/2}^\sigma+f_{i+1/2}^\sigma}.
\end{eqnarray}
In this study, we used a ``Harmonic'' interpolation ($\sigma=1$) for the energy 
density ($f_{i+1/2}^\sigma$) as
\begin{eqnarray}
f^\sigma_{i+1/2}\equiv \Biggl[ 
\biggl( \frac{E_{(\varepsilon_{i})}}{\varepsilon_{i}^3}  \biggr)^{1-r_{i+1/2}}
\biggl( \frac{E_{(\varepsilon_{i+1})}}{\varepsilon_{i+1}^3}  \biggr)^{r_{i+1/2}}
 \Biggr]^\sigma,
\end{eqnarray}
with $r_{i+1/2}=(\varepsilon_{i+1/2}-\varepsilon_{i})/(\varepsilon_{i+1}-\varepsilon_{i})$ (see \citet{BMuller10} for more details).
Like these, just only by evaluating an appropriate radiation momenta $X_{i+1/2}$ at the energy bin surface,
we can simultaneously vanish Eqns.(\ref{eq:EnergyShiftNumber=0})-(\ref{eq:EnergyShiftEnergy=0}) numerically
and maintain energy, momentum and number conservations.

\subsection{Implicit Time Update}
\label{sec:Implicit Time Update}
After the explicit update, we solve the following simultaneous equation
(e.g., Eq. (\ref{eq:ImpEq}));
\begin{eqnarray}
\label{eq:f(P)=0}
{\bf f}({\bf P}^{n+1})\equiv\frac{{\bf Q}({\bf P}^{n+1})-{\bf Q}^\ast}{\Delta t}
+{\bf S_{\rm adv,e}}({\bf P}^{n+1})+{\bf S_{\rm \nu m}}({\bf P}^{n+1})=0,
\end{eqnarray}
where ${\bf Q}$, ${\bf S_{\rm adv,e}}$ and ${\bf S_{\rm \nu m}}$ are expressed 
 in terms of the primitive variables ${\bf P}$
\begin{eqnarray}
\label{eq:NewtonIterationPrimitive}
{\bf P}=\left[
\begin{array}{c}
\rho\\ u_i \\ s\\ Y_e\\ E_{(\nu,\varepsilon)}\\ {F_{(\nu,\varepsilon)}}_i 
 \end{array}
\right],
\end{eqnarray}
at $(n+1)$-th time step.
Obviously, the baryon number density does not change in this step, i.e., $\rho^\ast=\rho^{n+1}$.
To get the solutions of the above simultaneous equation, we employ the 
Newton-Raphson iteration scheme with the inversion of the following matrix 
until a sufficient convergence is achieved;
{\setlength\arraycolsep{2pt}
\begin{eqnarray}
\label{eq:NewtonIteration}
\frac{\partial {\bf f}({\bf P}^I)}{\partial {\bf P}}\delta{\bf P}^I&=&-{\bf f}({\bf P}^I), \\
{\bf P}^{I+1}&=&{\bf P}^I+\delta{\bf P}^I,
\end{eqnarray}}
for $I=0,1,2.....,$ with the initial condition ${\bf P}^0={\bf P}^\ast$.
As for the convergence criteria for the Newton-Raphson iteration, we monitor
\begin{eqnarray}
\label{eq:convergence}
\frac{|\delta{\bf P}^{I}|}{|{\bf P}^{I}|}<tol.
\end{eqnarray}
where $tol$ represents a tolerance and we typically set $tol=10^{-8}$.

We note that, even if we evaluate the advection terms in energy space explicitly in time as ${\bf S_{\rm adv,e}}^n$,
Eqs.(\ref{eq:f(P)=0})-(\ref{eq:convergence}) do not change except the term ${\bf S_{\rm adv,e}}({\bf P}^{n+1})$
in Eq.(\ref{eq:f(P)=0}) is now moved to the explicit part.

\begin{figure}[htpb]
\begin{center}
\includegraphics[width=110mm]{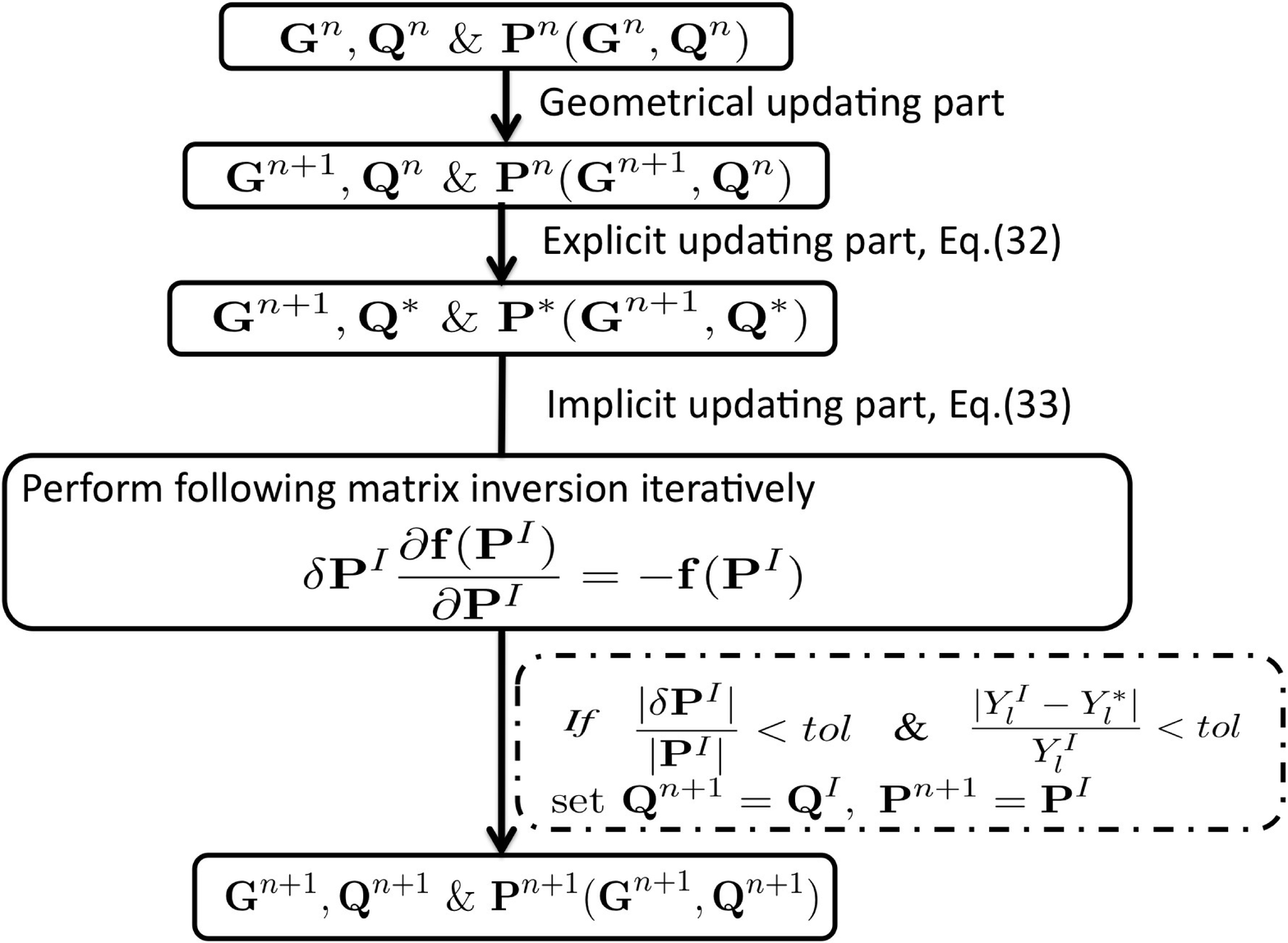}
\end{center}
  \caption{A flowchart to visualize how to update variables 
({\bf G}, {\bf Q}, and {\bf P}) from $n$-th 
to $(n+1)$-th time step (see text).}
\label{fig:Flowchart.eps}
\end{figure}

Our method for time update from $n$-th to $(n+1)$-th time step is 
summarized in order as follows and schematically drawn 
in Figure \ref{fig:Flowchart.eps}.
\begin{enumerate}
\item  {\tt Geometrical Update}. We first evolve all the BSSN and the gauge variables
{\bf G=}\{$\tilde\gamma_{ij}$, $\tilde A_{ij}$, $\phi$, $K$, $\tilde\Gamma^i$, $\alpha$, $\beta^i$\}
from $n$-th to $(n+1)$-th time step.
\item {\tt Explicit Update}. In the second step, all the advection in space $(\bf S_{\rm adv,s})$ and
gravitational source $(\bf S_{\rm grv})$ terms are added to obtain the 
 fractional time-step values ${\bf Q}^\ast$
and their consistent primitive variables ${\bf P}{\bf (G^{n+1}},{\bf Q^\ast})=(\rho,u_i,s,Y_e,E_{(\nu,\varepsilon)}, {F_{(\nu,\varepsilon)}}_i)$.
Advection of total lepton number (Eq.(\ref{eq:GRYl})) is also performed simultaneously to evaluate $Y_l^\ast$.
\item {\tt Implicit Update}. Finally, quantities
 in the fractional time-step (${\bf Q}^\ast$) 
 are implicitly updated to those in the $(n+1)$-th time-step (${\bf Q}^{n+1}$) by the Newton-Raphson iteration until a sufficient convergence is obtained 
with a constraint that the local
lepton fraction does not change (i.e. $Y_l^{n+1}=Y_l^\ast$ since $\rho^{n+1}=\rho^\ast$) in the diffusion region.
\end{enumerate}

Here we shortly summarize some 
technical details to achieve a 
sufficient lepton number conservation in practical 
core-collapse simulations.
As we have already mentioned in Sec. \ref{sec:Conservation of Energy and Lepton Number}, neutrino number conservation is formally satisfied 
by solving the energy-momentum conservation equations (\ref{eq:rad1})-(\ref{eq:rad2}).
Especially, in the neutrino trapping regime, solving the energy conservation
equation (\ref{eq:rad1}) is practically identical to solving the neutrino number
conservation equation (\ref{eq:LeptonNeutrino}) which was
proven in Sec. \ref{sec:Neutrino Number Transport in Diffusion Limit}.
However, in the finite difference method, it does not 
guarantee a perfect match between equations (\ref{eq:rad1})-(\ref{eq:rad2})
and equation (\ref{eq:LeptonNeutrino}) due to discretization error.
To minimize the difference, we add the following constraint to Eq.(\ref{eq:convergence})
\begin{eqnarray}
\label{eq:Ylcheck}
\frac{|Y_l^{I}-Y_l^\ast|}{Y_l^{I}}<tol,
\end{eqnarray}
as another criterion to exit the Newton-Raphson iteration.
Because of this additional criterion, the lepton number conservation is also satisfied when
the Newton-Raphson iteration converges.
Note that the local baryon number do not change in Eq.(\ref{eq:f(P)=0}), i.e., $Y_l^\ast=Y_l^{n+1}$ should be met
if the lepton number is conserved.

We also adopt another prescription in which $Y_l^{n+1}$ is used to achieve a better convergence for the Newton-Raphson method.
In some cases, $Y_e^{I}$ happens to go beyond the range of the 
employed EOS table during the iteration, especially when the Jacobian matrix is not well evaluated.
In those cases, we use $Y_l^{n+1}$ to reset $Y_e^{I}$ as below
\begin{eqnarray}
\label{eq:Yereset}
Y_e^I=Y_l^{n+1}-Y_{\nu_e}^I+Y_{\bar\nu_e}^I.
\end{eqnarray}
With this reset, we found that $Y_e^I$ hardly goes beyond the range of the EOS table and
also does not take unreasonable value for the supernova core.
In addition, the number of the Newton-Raphson iteration can sometime be reduced.
This is because the reset value $Y_e^I$ automatically satisfies the lepton number conservation $Y_l^{I}=Y_e^I+Y_{\nu_e}^I-Y_{\bar\nu_e}^I=Y_l^{n+1}$ which leads to the faster convergence.

Without these two prescriptions,
 we observed that the central lepton fraction at bounce deviates 
maximally $\sim0.05$ from the value immediately after neutrino trapping sets 
in ($\rho_c \gtrsim 10^{12}\,\, {\rm g}\,{\rm cm}^{-3}$).
Note that these ad-hoc constraints are introduced just to find a
more efficient path toward convergence in the implicit time update.
 The criterion for convergence is solely given by 
Eq. (\ref{eq:convergence}).
Since we do not include Eq.(\ref{eq:GRYl}) into Eq.(\ref{eq:f(P)=0}),
there is no inconsistency between the number of simultaneous equations 
$\bf f(P)$ and the unknown variables $\bf P^{n+1}$.

\section{Radiation Tests}
\label{Sec:Radiation Tests}
In this section, we show results of several simple test problems to check the 
validity of our M1 radiation transport code. Except for Sec.4.4, a flat 
spacetime is assumed throughout this section.

\subsection{Diffusion Wave Test}
\label{Sec:Diffusion Test}
We first perform a diffusion wave test, by which we 
check the validity of our flux implementation in 
the diffusion limit. Following \cite{Pons00}, a Dirac $\delta$-function 
type radiation source 
is initially located at $r=0$. Then we follow the diffusion of the source 
into the optically thick medium with zero absorptivity ($\kappa_a=0$)
 and high scattering opacity ($\kappa_s=10^2$, $10^5$).
 The source term in Eqs.(\ref{eq:rad1})-(\ref{eq:rad2}) thus becomes
 \begin{eqnarray}
S^\mu=-\kappa_s H^\mu.
\end{eqnarray}
The 3D computational domain is 
covered by 100 equidistant Cartesian zones 
in each direction (${\bf x}\in[-0.5,0.5]$) for a model with $\kappa_s=10^2$. This corresponds to 
a Peclet number $Pe=\kappa_s\Delta x=1$.
While, in the other model with $\kappa_s=10^5$, we raise the nested level to 2 to
achieve sufficient resolution at the center and to check that the nested structure
does not interfere the radiation propagation in the diffusion limit.
In this model, $32^3$ numerical cells cover the base domain ${\bf x}\in[-0.5,0.5]$ with 2 nested level.
The minimum grid width and Peclet number at the center are thus $\Delta x=1/128\sim0.78$
and $Pe=10^5/128\sim780$, respectively. Since $Pe$ is much higher than unity,
$\epsilon$ in Eq.(\ref{eq:AuditEpsilon}) approaches zero which enables us to
check whether the radiation transport in the diffusion limit is solved appropriately.

\begin{figure}[htpb]
\begin{center}
\includegraphics[width=90mm,angle=-90.]{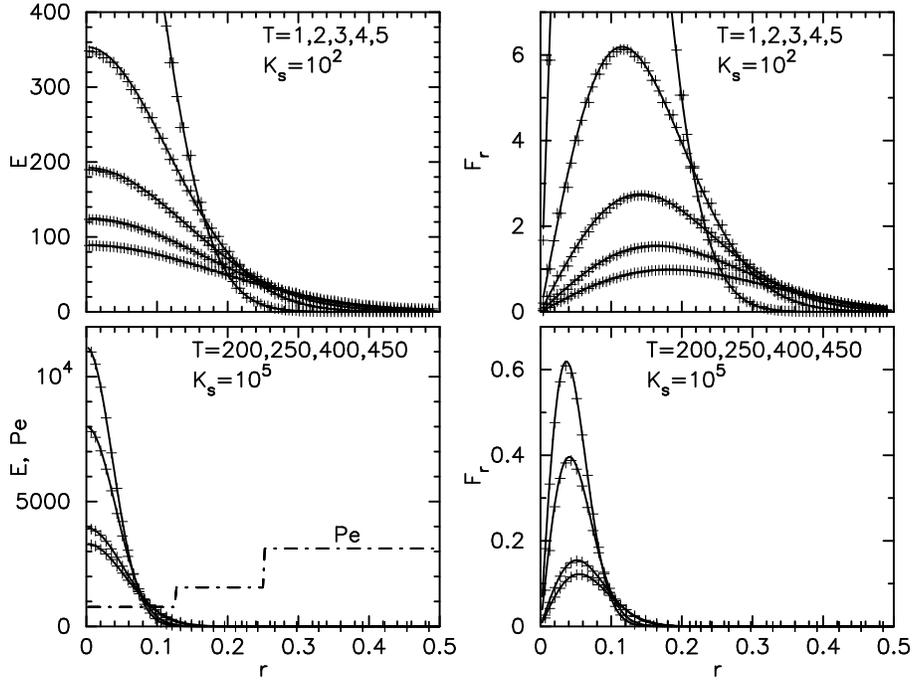}
\end{center}
  \caption{Diffusion tests with $\kappa_s=10^2$ ($Pe=\kappa_s\Delta x=1$) in upper two panels
  and with $\kappa_s=10^5$ ($Pe\ga780$) in bottom twos.
Our results (crosses) for the evolutions of the energy density (left panels) 
and radial energy flux (right panels) are compared 
with the analytical ones (solid lines).
  Note that the simulations start at $T=1$ and 200 for models with $\kappa_s=10^2$ and $\kappa_s=10^5$, respectively.
  The model with $\kappa_s=10^5$ is performed with two nested level structure and
  we plot the spatial profile of $Pe$ in the bottom-left panel for reference.
  Note that we reduce number of plots for our results (crosses) to avoid messy overplots.}
\label{fig:DiffTest.eps}
\end{figure}

The analytical solution of the zeroth and first order radiation momenta 
profiles at time $T$ \citep{Pons00} is given as
\begin{eqnarray}
E(r,T)&=&\left(\frac{\kappa_s}{T}\right)^{3/2}{\rm exp}\left(\frac{-3\kappa_s r^2}{4T}\right),\\
F_r(r,T)&=&\frac{r}{2T}E,
\end{eqnarray}
 respectively.
From Figure \ref{fig:DiffTest.eps}, it can be seen that our code can 
reproduce the analytical results quite well. 

\subsection{Shadow Casting Test}
\label{Sec:Shadow Test}
Next, we move on to a shadow casting 
test to check the ability of our code whether anisotropic radiation 
field can be appropriately evolved in the free streaming limit.
The initial setup is essentially the same as in \citet{Kanno13}.
 We set a perfect absorbing region
at $x\in[2,3]$ and $\{y,z\}\in[-2,2]$ inside the numerical domain $x\in[0,12]$ and $\{y,z\}\in[-4,4]$.
We impose a constant radiation $(E,F_x)=(1,f_{\rm max})$ from the left boundary.
Numerical resolution is $\Delta x(=\Delta y=\Delta z)=12/192$ and we set $f_{\rm max}=0.999$.
Within the perfect absorbing region, the absorbing opacity is set as $\kappa_a\Delta x\sim10^{10}$,
otherwise  $\kappa_a=0$.
 The source term in Eqs.(\ref{eq:rad1})-(\ref{eq:rad2}) thus becomes
 \begin{eqnarray}
\label{eq:ShadowTestSource}
S^\mu=-\kappa_a\left( (J-J^{\rm eq})u^\mu+H^\mu\right),
\end{eqnarray}
with $J^{\rm eq}=0$ and $u^\mu=(1,0,0,0)$.
\begin{figure}[htpb]
\begin{center}
\includegraphics[width=80mm,angle=-90.]{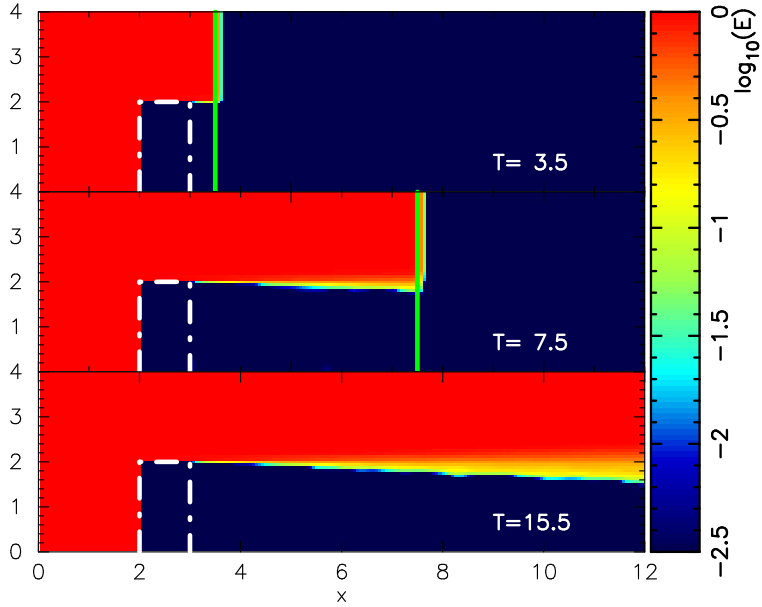}
\end{center}
  \caption{Shadow test at different time slices $T=$3.5, 7.5 and 15.5 from top to bottom.
  Color contour represents radiation energy density in logarithmic scale.
The absorbing region is expressed by white-dashed line and the expected radiation shock
 front (i.e., $x=T$) is denoted by a vertical green line.}
\label{fig:ShadowTest}
\end{figure}
In Fig. \ref{fig:ShadowTest}, we show three different time snapshots 
($T$=3.5, 7.5 and 15.5) of the radiation energy density in a 
logarithmic scale.
The absorbing region is expressed by white-dashed line and the radiation
 shock front ($x=T$) is denoted by vertical green line.
From this figure, we see that the absorbing condition works 
appropriately in our scheme and the radiation front propagates with the light 
speed $v=1$. Furthermore, the region beyond the absorbing box is barely
 contaminated by the radiation. These features are in good agreement with 
\citet{Kanno13}.

\subsection{Propagation in Free Streaming Regime}
\label{Sec:Propagation in Free Streaming Regime}
The next test is a spherical expansion of radiation field
 from a point-like source into optically thin medium.
The aim of this test is to check whether our code using
 the Cartesian coordinates can maintain
 the sphericity of the field during the expansion.
By following \citet{just15}, we define the radiation source $\mathcal S$ with radius $r_{\mathcal S}=1.5$
which centres at the origin of the 2D Cartesian domain with ${\bf x}\in [-7.5,7.5]$.
We also define the purely absorbing region $\mathcal A$ centered at ${\bf x}=(3.5,0)$ with radius $r_{\mathcal A}=1$.
Inside those two regions, the absorptivity $\kappa_a(\bf x)$ and equilibrium energy density $J^{\rm eq}(\bf x)$ are set as
 {\setlength\arraycolsep{2pt}
\begin{eqnarray}
\kappa_a(\bf x)=\left\{ \begin{array}{ll}
10{\rm exp}\{-(4\sqrt{x^2+y^2}/r_{\mathcal S})^2\} &  ,\bf x\in \mathcal S\\
10 &  ,\bf x\in \mathcal A,
\end{array} \right.
\end{eqnarray} 
and
 {\setlength\arraycolsep{2pt}
\begin{eqnarray}
J^{\rm eq}(\bf x)=\left\{ \begin{array}{ll}
10^{-1} &  ,\bf x\in \mathcal S\\
0 &  ,\bf x\in \mathcal A,
\end{array} \right.
\end{eqnarray} 
respectively.
In other regions, we set $\kappa_a(\bf x)=0$ and $J^{\rm eq}(\bf x)=0$.
We neglect the scattering opacity $\kappa_{\rm s}(\bf x)$.
The source term in Eqs.(\ref{eq:rad1})-(\ref{eq:rad2}) is thus the same as Eq.(\ref{eq:ShadowTestSource}).
 Regarding the initial setup for the zeroth and first order momenta of radiation field, we assume an arbitrary dilute radiation field as
{\setlength\arraycolsep{2pt}
\begin{eqnarray}
E&=&10^{-9}\\
 |F|/E&=&10^{-10}.
\end{eqnarray}}
The 2D numerical domain is covered by 196 equally distant Cartesian zones in each direction with two nested levels,
i.e. $\Delta x$ varies from $15/784$ to $15/196$.

In this test, we use following formula for the variable Eddington factor $\chi$ \citep{levermore84}
\begin{eqnarray}
\chi=\frac{3+4\bar F^2}{5+2\sqrt{4-3\bar F^2}}.
\end{eqnarray}
Only in this propagation test,
we take two possible options for evaluating the characteristic wave speeds of radiation field
in the optically thin limit $\lambda_{\pm}$.
Assuming a flat spacetime and using a following closure relation in the optically thin limit
\begin{eqnarray}
\label{eq:Pij_thin}
P^{\mu\nu}=E\frac{F^\mu F^\nu}{F_kF^k},
\end{eqnarray}
the first one is \citep{Shibata11}
\begin{eqnarray}
\label{eq:Lambda_Shibata11}
\lambda^{{\rm S},i}_{\pm}={\rm max/min}\left[ \frac{EF^i}{F_kF^k},\ \pm \frac{F^i}{\sqrt{F_kF^k}} \right],
\end{eqnarray}
and the second one is \citep{Skinner13}
\begin{eqnarray}
\label{eq:Lambda_Skinner13}
\lambda^{{\rm SO},i}_{\pm}=\left\{ \mu^i \bar F \pm \sqrt{  \frac{2}{3}\left(
4-3\bar F^2-\sqrt{4-3\bar F^2}\right)+2{\mu^i}^2\left(2-\bar F^2-\sqrt{4-3\bar F^2}\right)  } \right\}/\sqrt{4-3\bar F^2}.\nonumber \\
\end{eqnarray}
Here, $\mu^i\equiv \cos{\theta}$ is determined by the angle $\theta$ which defines the orientation of the energy flux $\bf F$
relative to the interface normal $x^i$.
\textcolor{red}{Note that, in the rest of sections, we use only Eq.(\ref{eq:Lambda_Shibata11}) for the free streaming part.}

Another thing to mention is that we limit the flux factor $f\equiv \sqrt{\gamma^{ij}{F_iF_j}}/E$ to be less than a maximum allowed value $f_{\rm max}$
by modifying the first order moment as
\begin{eqnarray}
\label{eq:FoverE_Modify}
F_i\rightarrow  {\rm min}(f_{\rm max}/f,1)F_i,
\end{eqnarray}
in every time step.
Note that setting $f_{\rm max}<1$ means that we add contribution from the isotropic
radiation pressure $P^{ij}_{\rm thick}$ according to Eq.(\ref{neu_p}).
The reason of this modification will be discussed later in this section.

In Fig. \ref{fig:FreeStreamingTest}, we show results for three different models
taking different values for $f_{\rm max}$ and using different evaluation formulae for $\lambda^{\rm S/SO}_{\pm}$.
Upper two panels (a,b) and bottom one (c) are models using Eq. (\ref{eq:Lambda_Shibata11})
and Eq. (\ref{eq:Lambda_Skinner13}), respectively.
In panels (a) and (c), we take $f_{\rm max}=1$, while  $f_{\rm max}=0.93$ is taken in panel (b).
The color represents the energy density $E$ in a logarithmic scale at four different time slices ($T=1,3,5,7$).
Inner two squares represent the boundary of nested structure and the dotted circle shows the purely absorbing region $\mathcal A$.
\begin{figure}[htpb]
\begin{center}
\includegraphics[width=90mm,angle=-90.]{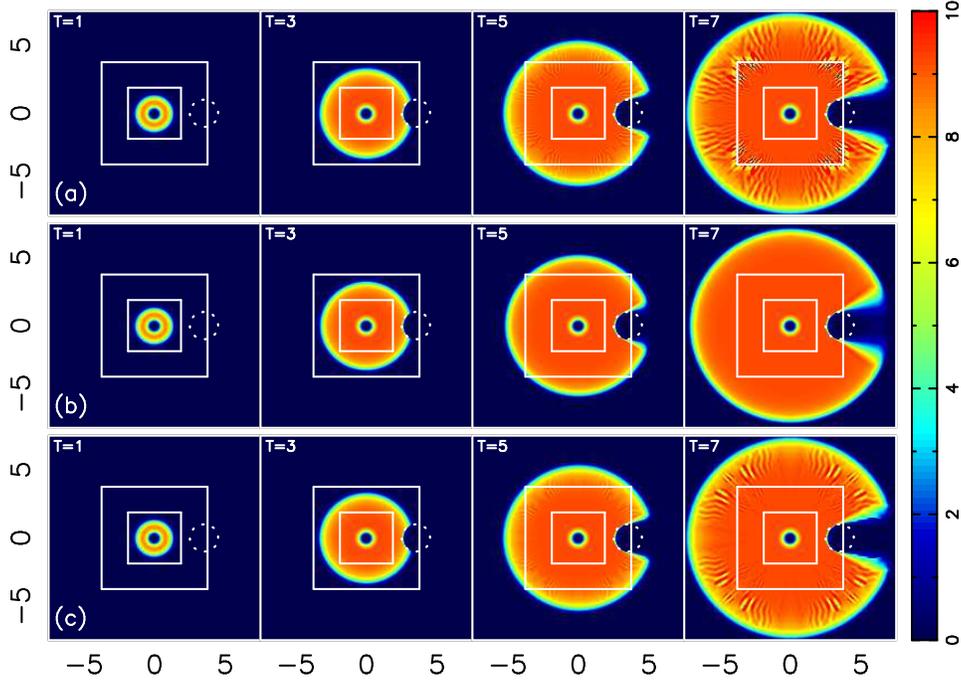}
\end{center}
  \caption{Time snapshots of spherical explosion test in the optically thin medium with different sets of $(f_{\rm max},\lambda_{\pm})$.
  Models (a) and (c) (top and bottom rows) are models with taking $f_{\rm max}=1$, while $f_{\rm max}=0.93$ is taken in model (b) (middle row).
  Eqs. (\ref{eq:Lambda_Shibata11}) and (\ref{eq:Lambda_Skinner13}) are used in models (a-b) and (c), respectively.
  Color scale represents radiation energy density on a linear scale in arbitrary unit.
  Inner two squares represent the boundary of nested structure and the dotted circle shows the purely absorbing region $\mathcal A$.
	}
\label{fig:FreeStreamingTest}
\end{figure}
From models (a) and (c), which use $f_{\rm max}=1$, we find non-isotropy which can be seen
as corrugated patterns behind the radiation front, obviously with the exception of the absorbing circle $\mathcal A$ and its shadowing region.
Furthermore in model (a) at $T=7$, another remarkable violation of isotropy is seen at $45^\circ$ away from the coordinate axises.
Note that, from snapshot at $T=7$ in panel (a), this violation seems to be associated with the nested structure.
We, however, confirmed that this is not the artifact of that by performing a run with zero nested level.
By comparing models (a) and (c), the violation seen at $45^\circ$ is eliminated in model (c).
Model (c) takes into account the propagation angle ($\mu^i$) relative to the interface normal
more explicitly than model (a) when we evaluate $\lambda^{\rm SO}_{\pm}$.
We therefore consider a more accurate evaluation in the characteristic wave speeds $\lambda_{\pm}$ for the numerical flux
(Eqs (\ref{eq:AuditEnergyFlux})-(\ref{eq:AuditMomentumFlux}))
is the key to follow the radiation in optically thin limit properly.
While in panel (b), in which we use $f_{\rm max}=0.93$, we do not see any spherical symmetry breaking.

We also apply this test to a supernova core profile.
We fix the hydrodynamical background, which has a typical core profile at $T_{\rm pb}\sim100$ ms,
and follow only the propagation of neutrino radiation.
The neutrino transport part is identical to our practical calculation reported later in Sec. \ref{sec:Core Collapse of a 15M star}
and all relevant neutrino-matter interactions, gravitational redshift and Doppler shift  terms are taken into account.
There is thus a transition from optically thick to thin regime.
We calculate two models with different values for $f_{\rm max}$.
$\lambda^{\rm S,i}_{\pm}$ (Eq.\ref{eq:Lambda_Shibata11}) is used in both models to evaluate the characteristic wave speeds.
In Fig. \ref{fig:FoverE.eps}, we show the radial profiles of (electron-type) neutrino luminosity for two cases: one is 
calculated with $f_{\rm max}=0.999$ (black diamonds) and
the other is with $f_{\rm max}=0.93$ (red filled triangles). 
\begin{figure}[htpb]
\begin{center}
\includegraphics[width=70mm,angle=-90.]{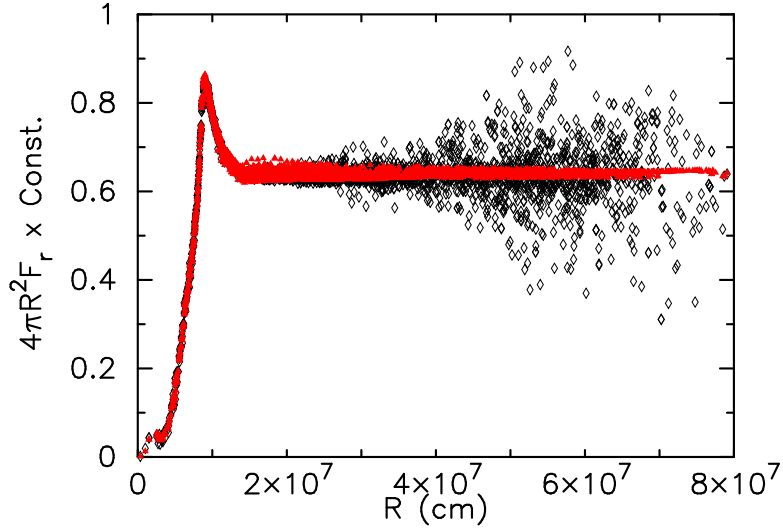}
\end{center}
  \caption{Arbitrary normalized surface integrated local neutrino luminosities 
measured by an Eulerian observer.
Background profile is a stationary and spherically symmetric SN core taken at post bounce time 100ms.
Models are with two different maximum allowed flux factors,
$f_{\rm max}=0.999$ (black diamonds) and $=0.93$ (red filled triangles).
}
\label{fig:FoverE.eps}
\end{figure}
As can be clearly seen, beyond the shock position ($R\gtrsim100$ km) where it is optically thin regime,
a large scatter is seen at $R \gtrsim 300-400$ km for model with $f_{\rm max}=0.999$.
Spatial location of those highly deviated radiation are again concentrated along $\sim45^\circ$ away from the coordinate axises
as the same as in the previous simple propagation test (panel (a) in Fig. \ref{fig:FreeStreamingTest}).
Meanwhile, in model with $f_{\rm max}=0.93$, there is little deviation and
the surface integrated local luminosity stays nearly constant in the transparent region as it should be.

From these two tests, we are currently enforced to set a ceiling value for the flux factor $f\le f_{\rm max}=0.93$
in the practical core-collapse simulation to follow spherical-like propagation stably.
Since $f_{\rm max}=1$ is the physically correct value, one may suspect that
our modification can be an obstacle for comparison of the emergent radiation profile with previous studies.
From previous fully relativistic Boltzmann transport simulation \citep{Liebendorfer01}, however,
it was shown that the flux factor higher than, e.g., $\sim0.93$ appears only beyond $R\gtrsim500-1000$ km.
Therefore even if we evaluate the emergent neutrino profile at the same radius $R=500$ km as previous studies,
it is altered only slightly $\sim$ a few \% and the modification cannot be a significant obstacle in this study.
Obviously, the SN hydrodynamics itself is not affected by our artificial treatment
since it is applied only to the optically thin region.

\subsection{Gravitational Redshift and Doppler Shift}
\label{Sec:Gravitational Redshift and Doppler Shift}
To check the energy coupling terms for gravitational redshift and Doppler shift,
$\sqrt{\gamma}\alpha \partial_\varepsilon \bigl(\varepsilon \tilde M^\mu_{(\varepsilon)} n_\mu\bigr)$
and $\sqrt{\gamma}\alpha \partial_\varepsilon\bigl(\varepsilon \tilde M^\mu_{(\varepsilon)} \gamma_{i\mu}\bigr)$ 
in Eqs.(\ref{eq:rad1})-(\ref{eq:rad2}),
we repeat the same tests performed in \cite{BMuller10,O'Connor14}.
We consider the propagation of radiation from a sphere with radius $R=10$ km
through a curved space-time and sharp velocity profile in spherical symmetry.
Regarding the sharp velocity profile, we take the following one
\begin{eqnarray}
u_r=\left\{ \begin{array}{cl}
0, & r\le70{\rm km} \\
-0.2c\left(\frac{r-70{\rm km}}{10{\rm km}}\right), & 70{\rm km}\le r\le80{\rm km} \\
-0.2c\left(\frac{80{\rm km}}{r}\right)^2, & r\ge80{\rm km}. \\
\end{array} \right.
\end{eqnarray} 
Here, $u_r$ is the radial component of $u_i$.
As for the curved space-time, we take the same energy density profile for matter
which is used in the second test problem in Sec. \ref{Sec:Propagation in Free Streaming Regime} and solve the Hamiltonian constraint
to obtain the conformal factor $\psi$ and lapse $\alpha$.
When we solve the Hamiltonian constraint,
we assume zero velocity profile ($v_i=0=\beta_i$) and the conformally flat approximation
($\gamma_{ij}=\psi^4\tilde\gamma_{ij}=\psi^4\delta_{ij}$).
As for the initial neutrino profile within the radiating sphere, we take
zero chemical potential $\mu_\nu=0$ and a temperature of 5 MeV, i.e., $\langle \varepsilon\rangle\sim15.7$ MeV,
in the free streaming limit $E=\sqrt{\gamma_{ij}F^iF^j}$.
Here $\langle \varepsilon\rangle$ is the mean energy of neutrinos as measured in the comoving frame.
Outside of the central radiating sphere, we choose an arbitral dilute radiation field with $\langle \varepsilon\rangle\sim15.7$ MeV.
We do not evolve neutrinos inside the radiating sphere and follow the propagation only at $R\ge10$ km.
The 3D computational domain is a cubic box with 3000 km width
(i.e. the outer boundary is at the radius of 1500 km from the origin)
and nested boxes with 7 refinement levels are embedded.
Each box contains $64^3$ cells and the minimum grid size near the origin is $\Delta x=366$m.
As for the energy grid of the neutrino radiation field, we use logarithmically spaced 20 energy bins $(N_\varepsilon=20)$
which center from $\varepsilon=1$ MeV to 300 MeV.
We calculate two models with and without the sharp velocity profile.
Both models are calculated in the curved space-time.

For given velocity and space-time profiles, the free streaming neutrinos propagate with
obeying the following relations \citep{BMuller10,O'Connor14}
{\setlength\arraycolsep{2pt}
\begin{eqnarray}
W\alpha (1+v_r)\langle\varepsilon\rangle&=&{\rm const,}
\label{eq:energy_analytic_profile}\\
\sqrt{\gamma}\alpha r^2 \gamma^{rr}\int d\varepsilon F_{r}\equiv L_{\rm eul}&=&{\rm const,}
\label{eq:Leul}
\end{eqnarray}}
where $W$ is the Lorentz factor and $v_r=u_r/u^t$.
$\sqrt{\gamma}=\psi^6$ and $\gamma^{rr}=\psi^{-4}$ are the determinant and $rr$-component of the three metric, respectively,
in the conformally flat approximation.
$L_{\rm eul}$ is the luminosity measured by an Eulerian observer.
Eq.(\ref{eq:Leul}) is derived from the stationary solution of Eq.(\ref{eq:rad1}) with neglecting the source terms.

\begin{figure}[htpb]
\begin{center}
\includegraphics[width=160mm,angle=-90.]{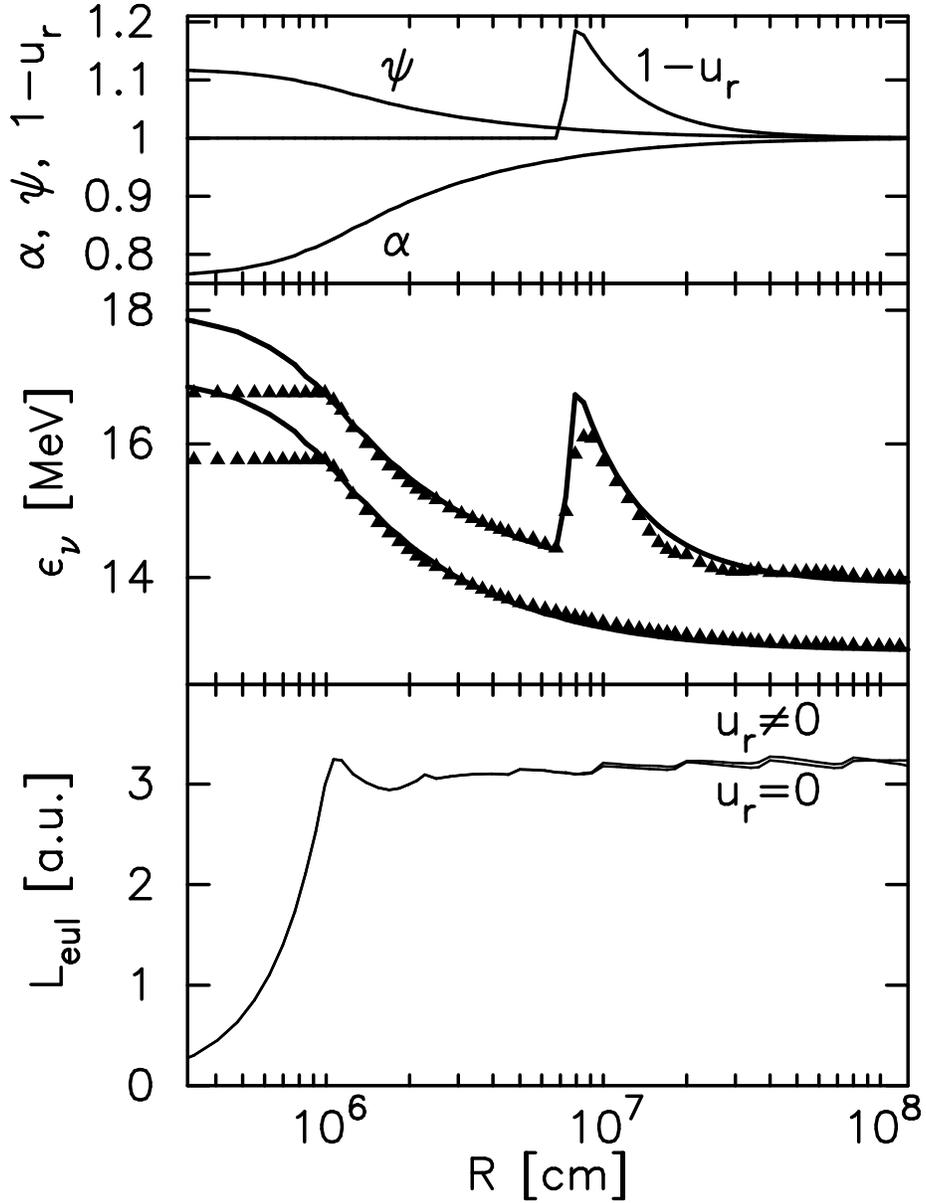}
\end{center}
  \caption{In the top panel, we show background profiles for $\psi$, $\alpha$ and $1-u_r$.
  The mean energy $\langle \varepsilon\rangle$ and Eulerian luminosity $L_{\rm eul}$ profiles are plotted in the middle and bottom panels, respectively.
  Two test cases, with ($u_r\ne0$) and without ($u_r=0$) the shock profile, in the curved space-time are plotted.
  In the middle panel, our results are denoted by filled triangles and the solid lines are analytical results.
  Note that $\langle \varepsilon\rangle$ for the model with the shock profile is shifted upward with 1MeV to avoid the overlapping of plots.
    }
\label{fig:EnergyShift}
\end{figure}
In the top panel of Fig. \ref{fig:EnergyShift}, we show the background profiles for $\psi$, $\alpha$ and $1-u_r$.
$\langle \varepsilon\rangle$ and $L_{\rm eul}$ are plotted in the middle and bottom panels, respectively.
In the middle panel, our results (filled triangles) and analytical ones (solid lines) are plotted.
The analytical expression for $\langle \varepsilon\rangle$ is derived from Eq.(\ref{eq:energy_analytic_profile})
where the constant in the right hand side is evaluated at $R=10$ km.
Numerical results are taken after they settle down in nearly stationary.

As seen in the middle panel, our results reproduce the analytical ones quite well except at the shock front.
We consider that the less agreement at the shock front originates from the coarse spatial resolution.
The finite difference expression for the term $\nabla_\alpha u_\beta$ in the energy coupling terms
cannot capture the sharp velocity profile with enough accuracy.
From the bottom panel, we can find the Eulerian luminosity stays nearly constant with a few percent deviation beyond $R\ga 20$ km.
Note that the zig-zag pattern seen in $L_{\rm eul}$ is an artifact of the mesh refinement.
}}

\section{Core Collapse of a 15$M_{\odot}$ Star}
\label{sec:Core Collapse of a 15M star}
In order to confirm the validity of our new supernova code, 
it is of primary importance to make a detailed comparison with the
 previous published results. We employ the data from 
 1D-GR neutrino transport simulations 
\citep{Liebendorfer05,Sumiyoshi05,BMuller10}
and from 2D-GR ones \citep{BMuller14}. We chose these models because
 all of them
 took the same progenitor and employed similar
 microphysics in the GR CCSN simulations.
\citet{Liebendorfer05} presented detailed comparison of two independent 
numerical codes, AGILE-BOLTZTRAN \citep{Liebendorfer04} and 
VERTEX-PROMETHEUS \citep{Rampp02}.
Since their results are available online\footnote{We used data which can be downloaded at
\hbox{http://iopscience.iop.org/0004-637X/620/2/840/fulltext/datafiles.tar.gz}},
we are able to make a detailed comparison with their data set. AGILE-BOLTZTRAN solves 
 the GR Boltzmann equation 
with the $S_n$ method
in spherically symmetric Lagrangian mesh, whereas VERTEX is an Eulerian 
code that solves the moment equations of a model Boltzmann equation
 by the VEF method in the Newtonian hydrodynamics plus a modified GR potential (VERTEX-PROMETHEUS)
and also in the conformally-flat GR hydrodynamics (VERTEX-CoCoNuT, \citet{Dimmelmeier02,BMuller10}). Our code is rather similar to VERTEX-CoCoNuT than AGILE-BOLTZTRAN except for the different geometrical solvers and the different
 coordinate systems are adopted.
In the following, we label the 
results of AGILE-BOLTZTRAN as ``ABG15'',
of VERTEX-PROMETHEUS as ``VXG15'' and of \cite{BMuller10} as ``BMG15''\footnote{
We read their data from the figures digitally and plot them in this paper.}.

\subsection{Numerical Setups}
\label{sec:Numerical Setups}

We employ a 15$M_{\odot}$ progenitor \citep[model ``s15s7b2'' in][]{WW95}
and follow core collapse, bounce and initial postbounce phase up to $\sim50$ ms.
We use the EOS of Lattimer \& Swesty (1991) (LS EOS) with an incompressibility 
parameter of $K=220$ MeV.
Even though our choice of $K=220$ MeV is different 
from the one ($K=180$ MeV) used in \citet{Liebendorfer05,Sumiyoshi05,BMuller10,BMuller14},
\cite{Thompson03} showed that differences in the neutrino profiles
among models with the different $K$ of LS EOS are a few \% around core bounce
 and less than $\sim10$ \% for the first $\sim200$ ms after bounce.
We thus consider that the different choice of $K$ barely disturbs the aim of 
  our comparison study.

 As for neutrino opacities, the standard weak interaction set in
  \citet{Bruenn85} and \citet{Rampp02} plus nucleon-nucleon bremsstrahlung \citep{Hannestad98} is taken into account 
(see Table 1 and
 Appendix \ref{sec:Neutrino Matter Interaction Terms} for more detail).
 For simplicity, we neglect higher harmonic angular dependence 
 of the reaction angle when we calculate the source terms
for neutrino electron scattering, thermal pair production and annihilation of neutrinos, and nucleon-nucleon bremsstrahlung (i.e.,  $B^{0,\mu}_{(\varepsilon),\rm nes/tp}$ and $C^{1,\mu}_{(\varepsilon),\rm nes/tp}$ are set to be 0, see Apps. \ref{sec:Neutrino Electron Scattering} and \ref{sec:Thermal Pair Production and Annihilation of Neutrinos}).

\begin{table}[htpb]
\begin{center}
\begin{tabular}{c cc}
\hline\hline
Process & reference & summarised in\\
\hline
$n\nu_e\leftrightarrow e^-p$ &  \cite{Bruenn85,Rampp02} & App.\ref{sec:Neutrino Absorption and Emission}\\
$p\bar{\nu}_e\leftrightarrow e^+n$ &   \cite{Bruenn85,Rampp02} & App.\ref{sec:Neutrino Absorption and Emission}\\
$\nu_e A\leftrightarrow e^-A'$ &  \cite{Bruenn85,Rampp02} & App.\ref{sec:Neutrino Absorption and Emission}\\
$\nu p\leftrightarrow \nu p$ &  \cite{Bruenn85,Rampp02} & App.\ref{sec:Isoenergy Scattering of Neutrinos}\\
$\nu n\leftrightarrow \nu n$   &  \cite{Bruenn85,Rampp02} & App.\ref{sec:Isoenergy Scattering of Neutrinos}\\
$\nu A\leftrightarrow \nu A$   &  \cite{Bruenn85,Rampp02} & App.\ref{sec:Isoenergy Scattering of Neutrinos}\\
$\nu e^{\pm}\leftrightarrow \nu e^{\pm}$   &  \cite{Bruenn85} & App.\ref{sec:Neutrino Electron Scattering}\\
$e^-e^+\leftrightarrow \nu \bar\nu$ &  \cite{Bruenn85} & App.\ref{sec:Thermal Pair Production and Annihilation of Neutrinos} \\
$NN\leftrightarrow \nu \bar\nu NN$  & \cite{Hannestad98} & App.\ref{sec:Nucleon-nucleon Bremsstrahlung}\\
\hline
\end{tabular}
\caption{The opacity set included in this study and their references.
Note that $\nu$, in neutral current 
reactions, represents all species of neutrinos ($\nu_e,\bar\nu_e,\nu_x$) with $\nu_x$ 
 representing heavy-lepton neutrinos (i.e. 
$\nu_{\mu}, \nu_{\tau}$ and their anti-particles).}
\label{tb:NeutrinoInteraction}
\end{center}
\end{table}

In this study we investigate two models, 1DG15 and 3DG15, both of which are
 computed in the 3D
 Cartesian coordinates.
While model 3DG15 is calculated without any artificial constraints, 
 model 1DG15 is considered to mimic 1D model by artificially suppressing
the non-radial component of the flow velocity $u_i$ as 
\begin{eqnarray}
\label{eq:1DVr}
u_i=\frac{x^j u_j}{|x|^2}x^i.
\end{eqnarray}
Although this artificial elimination could potentially lead to the shift 
of the kinetic energy into the thermal one,
our previous study \citep{KurodaT12} showed that 
the violation of the momentum constraint is negligible during the 
early postbounce phase ($T_{\rm pb}\lesssim 100$ ms where $T_{\rm pb}$ denotes the postbounce time). 

The 3D computational domain is a cubic box with 8000 km width (i.e. the outer boundary is at the radius of 4000 km from the origin)
and nested boxes with 5 refinement levels, at the beginning of calculation, to
8 refinement levels, when the central rest mass density reaches $5\times10^{13}$ g cm$^{-3}$,
are embedded without any spatial symmetry.
Each box contains $64^3$ cells and the minimum grid size near the origin at bounce is $\Delta x=488$m.
In the vicinity of the stalled shock front $R\sim100$ km, our resolution 
achieves $\Delta x\sim 3.6$ or 7.2 km,
i.e., effective angular resolution becomes 3.6km/100km $\sim2^\circ$ or 
$\sim4^\circ$, which is considered to be rather too coarse resolution to follow 
a nearly spherical structure in the Cartesian grids.
Compared to recent 3D-GR studies \citep{Ott12a,KurodaT14}, the resolution is still approximately two times coarser. The time step 
size is $\Delta t=10^{-7}$s which corresponds to $\sim$0.06 in the CFL number.
The main reason for taking such a small timestep is to get 
 a rapid convergence during the iteration in our implicit time update part (see, Sec.\ref{sec:Implicit Time Update}), which is not restricted by the 
CFL condition (of the explicit update in a pure hydrodynamics case).
As for the energy grid of the neutrino radiation field, we use logarithmically spaced 20 energy bins $(N_\varepsilon=20)$
which center from $\varepsilon=1$ MeV to 300 MeV.

\subsection{Results}
\label{sec:Results}

In this section we start to make a detailed comparison 
 firstly before bounce (section \ref{bef_bounce}) and then after bounce 
(section \ref{aft_bounce}) between our code
(models 1DG15 and 3DG15), AGILE-BOLTZTRAN (model 
ABG15), VERTEX-PROMETHEUS (model VXG15), and (partly) from VERTEX-COCONUT
 (model BMG15). In the following, we call the results from the latter 
 three codes as the {\it reference} results.

\subsubsection{Before bounce}\label{bef_bounce}
\begin{figure}[htpb]
\begin{center}
\includegraphics[width=95mm,angle=-90.]{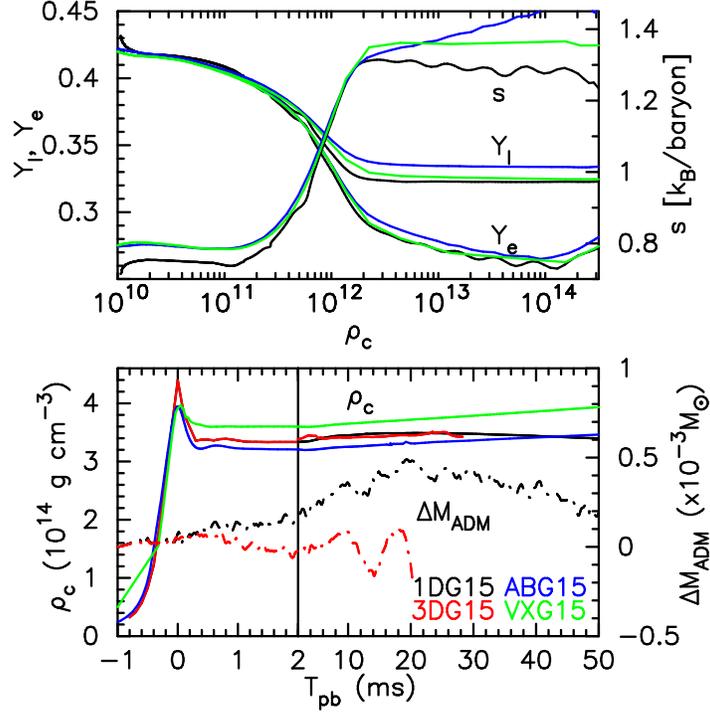}
\end{center}
  \caption{Upper: the central (matter) entropy $s$, electron fraction $Y_e$ and
total lepton fraction $Y_l = Y_e + Y_{\nu}$ as a function of central density $\rho_{\rm c}$ for model 1DG15 (black),
ABG15 (blue) and VXG15 (green). Lower: Comparison of 
postbounce evolution of the central density (solid) between 
 the four models and
the deviation of the ADM mass $\Delta M_{\rm ADM}$ (dash-dotted) for our two
 models with (model 1DG15) or without 
the artificial elimination of the non-radial
 velocity (model 3DG15, see text).}
\label{fig:RhoC.eps}
\end{figure}

In the upper panel of Fig. \ref{fig:RhoC.eps}, we plot 
the central (matter) entropy $s$, electron fraction $Y_e$ and the
total lepton fraction $Y_l = Y_e + Y_{\nu}$ as a function of the central (rest-mass) density $\rho_{\rm c}$ 
for model 1DG15 (black), ABG15 (blue), and VXG15 (green), respectively.
The lower panel of Fig. 1 shows the evolution of $\rho_{\rm c}$
and the deviation of the ADM mass $\Delta M_{\rm ADM}$ from its initial value.
 Here $M_{\rm ADM}$ is given by
\begin{eqnarray}
\label{eq:Madm}
M_{\rm ADM}&=&\int dx^3
 \left[\Bigl(S_0+\int d\varepsilon E_{(\nu,\varepsilon)}\Bigr)e^{-\phi}
 +\frac{e^{5\phi}}{16\pi}\Bigl(\tilde A_{ij}\tilde A^{ij}-\frac{2}{3}K^2-\tilde{\gamma}^{ij}\tilde{R}_{ij}e^{-4\phi}\Bigr)
\right]\nonumber \\
&+&\int d\sigma \left[ e^{-\phi}\Bigl(S^i+\int d\varepsilon{F_{(\nu,\varepsilon)}}^i\Bigr)\cdot \hat{\bf n} \right],
\end{eqnarray}
where the second line denotes energy loss due to momentum 
and neutrino energy flux through the numerical boundary
and $\hat{\bf n}$ represents a unit normal vector to the surface element $d\sigma$.
In the above surface integration, we neglect energy loss due to 
gravitational wave emission since it is negligibly small 
($\sim10^{-11}M_\odot c^2$, e.g., \citet{Scheidegger10})
compared to the violation
 of the ADM mass in the CCSN environment (e.g., \citet{Kotake13} for a review).

The top panel of Fig. 1 shows that the neutrino trapping starts 
$\rho_{\rm c}\sim2\times10^{12}$ g cm$^{-3}$ for model 1DG15
 and the lepton fraction remains constant with $Y_l=0.323$ afterward.
The evolution of our central $Y_e$ and $Y_l$ (black lines) is
 quantitatively in good agreement 
with VXG15 (green line, see also \citet{BMuller10,Buras06a}) and 
ABG15 (blue line).
As already explained in Sec. \ref{sec:Conservation of Energy and Lepton Number},
we solve the advection equation (Eg. (\ref{eq:GRYl})) 
of the total lepton number (density) $\rho Y_l$ as a 
 constraint to ensure the lepton number conservation. Because of the treatment, 
 the evolution of the lepton/electron fraction is in excellent agreement 
with the reference results. 
As already pointed out by \citet{O'Connor14}, this also relies
 on the accurate implementation of inelastic neutrino-electron scattering, 
 energy-bin coupling, and the appropriate closure relation.

After the core deleptonization ceases (i.e., $Y_l$ stays nearly constant with 
 increasing central density),
 the inner core evolves almost adiabatically and
the entropy remains nearly constant as it should be.
 A small breaking of the adiabaticity (decrease by
$3.8 \%$ in the central entropy) is seen 
in model 1DG15 before bounce (see also \citet{O'Connor14}). However,
 the (time-averaged) value $s\sim1.3$ $k_{\rm B}$ ${\rm baryon}^{-1}$ is roughly in good agreement with the reference results (see also \citet{Liebendorfer05,Sumiyoshi05,Buras06a}, and \citet{BMuller10}) and this would not have a big impact 
on the subsequent core evolution due to the short simulation time in this study.

As for the total energy conservation (bottom panel of Fig.3), it is maximally 
violated with the amount of $\Delta M_{\rm ADM}\sim4\times10^{-4}M_\odot$
for model 1DG15 (i.e. $\sim 8 \times 10^{50}$ erg). The violation at bounce 
is slightly worse than that ($\Delta_{\rm ADM} \sim 5 \times 10^{50}$ erg) of the 
VERTEX-CoCoNut code \citep{BMuller10},
 which should be improved and kept much smaller in more precise CCSN modelling.
As one would anticipate, the violation is bigger for model 1DG15
 (dashed black line) with the artificial
 elimination of the non-radial velocity than for model 3DG15 without 
(dashed red line). In our previous study (with the gray M1 scheme,
 \citet{KurodaT12}),
  the violation of the ADM mass is typically one order-of-magnitude smaller
 than that for the corresponding 3D model with (approximately)
 twice higher resolution. Because 
the computational time for the implicit update in the current code is very 
 expensive (using our best available resources),
we are now forced to employ a quite coarse resolution.
 It is important to clarify how the violation of the total energy conservation 
would be improved with increasing numerical resolution. We leave this for future
 work.

\begin{figure}[htpb]
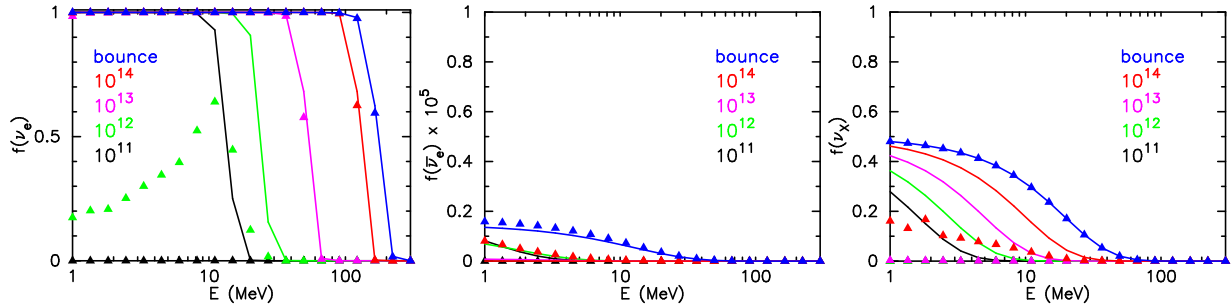

\begin{center}
\includegraphics[width=40mm,angle=-90.]{f4.eps}
\includegraphics[width=40mm,angle=-90.]{f5.eps}
\includegraphics[width=40mm,angle=-90.]{f6.eps}
\end{center}
  \caption{Neutrino distribution function $f(\nu,\varepsilon)$ (filled triangles)
and Fermi-Dirac distribution function at equilibrium
 (solid lines) at the innermost mesh for model 1DGR.
Lines and triangles are color coded according to the infall phase.
Note that $f(\nu,\varepsilon)$ for anti-electron neutrino ($\bar{\nu}_e$)
is multiplied by $10^5$ for comparison.}
\label{fig:DistF.eps}
\end{figure}

 Fig. \ref{fig:DistF.eps} shows a spectral shape of the 
neutrino distribution function $f(\nu,\varepsilon)$ (filled triangles),
\begin{eqnarray}
f{(\nu,\varepsilon)}=\frac{J_{(\nu,\varepsilon)}}{4\pi\varepsilon^3},
\end{eqnarray}
  which is estimated at the innermost grid point of model 1DG15 when
 the central density $\rho_c$ reaches $10^{11,12,13,14}$ g cm$^{-3}$,
 and at bounce, respectively. Solid curves represent
 the Fermi-Dirac distribution at equilibrium. 
From the left panel, it can be seen that 
  $\beta$-equilibrium is achieved 
 for electron-type neutrino ($\nu_e$) at 
$10^{12} {\rm g}\,{\rm cm}^{-3} \lesssim \rho_c \lesssim 10^{13} {\rm g}\,{\rm cm}^{-3}$,
which is consistent with the neutrino trapping density as shown in Fig.
 \ref{fig:RhoC.eps}. In \cite{Bruenn85}, the $\beta$-equilibrium for $\nu_e$
 was obtained after $\rho_c=2.46 \times 10^{12}$ g cm$^{-3}$ in their 
corresponding model to ours (``standard'' model).
Note that at bounce, electron type neutrinos, in 
the low energy range $\la5$ MeV, slightly violate the fermi blocking, i.e.,
the distribution function $f{(\nu_e)}$ exceeds one.
The excess, however, is $\la10^{-5}$ and we consider that it is a 
negligible amount.
 As shown from the middle ($\bar\nu_e$) and right panel ($\nu_X$) of Fig.4,
 other neutrino species are thermalized only after $\rho_{\rm c}$ 
exceeds $10^{14}$ g cm$^{-3}$. These features are quantitatively 
consistent with \citet{bruenn97,Rampp02}. It may be surprising that 
regardless of the use of 
 different EOS and different hydrodynamics codes, the trapping density of
 $\nu_e$ ($\rho_{\rm trap} = 2 \sim 3 \times 10^{12}$ g cm$^{-3}$)
in modern simulations (e.g., top panel of Fig.3) is in good agreement
 with the pioneering work in the 1980's \citep{Bruenn85}.
 It should be also
noted that for the more 
accurate determination of the core deleptonization 
 the improved electron capture rates 
\citep{Langanke00,juoda} need to be implemented as in \citet{langanke03,lentz12a,lentz12b}.

So far, we have shown that our M1 scheme can capture several important
 phenomena regarding deleptonization, such as the neutrino trapping and
 the conservation of the lepton fraction in
 the diffusion region.
As already denoted in Sec. \ref{sec:Radiation Hydrodynamics},
this is not trivial in the finite difference method especially
 when one transports
conservative radiation moments $(E_{(\varepsilon)},{F_{(\varepsilon)}}_i )$
instead of the corresponding comoving variables of
 $(\mathcal J_{(\varepsilon)},{\mathcal H_{(\varepsilon)}}_i )$.
A key is to find an appropriate 
Eddington factor $\chi_{(\varepsilon)}$
by which the neutrino energy flux approaches 
${F_{(\varepsilon)}}^i \rightarrow 4/3E_{(\varepsilon)}v^i$ in the diffusion limit
(see, Eq.(\ref{eq:SlowMotionlimitH})).
Since this relation can be achieved only when $P^{ij}_{(\varepsilon)}=E_{(\varepsilon)}\gamma^{ij}/3$ holds,
$\chi_{(\varepsilon)}$ and $\bar F$ should approach 1/3 and 0 
(e.g., our closure relation (Eq. (\ref{neu_p}))) in the limit,
 respectively.

\begin{figure}[htpb]
\begin{center}
\includegraphics[width=60mm,angle=-90.]{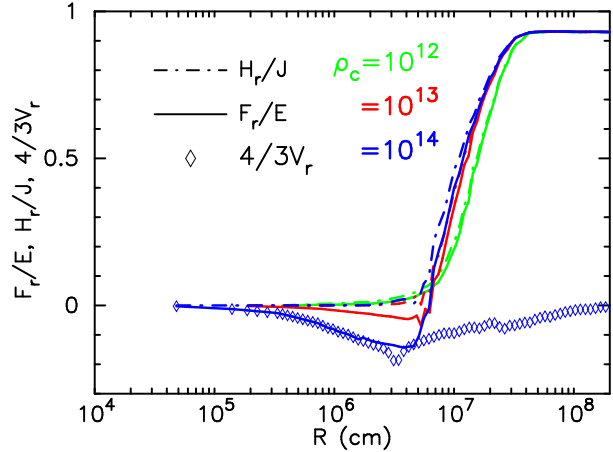}
\end{center}
  \caption{Radial profiles of $\langle F_r\rangle/\langle E\rangle$ (solid lines)
and $\langle H_r\rangle/\langle J\rangle$ (dash-dotted lines) for electron-type 
 neutrino ($\nu_e$)
when the central density reaches $\rho_c=10^{12,13,14}$ g cm$^{-3}$ in
 model 1DG15.
$V_r$ denotes the radial velocity of matter (see text).}
\label{fig:JEoverHF.eps}
\end{figure}

To show that both $\bar F=\sqrt{h_{\mu\nu}H^\mu_{(\varepsilon)} 
H^\nu_{(\varepsilon)}/J^2_{(\varepsilon)}} $ and ${F_{(\varepsilon)}}^i$ actually 
approach 0 and $4/3E_{(\varepsilon)}v^i$ in the opaque region,
we plot in Fig. \ref{fig:JEoverHF.eps} the radial profiles of $\langle F_r\rangle/\langle E\rangle$ (solid lines)
and $\langle H_r\rangle/\langle J\rangle$ (dash-dotted lines)
 at different $\rho_c$.
Here  $\langle X\rangle\equiv\int d\varepsilon X$ represents the 
energy integration of $X$.
At $\rho_{\rm c}=10^{12}$ g cm$^{-3}$, both solid and dash-dotted green lines
almost coincide.
However, as the infalling matter velocity comes closer to be 
relativistic ($\rho_{\rm c} 
\gtrsim 10^{13}$ g cm$^{-3}$),
both lines start to split especially within $R\lesssim70$ km. In the 
 central region ($R\lesssim70$ km), $\langle H_r\rangle/\langle J\rangle$ 
approaches 0 towards the center, whereas 
$\langle F_r\rangle/\langle E\rangle$ becomes negatively 
large with the peak being around $R \sim 30$ km and then converges to
 zero to the center. We also plot the radial velocity of matter ($V_r$) 
measured in the Eulerian frame 
and multiplied by $4/3$ (blue diamonds) at $\rho_{\rm c}=10^{14}$ g cm$^{-3}$.

Because of our appropriate evaluation for the Eddington factor, 
the flux factor measured in the Eulerian frame
(approximated here by $\langle F_r\rangle/\langle E\rangle$) 
 nicely matches with $4V_r/3$ in the optically thick region.
 This neutrino advection is essentially important
 for the radiation energy ($E_{(\varepsilon)}$) to move with the same velocity 
$v^i$ with matter in the opaque region
(see, Eqs. (\ref{eq:SlowMotionlimitH})-(\ref{eq:SlowMotionlimit:eq:rad1_3})).
Here we shortly comment on the definition of the flux factor $\bar F$.
In our previous study \citep{KurodaT12}, we employed the definition
$\bar{F}\equiv\sqrt{\gamma^{ij}F_iF_j}/E$ which is one of the 
candidates for $\bar F$ \citep{Shibata11}.
 As can be clearly seen from the split
  between the solid and dashed-dotted lines in Fig. \ref{fig:JEoverHF.eps},
 we show that our previous choice of the flux factor 
is not adequate because 
 the optically thick medium moves (albeit mildly) relativistically in the 
collapsing core.

\subsubsection{After bounce}\label{aft_bounce}

\begin{figure}[htpb]
\begin{center}
\includegraphics[width=100mm,angle=-90.]{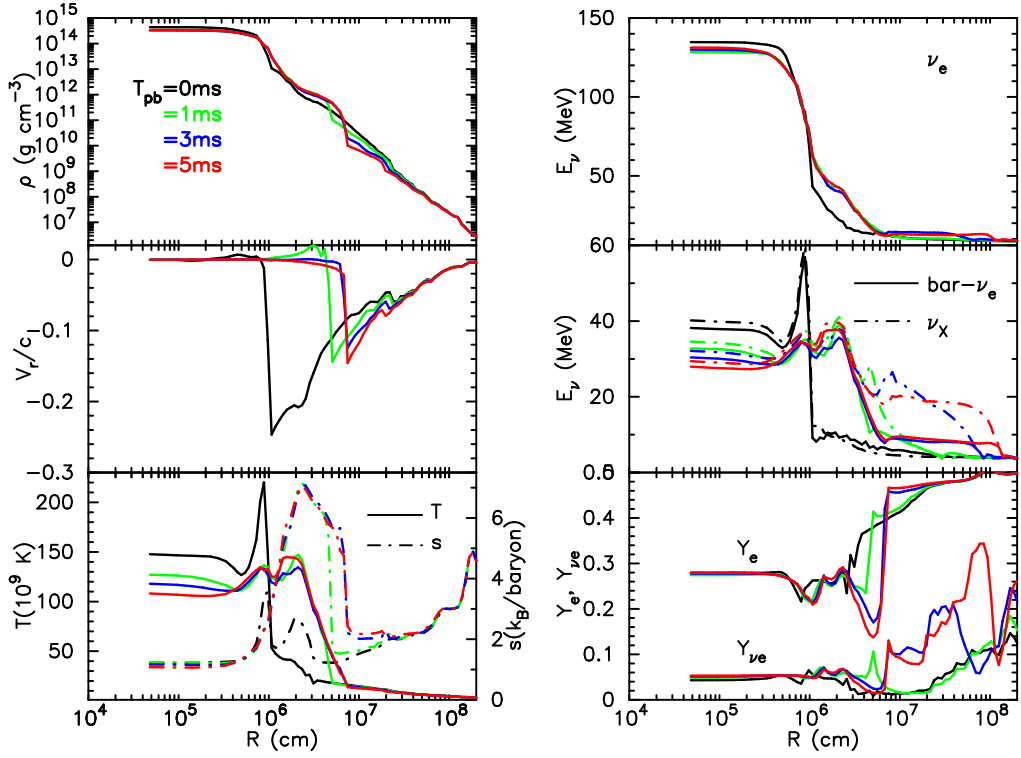}
\end{center}
  \caption{Radial profiles of the rest mass density $\rho$,
the radial velocity $v_{\rm r}$ normalised by the speed of light $c$, the matter temperature $T$, the entropy $s$,
the averaged neutrino energy $E_\nu$ and the electron/neutrino fractions $Y_e/Y_l$ at selected time slices shortly after bounce $T_{\rm pb}$ = 0, 1, 3, 5 ms.
 Note in this plot that profiles only along the $x$-axis of model 1DG15 are 
 shown (because model 3DG15 shows almost the same profiles).}
\label{fig:ProfilePB.eps}
\end{figure}

In Fig. \ref{fig:ProfilePB.eps}, we show
the radial profiles of various quantities (the rest mass density $\rho$,
 radial velocity $v_{\rm r}$, matter temperature $T$, entropy $s$,
 averaged neutrino energy $E_\nu$ and electron/neutrino fraction $Y_e/Y_l$)
 at the selected time slices shortly after bounce.
Here, we define the average neutrino energy $E_\nu$ as
\begin{eqnarray}
E_\nu\equiv \frac{\int d\varepsilon \,\varepsilon^3\,f{(\nu,\varepsilon)}}{\int d\varepsilon\, \varepsilon^2 f{(\nu,\varepsilon)}}.
\end{eqnarray}

When the central density exceeds nuclear saturation densities (top left 
 panel of Fig.\ref{fig:ProfilePB.eps}), the core bounces
 because of the strong repulsive force of nuclear matter. Then
 the bounce shock is formed (green line, 
left middle panel of Fig.\ref{fig:ProfilePB.eps}) which propagates outward
with dissociating infalling heavy nuclei into free protons and neutrons.
Production of enormous free protons/neutrons significantly enhances electron 
capture process, $e^-p\rightarrow\nu_en$,
behind the stalling shock.
Immediately after bounce, those high-energy neutrinos 
(top right panel) are still trapped inside the 
optically thick medium behind the shock. The medium, however, quickly 
becomes transparent to neutrinos and those 
neutrinos are liberated suddenly
as a burst. This neutronization burst enhances further electron capture
 behind the shock due to the
continuous deviation from the $\beta$-equilibrium, leading 
 to a characteristic trough in the $Y_e$ profile seen at $R\sim50$ km
 behind the shock (right bottom panel of Fig.\ref{fig:ProfilePB.eps}). 
Due to the energy loss by the photodissociation of the 
 iron nuclei and the rapid neutrino leakage,
the bounce shock stalls at $R \sim 70$ km within 
$T_{\rm pb}\sim 3$ ms (left panels of Fig.\ref{fig:ProfilePB.eps}).
Such dynamical features are commonly seen in the reference models 
 \citep{Liebendorfer05,BMuller10}. This indicates
that our M1 scheme can capture the basic behaviours of the 
 neutrino propagation from the optically thick to thin medium (otherwise
 it would result in either the absence
or the different position of the $Y_e$ trough).

\begin{figure}[htpb]
\begin{center}
\includegraphics[width=60mm,angle=-90.]{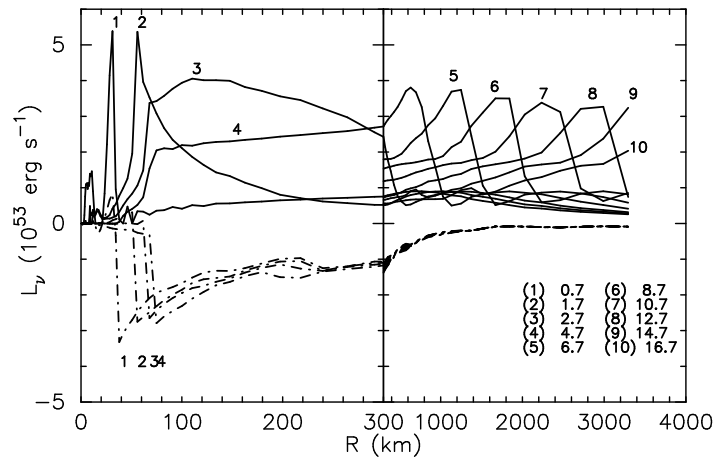}
\end{center}
  \caption{Radial profiles of the (electron-type) neutrino energy flux (solid lines)
and the radial velocity (dash-dotted lines).
Numbers beside each line (1,2,3..) correspond to time slices, which 
are denoted in the lower right part with $T_{\rm pb}$ in ms (0.7, 1.7, 2.7..).
For the radial velocity in left panel, we plot only the first four time slices.}
\label{fig:EnergyFlux.eps}
\end{figure}

How the neutronization burst is produced is more clearly 
 depicted in Fig. \ref{fig:EnergyFlux.eps} where
we plot the radial profiles of the neutrino 
($\nu_e$) energy flux (solid lines) and the radial velocity of matter 
(dash-dotted lines).
At $T_{\rm pb}=0.7$ ms, enormous neutrinos are still trapped and confined behind
 the shock, which is shown as a
 sharp peak in the energy flux around $10\lesssim R\lesssim20$ km.
At $T_{\rm pb}=1.7$ ms, these neutrinos overtake the shock front 
because of the lowering opacity outward.
Then the pulse of the neutrino burst propagates freely to the optically thin region (time label larger than 4)
and eventually emerges out of the computational domain (labels 9 and 10).
These profiles are consistent with the results of \citet{Bruenn91}, \citet{Thompson03}, and \citet{Rampp02}.

\begin{figure}[htpb]
\begin{center}
\includegraphics[width=80mm,angle=-90.]{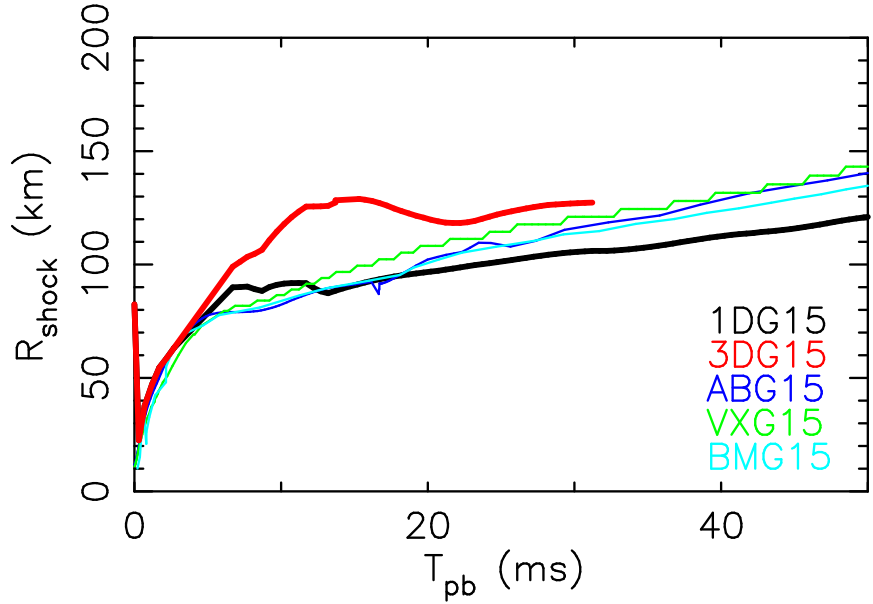}
\end{center}
  \caption{Time evolution of the shock radius for five different models 
1DG15, 3DG15, ABG15, VXG15, and BMG15.
  As for our models, we plot the angle averaged values.}
\label{f8}
\end{figure}
Now we move on to make the code comparison in the postbounce phase.
 Fig.\ref{f8} compares the evolution of the shock radii for five models;
 two from our code (1DG15 (black line) and 3DG15 (red line)), one 
from AGILE-BOLTZTRAN (ABG15, blue line), and two from VERTEX 
 with PROMETHEUS (VXG15, green line) and with CoCoNuT (BMG15, 
light blue line). Note that 
the final simulation time ($T_{\rm sim}$) is 50 ms for model 1DG15
 (black line), whereas 
 $T_{\rm sim}$ is 32 ms for model 3DG15 (red line) 
simply limited by our available 
 computational resources.

The deviation seen in model 3DG15 (red line in Fig. \ref{f8}) from the rest of the 
 1D models is remarkable especially after $T_{\rm pb} \sim 5$ ms. This is because 
the bounce shock expands more energetically in 3D pushed primarily by 
prompt convection behind the shock.
Using the same progenitor, \citet{BMuller12a} showed that
 the average shock radius becomes larger in 2D (their model G15) 
than in 1D (their model G15-1D) at $T_{\rm pb} \sim 70$ ms
because of the hot-bubble convection starts which is seeded during the deceleration of the prompt shock.
The Cartesian coordinates have the intrinsic quadrupole perturbation
and it affects significantly on the growth of the prompt convection.
The post bounce time, when the significant multi dimensionality appears,
thus differ between our model and the ones using the spherical polar coordinates.
In addition, our coarse numerical resolution might also lead to the earlier appearance of the initial convection
because of even larger seed perturbations.

The larger shock radius in model 3DG15 than that in 1DG15 is also consistent with 
our previous result with leakage scheme (\citet{KurodaT12}, see also
 \citet{couch13a,Hanke12} 
for extensive discussion
 about the dimensional dependence on the postbounce dynamics).
Now let us focus on our (pseudo-)1D model (black line in Fig. \ref{f8}).
The shock radius of our code is in good agreement with the reference results
  exceptionally before $T_{\rm pb} \lesssim 20$ ms, whereas the shock radius
tends to be smaller until the end of the simulation time. 
We consider that the difference could primarily come from the use of the Cartesian
coordinates with low numerical resolution and not from the neutrino-matter interaction terms.
This is because our calculation in 1D spherical coordinates with using the same neutrino-matter interaction terms
shows a good agreement in the shock evolution (see, Appendix \ref{sec:Comparison in 1D Spherical Coordinate}).
We give a more detailed discussion elsewhere below.


 \begin{figure}[htpb]
\begin{center}
\includegraphics[width=80mm,angle=-90.]{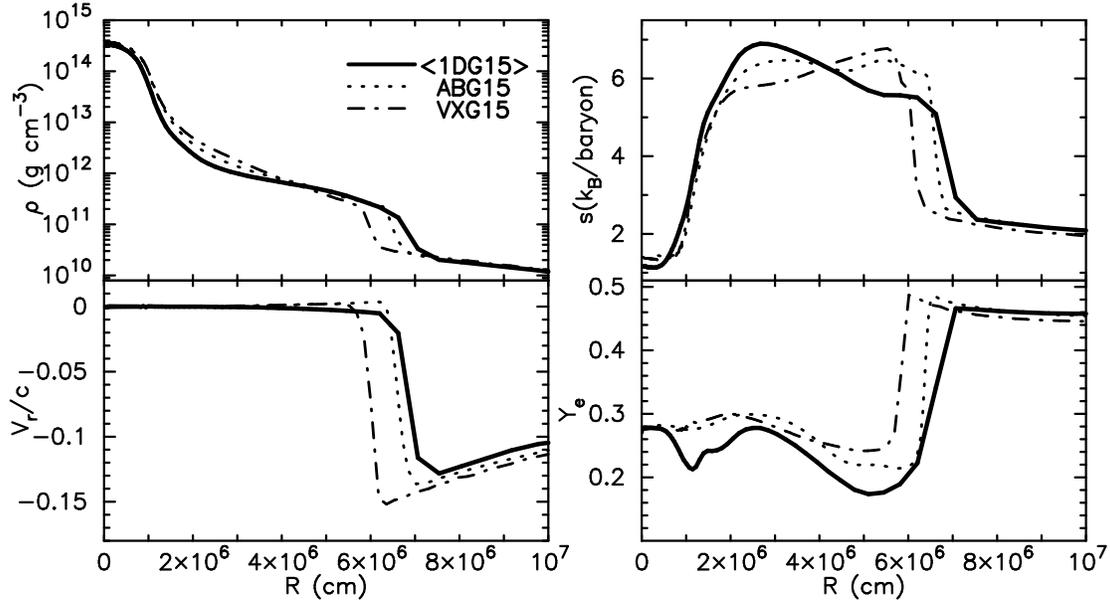}
\end{center}
  \caption{In the clockwise direction from the top left panel, we show
 radial profiles of the (angle averaged) rest mass density $\rho$,
  entropy $s$, electron fraction $Y_e$, and radial velocity $V_{\rm r}$ 
  at $T_{\rm pb}$ = 3 ms
  for models ``$\langle$1DG15$\rangle$'', ``ABG15'', and ``VXG15''.}
\label{f9}
\end{figure}

\begin{figure}[htpb]
\begin{center}
\includegraphics[width=80mm,angle=-90.]{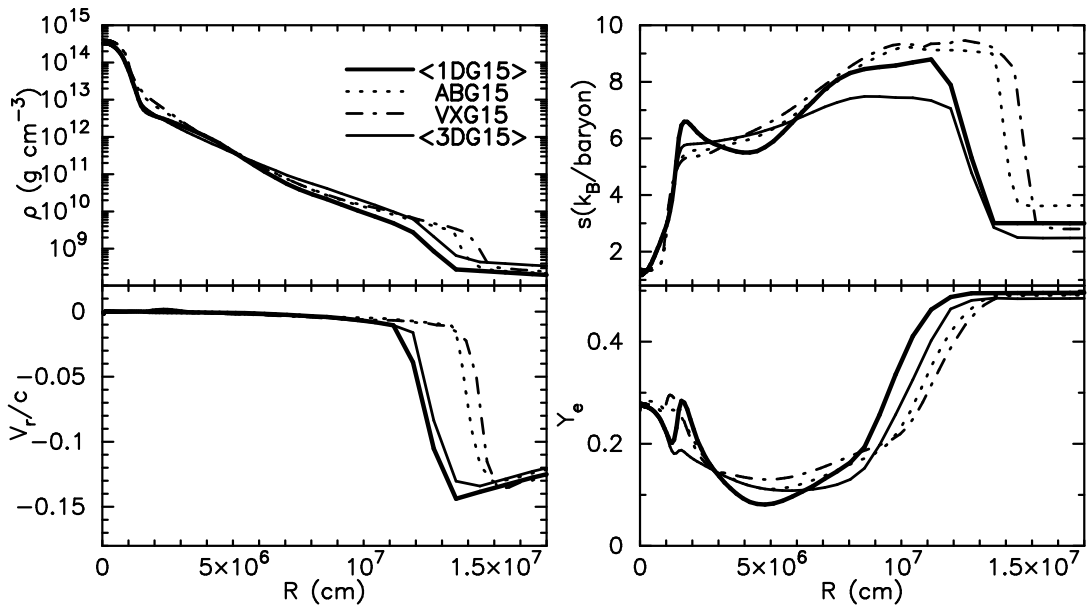}
\end{center}
  \caption{Same as Fig. \ref{f9}, but at $T_{\rm pb}$ = 50 ms
  for models ``$\langle$1DG15$\rangle$'', ``ABG15'' and ``VXG15''.
  ``$\langle$3DG15$\rangle$'' at $T_{\rm pb}$=32 ms is also plotted as a reference by a thin solid line.}
\label{f10}
\end{figure}

In Fig. \ref{f9} and \ref{f10}, we show various quantities 
 from our code (labeled by  ``$\langle$1DG15$\rangle$''
 and ``$\langle$3DG15$\rangle$'' only in Fig. \ref{f10}), AGILE-BOLTZTRAN (``ABG15''),
 and VERTEX-PROMETHEUS (``VXG15'') at two different postbounce times.

At 3 ms after bounce (Fig.\ref{f9}), the (angle-averaged) radial 
position of the stalled shock (bottom left panel) is $R \sim 70$ km for
 model 1DG15 (thick solid line). 
 As seen, the velocity profile matches more closely the profile from AGILE-BOLTZTRAN
 (ABG15) than from VERTEX-PROMETHEUS (VXG15). This is also the case for the
 profiles of the density (top left panel), entropy (top right panel), and 
 $Y_e$ (bottom right panel). It should be noted that the more 
recent results from the VERTEX-PROMETHEUS code with an improved GR potential
 \citep{Marek06} agree very well with the AGILE-BOLTZTRAN code, hence 
 with our code. Therefore we mainly compare to model ABG15 in
 the following.

Looking at Fig. \ref{f9} more closely, one can see that 
the profiles of our entropy (top right panel)
 and $Y_e$ (bottom right panel) differ appreciably from model ABG15 
especially in the region behind the stalled shock ($R \lesssim 70$ km) and
 above the unshocked inner core ($R \gtrsim 10$ km).
Let us remark that the early postbounce evolution starting from 
the shock formation, followed by the emergence of the neutronization burst, 
until the shock stall is numerically most challenging.
The code difference from the shock capturing scheme as well as the 
treatment of GR, the accuracy of the neutrino transport schemes 
could potentially impact the radiation-hydrodynamics evolution at the transient phase
 (i.e., 3 - 5 ms after bounce).
Another remarkable difference is seen in the electron fraction.
Among three results, only our result shows negative bump in $Y_e$ profile at $10\lesssim R\lesssim20$ km.
However this bump disappears in our 1D spherically symmetric test problem
in which we solve the advection terms in energy space ${\bf S_{\rm adv,e}}$ explicitly in time
(see, Appendix \ref{sec:Comparison in 1D Spherical Coordinate} for more detail).

Regarding the shock capturing, AGILE-BOLTZTRAN
uses an artificial viscosity type with
 the second order accuracy in space, whereas our code employs the approximate Riemann solver 
(HLLE scheme like the VERTEX code) with the second-order accuracy in space both for radiation-hydrodynamical
and geometrical variables.
Thus the hydrodynamics part of our code is slightly more accurate than AGILE-BOLTZTRAN.
On the other hand, the use of the approximate closure relation 
 apparently falls behind the Boltzmann code especially 
in the semi-transparent region. Above all, the use of 
the Cartesian coordinates, which is very common
 in full-GR simulations\footnote{See, however, \citet{gual} about recent
report of the code development of numerical 
relativity in the polar coordinates.}
 makes the comparison
 to the genuine ``1D'' results of the reference models (based on the 1D 
 Lagrangian code (AGILE) and the multi-D code using the polar coordinates
 (VERTEX)) even more challenging. 



At $T_{\rm pb}=50$ ms (Fig.\ref{f10}), 
the differences between 
 our pseudo-1D model (1DG15, thick line) and the reference results
 still remain to be seen rather remarkably in the postshock
 region ($R \gtrsim 100$ km), however this is not surprising
 given the different 
 shock evolution (Fig. \ref{f8}). Here we consider that the numerical
 resolution in the postshock region sensitively affects the shock
 evolution. In the current resolution, 
the typical grid size of our nested box is $\sim$ 7.8 km 
at $120\lesssim R\lesssim 240$ km ($\sim 4^\circ$ resolution). 
As shown in Fig.\ref{f8} (red line),
 the deviation of the shock radius from the reference models 
becomes remarkable at $T_{\rm pb} \gtrsim 20$ ms, which roughly coincides
 with the time when the shock reaches to the coarser level of the nested grid.
 There the shock front is resolved only by a few grid cells.
We consider that at least a factor 
of two or more higher resolution is required to reproduce 1D results,
 i.e., to recover the sphericity of the system in the 
Cartesian coordinates. But this is not an easy task as we will discuss
the code performance in section \ref{sec:Conclusions}.


 From Fig. \ref{f10}, one can also see that 
the profile of the density, velocity, 
 entropy, and $Y_e$ at the central region ($R \lesssim 60$ km) agrees with 
 the reference results roughly within an accuracy of $\sim 10 \%$. 
Given the insufficient numerical resolution, this good agreement in the 
 semi-transparent region would support the validity of the prescribed
 closure scheme, which is in line with the recent results by \citet{O'Connor14} and \citet{just15}.
 Fig.\ref{f10} also shows that
 the profile of model 3DG15 (thin solid line) differs from that of model 1DG15 
 (thick solid line). 
   
  The entropy profile for model 3DG15 (thin line) behind the shock 
($R \lesssim 130$ km) becomes more flat compared to the 1D models, which 
 is due to convective mixing behind the shock. 
 As a result, the profiles of $Y_e$ and entropy in model 3DG15  
 become slightly closer to the reference results, though the similarity is just a coincidence and is not meaningful.





\begin{figure}[htpb]
\begin{center}
\includegraphics[width=140mm,angle=-90.]{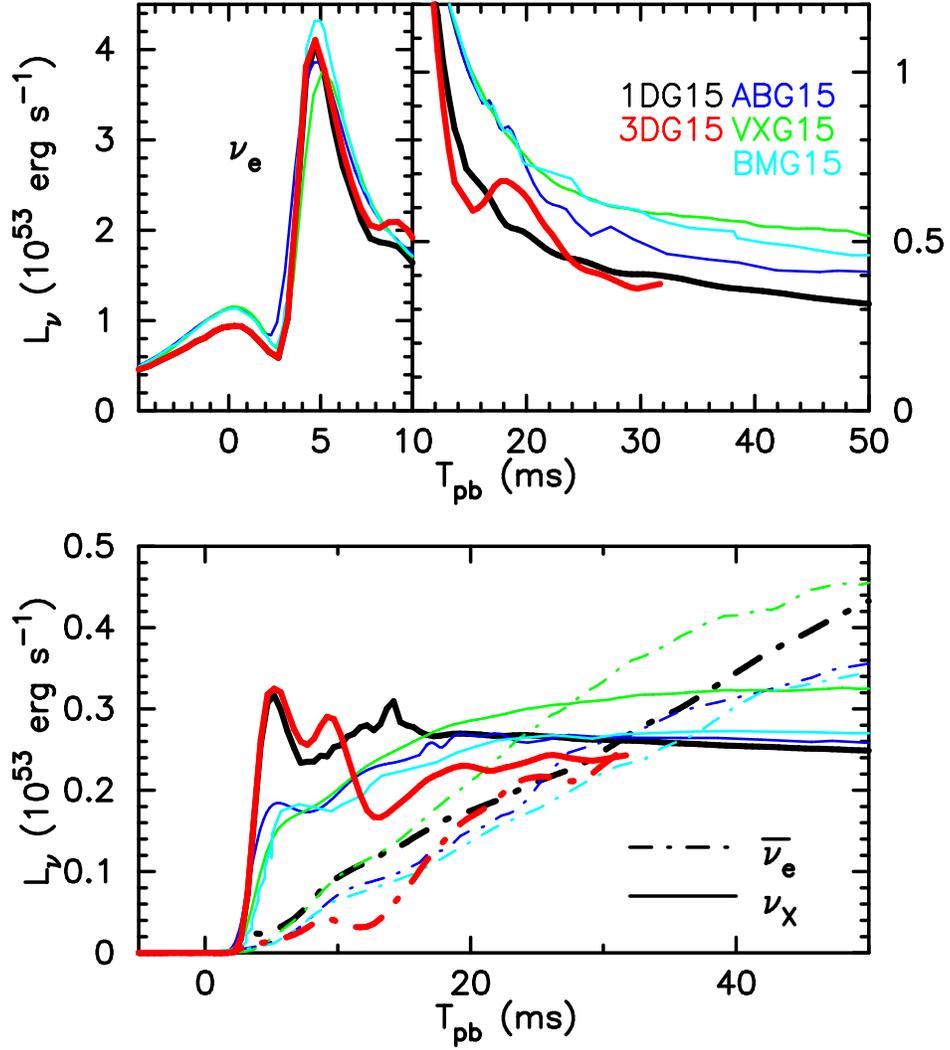}
\end{center}
  \caption{Same as Fig. \ref{f8} but for 
the neutrino luminosity $L_\nu$ of electron type neutrinos (top panel) and anti-electron type and heavy-type neutrinos (bottom panel)
  for the five different models.}
\label{fig:Lnu.eps}
\end{figure}

Fig.\ref{fig:Lnu.eps} shows the comparison of 
the neutrino luminosity $L_\nu$ which is the surface 
integral of the energy integrated comoving energy flux $H^{\mu}_{(\varepsilon)}$
through the surface of cubic box with 1000 km width
(i.e. approximately 500 km from the origin).
The peak luminosity, $L_{\nu_e,{\rm peak}}=4.1\times10^{53}$ erg s$^{-1}$, well 
agrees with $3.85\times10^{53}$ (ABG15), $3.8\times10^{53}$ (VXG15),
 and $4.3\times10^{53}$ (BMG15), respectively.
After the neutronization phase when the 1D core enters to 
a quasi-static phase ($T_{\rm pb} \gtrsim 20$ ms),
 the anti-electron and heavy-lepton neutrino luminosities show 
quite consistent behaviors with the reference models and the differences 
between our results and, e.g., ABG15 are as small as $\sim10$ \%.
 
Regarding the $\nu_e$ luminosity, we see a systematically lower value than the other three reference models.
The difference reaches $\sim30$ \% (or $\sim10^{52}$ erg s$^{-1}$) compared to ABG15 at $T_{\rm pb}=50$ ms.
As a consequence, $\nu_e$ and $\bar\nu_e$ luminosities become comparable at $T_{\rm pb}\sim45$ ms which is significantly
earlier than the reference values $T_{\rm pb}\sim70$ ms in ABG15, BMG15 and also in a model of \citet{O'Connor14}.
This means that the lepton number loss from PNS is less than those previous studies since
it can be measured roughly by the time integration of
$L_{\nu_e}/\varepsilon_{\nu_e}- L_{\bar\nu_e}/\varepsilon_{\bar\nu_e}$.
However, our 1D spherical test, albeit in the Newtonian limit, shown in Appendix \ref{sec:Comparison in 1D Spherical Coordinate},
does not exhibit such inconsistency and show quite reasonable neutrino profiles with a previous study.
We therefore consider that the reason for less agreement, especially seen in $L_{\nu_e}$,
mainly comes from the spatial advection term in the Cartesian coordinates and not from the local neutrino matter interaction terms.



\begin{figure}[htpb]
\begin{center}
\includegraphics[width=90mm,angle=-90.]{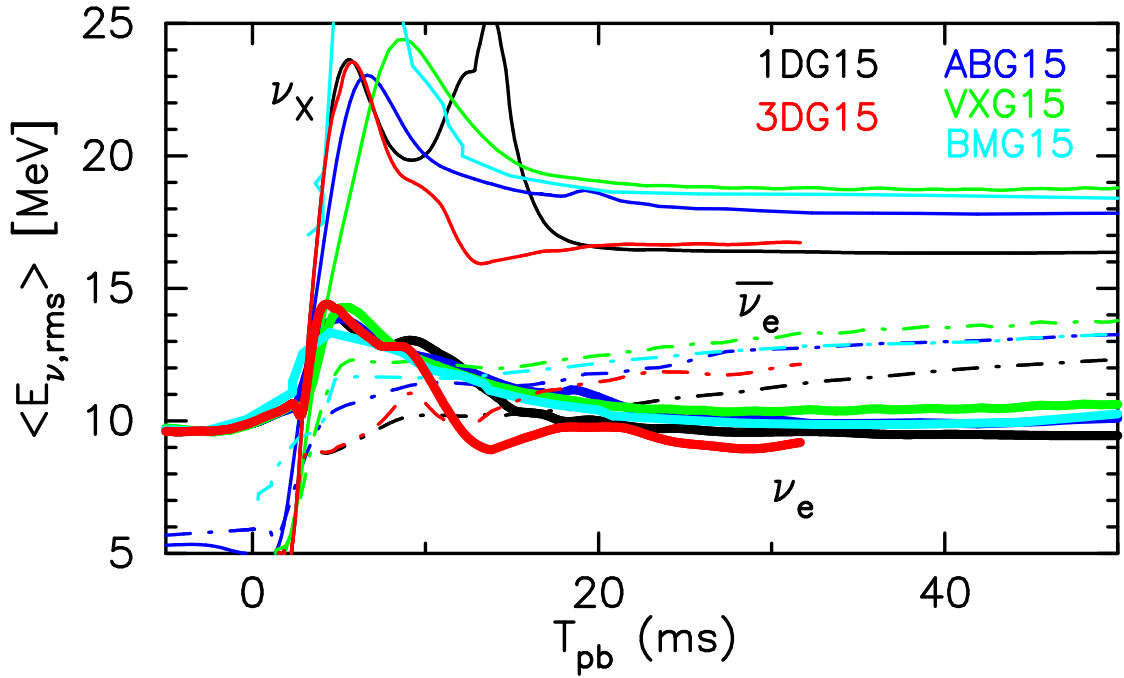}
\end{center}
  \caption{Time evolution of RMS energies of emergent
neutrinos $E_{\nu,{\rm rms}}$ for five different models.
Our results are measured at the surface of cubic box with 1000 km width.}
\label{fig:Enu.eps}
\end{figure}

We also compare the neutrino spectral difference in Fig. \ref{fig:Enu.eps}.
In the figure, we show time evolutions of the root-mean-square (RMS) energies of emergent
neutrinos $E_{\nu,{\rm rms}}$ measured at the surface of cubic box with 1000 km width.
For $E_{\nu,{\rm rms}}$, we use the same definition as in \cite{Liebendorfer05}
and it is defined as below.
\begin{eqnarray}
E_{\nu,{\rm rms}}\equiv \sqrt{\frac{\int d\varepsilon f{(\nu,\varepsilon)}\varepsilon^4}{\int d\varepsilon f{(\nu,\varepsilon)}\varepsilon^2}}.
\end{eqnarray}
Again in Fig. \ref{fig:Enu.eps}, we plot averaged value $\langle E_{\nu,{\rm rms}}\rangle$ over the surface of the cubic box.
From the figure, we find good agreement with the reference codes 
and differences are less than $\sim$1 MeV
for $\nu_e$ and $\bar\nu_e$ and $\lesssim$2 MeV for $\nu_X$
after the neutronization phase ceases $T_{\rm pb}\gtrsim20$ ms.
In our model 1DG15, we see a spurious second peak in $E_{\nu_X,{\rm rms}}$ profile
around $T_{\rm pb}\sim13$ ms.
However, the second peak disappears in our model 3DG15 and we thus consider that it is most likely due to our artificial treatment for
the matter velocity and is insignificant.

\section{Summary}
\label{sec:Conclusions}
In this paper, we have presented
 our newly developed multi-D full GR neutrino-radiation-hydrodynamic code.
 The code was designed to evolve the Einstein field
equation together with the GR radiation hydrodynamic equations
in a self-consistent manner while satisfying the Hamiltonian and momentum 
constraints. Using an M1 closure scheme, 
we solved spectral neutrino transport of the radiation energy and momentum 
in conservative way based on a truncated moment formalism. 
Beside the energy and momentum conservation, we paid particular 
attention to the lepton number conservation, especially in the neutrino trapping regime.
 
We explained formally that the neutrino number is transported appropriately
by solving only the energy and momentum conservation equations of the neutrino radiation field,
especially focusing on the neutrino trapping regime.
In addition, we showed that the advection terms in energy space are essential for reproducing the neutrino trapping.

To validate our new numerical code, we first made bottom-line tests such as
the diffusion test, shadow casting test and spherical propagation in free streaming regime to check the advection term in space.
In addition, as for the important factor in CCSN simulation, gravitational redshift and Doppler shift are
tested in a curved space-time with a sharp velocity profile.
Through these standard tests, we confirmed that our code is capable to reproduce analytical results accurately
and that all source terms other than the neutrino-matter interaction terms are correctly implemented.

We then performed a practical core collapse simulation
with which we could also confirm whether the above mentioned neutrino number conservation,
especially in the trapping region, is satisfied adequately.
We followed gravitational collapse, 
bounce and initial post bounce phase up to $T_{\rm pb}\sim50$ ms of a 
15M$_\odot$ star to make a detailed comparison with previous studies 
with Boltzmann neutrino transport.
Regarding neutrino opacities,
we currently employed baseline set where inelastic neutrino-electron scattering, 
thermal neutrino production via pair annihilation and 
nucleon-nucleon bremsstrahlung were included.
We started the code comparison before core bounce.
 Important features such as the evolution of the central $Y_l$ and entropy
 as well as the neutrino trapping density showed nice agreement
 with the reference models. Next we made the code comparison after 
bounce till the 
stall of the bounce shock. At this transient phase, we checked 
 that the M1 code can capture the overall evolution in the Boltzmann codes
 regarding the neutrino propagation from the opaque to the transparent region.
Considering our insufficient resolution and the difference coordinates
 used, the neutrino luminosity and the energy spectrum of our code showed 
 good agreements with the reference results.
The peak luminosity ($L_{\nu_e,{\rm peak}}=3.9\times10^{53}$ erg s$^{-1}$) showed
almost the same value with the reference models.
After the neutronization, the neutrino luminosities showed 
consistent values though there is a systematic lower shift in the electron-type neutrino luminosity.
 The RMS energies of the emergent neutrinos showed a similar level of the
 match with the reference results, which demonstrates the validity of our code.

In the end, we shortly discuss for the future run 
(using Exa-scale platforms) in order to check a numerical convergence of 
our 3D results and to get a more closer match with the Boltzmann results 
in the much longer postbounce phase. 
 As already mentioned, we were forced to adopt a low numerical
 resolution in this study (effective angular resolution behind the stalled 
 shock $\sim120$ km is $\sim$ 8km/120km $\sim $3.8$^\circ$) due to our 
limited computational resource.
 The code has been already tuned to get a 
high parallel efficiency ($\gtrsim 90$ \%, e.g., going from 2048 to 4096 cores)
with a very high performance efficiency ($\sim 32$ \%) which measures 
the ratio of the real performance of our code
 to the theoretical peak performance measured by the platform, Cray XC30 in our case.
However, it still takes $\sim$ 1.25 CPU days to follow $\sim$ 1 ms 
postbounce in model 3DG15 by the peta-flops machine occupying $\sim$ 10\% of its total resource (2048 processors).
 Just only comparing the angular resolution $\lesssim2^\circ$ of recent 
numerical studies \citep[e.g.,][]{Hanke13,BMuller14,Takiwaki14}, 
 the employed resolution in this study is approximately two times coarser.
 Currently, the wall clock time per each timestep in 
 the postbounce phase (the typical numerical time step $\Delta t=10^{-7}$ s) is $\sim5$ s if we use 4096 processors
 (i.e., we can follow the postbounce time of $\sim$1.7ms per day, which we write as 1.7ms/day in the following).
 With this performance, we estimate how much computational resource and how many computational
 time are required for the twice high resolution model
 to follow $\sim200$ ms after bounce, e.g., until the shock revival is expected to occur
for low-mass progenitor stars in 3D (e.g., \citet{Takiwaki14,melson15}).
 Since our current resolution at the 
origin $\Delta x\sim480$ m is marginally acceptable,
 we may just need to double the number of the 
numerical meshes in each nested box.
 
 If we are able to double the resolution with the fixed innermost mesh size
 and use $4096\times2^3\sim32,000$ processors,
 we can also follow the same postbounce evolution $\sim$1.7ms/day,
 or $\sim$3.4ms/day occupying $4096\times2^3\times2\sim64,000$ processors.
 Even if we can luckily take the latter case ($\sim$3.4ms/day), 
 we still need approximately two months to follow $\sim200$ ms after 
bounce occupying the $\sim 64,000$ processors.
 For more massive progenitors, the shock revival in 3D models
would be much delayed than in 2D ($T_{\rm pb} \gtrsim 600 $ ms, e.g., \citet{Marek06,BMuller12a,BMuller12b,Nakamura2D14}). 
This is apparently beyond the maximum computational time allocated to us 
in the K computer and surely needs Exa-scale platforms (in the next decade
 to come).

Before the advent of these next-generation supercomputers, it is noted 
 that we actually have lots of tasks to improve our code. We expect that 
we can still enhance the numerical efficiency 
especially when we get a convergent solution during the 
Newton-Raphson iteration in the implicit update. 
At present, the convergence becomes worse where the
 energy fluxes of neutrinos $F^\mu/H^\mu$
 become non-negligible (i.e., just above the neutrino sphere).
 We have confirmed that the neutrino-election scattering 
is one of the dominant factors which delays the convergence. 
With an eye towards
 actual application of this code in CCSN simulations, we plan to
 continue our code refinement, in which we not only need to 
 employ a more elaborate set of neutrino opacities 
(e.g., \citet{horowitz02,Burrows06,sumi08,gabriel12,tobias13,barti})
 with including the higher-order angular dependence of the
 reaction angle that was omitted for simplicity (e.g., 
Sec. \ref{sec:Numerical Setups}), but also to 
find a more efficient algorithm to deal with the 
resulting (more complicated) Jacobi matrix in the implicit update. 
This study is only our very first step towards a more realistic coding,
 which is indispensable for quantitative study of 
 stellar core-collapse and the explosion mechanism.

\acknowledgements
This work was supported by the European Research Council (ERC; FP7) under ERC
Advanced Grant Agreement N$^\circ$ 321263 - FISH.
TK acknowledges fruitful discussions with M. Liebend\"orfer, R. M. Cabez\'on,
 M. Hempel, and K.-C. Pan.
 We acknowledge A. Imakura for his useful 
advise for an efficient matrix inversion scheme.
TT and KK are thankful to K. Sato and S. Yamada 
for continuing encouragements.
Numerical computations were carried out on Cray XC30 at Center for Computational Astrophysics, National Astronomical Observatory of Japan.
 This study was supported by the Ministry of Education, Science and 
Culture of Japan (Nos. 24103006, 24244036,  26707013, and 26870823) 
and by HPCI Strategic Program of Japanese MEXT.

\appendix
\label{appA}
\section{Neutrino Matter Interaction Terms}
\label{sec:Neutrino Matter Interaction Terms}
In this appendix, we summarize neutrino matter interaction processes included in this study
which are: absorption and emission process
{\setlength\arraycolsep{2pt}
\begin{eqnarray}
\nu_en&\leftrightarrow& e^-p, \\
\bar\nu_ep&\leftrightarrow& e^+n,\\
\nu_eA&\leftrightarrow& e^-A',
\end{eqnarray}}
isoenergy scattering of neutrinos off nucleons and heavy nuclei
{\setlength\arraycolsep{2pt}
\begin{eqnarray}
\nu n&\leftrightarrow& \nu n,\\
\nu p&\leftrightarrow& \nu p,\\
\nu A&\leftrightarrow& \nu A,
\end{eqnarray}}
inelastic neutrino electron scattering
{\setlength\arraycolsep{2pt}
\begin{eqnarray}
\nu e&\leftrightarrow& \nu e,
\end{eqnarray}}
thermal neutrino pair production and annihilation
{\setlength\arraycolsep{2pt}
\begin{eqnarray}
e^-e^+&\leftrightarrow& \nu \bar\nu,
\end{eqnarray}}
and nucleon-nucleon bremsstrahlung
{\setlength\arraycolsep{2pt}
\begin{eqnarray}
NN&\leftrightarrow& NN\nu \bar\nu.
\end{eqnarray}}

Four vector neutrino matter interaction term $S^\mu$ is given by summation of
all these interaction terms as
\begin{eqnarray}
S^\mu\equiv S^{\rm nae,\mu}+S^{\rm iso,\mu}+S^{\rm nes,\mu}+S^{\rm tp,\mu}+S^{\rm brem,\mu},
\end{eqnarray}
where $S^{\rm nae,\mu}$, $S^{\rm iso,\mu}$, $S^{\rm nes,\mu}$, $S^{\rm tp,\mu}$ and $S^{\rm brem,\mu}$
are the four vector source terms of neutrino absorption and emission (nae),
isoenergy scattering of neutrinos off nucleons and heavy nuclei (iso),
inelastic neutrino electron scattering (nes),
thermal neutrino pair production and annihilation (tp)
and nucleon-nucleon Bremsstrahlung (brem), respectively.

We briefly summarize each interacting kernels, opacities and source terms in following subsections.
For more detailed expressions and explanations, the reader is referred to
\cite{Bruenn85,Hannestad98,Rampp02}.
Derivation of the four vector source term from these quantities is given by \cite{Shibata11}.

\subsection{Neutrino Absorption and Emission}
\label{sec:Neutrino Absorption and Emission}
We consider that neutrino absorption and emission: by free neutrons $\nu_en\leftrightarrow e^-p$,
by free protons $\bar\nu_ep\leftrightarrow e^+n$ and by heavy nuclei $\nu_eA\leftrightarrow e^-A'$.
The four vector source term $S^{\rm nae,\mu}$ is given by
\begin{eqnarray}
\label{eq:Snae}
S^{\rm nae,\mu}_{(\nu,\varepsilon)}=\kappa^{\rm nae}_{(\nu,\varepsilon)}\Biggl[ \biggl(\frac{4\pi\varepsilon^3}{{\rm exp}\{\beta(\varepsilon-\mu_\nu)\}+1}-J_{(\nu,\varepsilon)}\biggl) u^\mu-H^\mu_{(\nu,\varepsilon)}   \biggr],
\end{eqnarray}
where $\kappa^{\rm nae}$ is the opacity, $\beta=1/k_{\rm B}T$ with $k_{\rm B}$ the Boltzmann's constant and
$\mu_{\nu_e}=-\mu_{\bar\nu_e}=\mu_e-\mu_p+\mu_n$ is the chemical potential of neutrinos in thermal equilibrium with matter.
Here, $\mu_e$, $\mu_n$ and $\mu_p$ are the chemical potentials of electrons, neutrons and protons,
with including rest mass energy of each particle, respectively.
For each reaction ($\nu_en\leftrightarrow e^-p$, $\bar\nu_ep\leftrightarrow e^+n$ and $\nu_eA\leftrightarrow e^-A'$),
we first evaluate absorptivity $1/\lambda$ $[s^{-1}]$ and then emissivity $j$ $[s^{-1}]$.

Absorptivity for reaction $\nu_en\leftrightarrow e^-p$ is expressed as \citep{Bruenn85,Rampp02}
\begin{eqnarray}
\frac{1}{\lambda_\varepsilon}=\frac{c}{(\hbar c)^4} \frac{G_F^2}{\pi}(g_V^2+3g_A^2)\eta_{np}
\Bigl[1-F_{e^-}(\varepsilon+Q)\Bigr](\varepsilon+Q)\sqrt{(\varepsilon+Q)^2-m_e^2c^4},
\end{eqnarray}
where we employed $g_V=1$ and $g_A=1.23$
as form factors resulting from the virtual strong interaction processes.
$F_x(\varepsilon)\equiv[1+{\rm exp}(\beta(\varepsilon-\mu_x))]^{-1}$
is the Fermi distribution function of fermion $x$ with energy $\varepsilon$.
$G_F(=8.957\times10^{-44}$ MeV cm$^3$) is the Fermi constant.

For reaction $\bar\nu_ep\leftrightarrow e^+n$, 
\begin{eqnarray}
\frac{1}{\lambda_\varepsilon}=\left\{ \begin{array}{ll}
\frac{c}{(\hbar c)^4} \frac{G_F^2}{\pi}(g_V^2+3g_A^2)\eta_{pn}
\Bigl[1-F_{e^+}(\varepsilon-Q)\Bigr](\varepsilon-Q)\sqrt{(\varepsilon-Q)^2-m_e^2c^4},  & {\rm for}\ \varepsilon\ge m_ec^2+Q  \\
0, & {\rm otherwise,} \\
\end{array} \right.
\end{eqnarray} 
where $Q=m_nc^2-m_pc^2$ is the rest mass energy difference of the neutron and proton.

Finally for reaction $\nu_eA\leftrightarrow e^-A'$,
\begin{eqnarray}
\frac{1}{\lambda_\varepsilon}=\left\{ \begin{array}{ll}
e^{-\beta\Delta}\frac{c}{(\hbar c)^4} \frac{G_F^2}{\pi}g_A^2 \frac{2}{7}N_pN_h n_A
\Bigl[1-F_{e^-}(\varepsilon+Q')\Bigr](\varepsilon+Q')\sqrt{(\varepsilon+Q')^2-m_e^2c^4}, & {\rm for}
\ \varepsilon\ge m_ec^2-Q' \\
0, & {\rm otherwise,} \\
\end{array} \right.
\end{eqnarray} 
where $Q'\equiv\mu_n-\mu_p+\Delta$ is the mass difference between the initial and the final states through the reaction.
By following \cite{Bruenn85,Rampp02}, we employed $\Delta=3$ MeV and
\begin{eqnarray}
N_p=\left\{ \begin{array}{ll}
0, & {\rm for}\ Z<20 \\
Z-20, & {\rm for}\ 20<Z<28 \\
8, & {\rm for}\ Z>28, \\
\end{array} \right.\\
N_h=\left\{ \begin{array}{ll}
6, & {\rm for}\ N<34 \\
40-N, & {\rm for}\ 34<N<40 \\
0, & {\rm for}\ N>40, \\
\end{array} \right.
\label{eq:BlockHeavyNuclei}
\end{eqnarray} 
for the number of protons $N_p$ and holes $N_h$ in the dominant GT resonance for the electron capture.

As for the values $\eta_{pn}$ and $\eta_{np}$, we adopted ones proposed in
\cite{Bruenn85,Rampp02}
{\setlength\arraycolsep{2pt}
\begin{eqnarray}
\eta_{pn}&=&\frac{n_n-n_p}{{\rm exp}\Bigl(\beta(\mu_n-\mu_p-Q)\Bigr)-1},\\
\eta_{np}&=&\frac{n_p-n_n}{{\rm exp}\Bigl(-\beta(\mu_n-\mu_p-Q)\Bigr)-1},
\end{eqnarray}}
while we set 
{\setlength\arraycolsep{2pt}
\begin{eqnarray}
\eta_{pn}&=&n_p, \\
\eta_{np}&=&n_n,
\end{eqnarray}}
in the non-degeneracy regime where $\mu_n-\mu_p-Q<0.01$ MeV is met.

Once we evaluate the absorptivity, the emissivity $j$ and the opacity $\kappa$ are obtained as \citep{Bruenn85}
{\setlength\arraycolsep{2pt}
\begin{eqnarray}
j_{\varepsilon} =e^{\beta (\varepsilon-\mu_{\nu})}\frac{1}{\lambda_{\varepsilon}},
\end{eqnarray}}
and
\begin{eqnarray}
\kappa^{\rm nae}_{(\varepsilon)}=j_\varepsilon+\frac{1}{\lambda_\varepsilon}.
\end{eqnarray}

\subsection{Isoenergy Scattering of Neutrinos}
\label{sec:Isoenergy Scattering of Neutrinos}
The four vector source term for isoenergetic scattering of neutrinos off free nucleons and heavy nuclei is written as \citep{Shibata11}
\begin{eqnarray}
\label{eq:Siso}
S^{\rm iso,\mu}_{(\varepsilon)}=-\chi^{\rm iso}_{(\varepsilon)}H^\mu_{(\varepsilon)}.
\end{eqnarray}

To evaluate $\chi^{\rm iso}_{(\varepsilon)}$ from the isoenergetic scattering kernel
$R^{\rm iso}(\varepsilon,\omega)$,
we expand it into a Legendre series
in term of $\omega$ (here $\omega$ is the cosine of the scattering angle) up to the first order as
\begin{eqnarray}
R^{\rm iso}(\varepsilon,\omega)\approx \frac{1}{2}\Phi^{\rm iso,0}_{(\varepsilon)}+\frac{3}{2}\omega \Phi^{\rm iso,1}_{(\varepsilon)},
\end{eqnarray}
where $\Phi^{\rm iso,0}_{(\varepsilon)}$ and $\Phi^{\rm iso,1}_{(\varepsilon)}$ are
the zeroth and first order of scattering kernels, respectively.
After angular integration of the angular dependent source term with respect to $\omega$,
$\chi^{\rm iso}_{(\varepsilon)}$ is expressed as below \citep{Shibata11}
\begin{eqnarray}
\chi^{\rm iso}_{(\varepsilon)}=\varepsilon^2\biggl(\Phi^{\rm iso,0}_{(\varepsilon)}
-\Phi^{\rm iso,1}_{(\varepsilon)}\biggr).
\end{eqnarray}

For scattering process on free nucleon ($n/p$), zeroth and first order kernels become
{\setlength\arraycolsep{2pt}
\begin{eqnarray}
\Phi^{\rm iso,0}_{(\varepsilon)}&=&\frac{2\pi}{(h c)^3}\frac{G_F^2}{2\pi h}\eta_{NN}\Bigl({(h_V^N)}^2+3{(h_A^N)}^2\Bigr), \\
\Phi^{\rm iso,1}_{(\varepsilon)}&=&\frac{2\pi}{(h c)^3}\frac{G_F^2}{2\pi h}\eta_{NN}\Bigl({(h_V^N)}^2-{(h_A^N)}^2\Bigr).
\end{eqnarray}}
Obviously, $\Phi^{\rm iso,0/1}_{(\varepsilon)}$ has a dimension MeV$^{-2}$ s$^{-1}$ and $\chi^{\rm iso}_{(\varepsilon)}$
thus has a dimension s$^{-1}$.
In above, $N$ takes $n$ (neutron) or $p$ (proton) and $h_V^N$ and $h_A^N$ are defined as \citep{Bruenn85}
{\setlength\arraycolsep{2pt}
\begin{eqnarray}
h_V^n&=&-\frac{1}{2},\\
h_V^p&=&\frac{1}{2}-2{\rm sin}^2\theta_W,\\
h_A^n&=&-\frac{1}{2}g_A,\\
h_A^p&=&\frac{1}{2}g_A.
\end{eqnarray}}
where $\theta_W$ is the Weinberg angle and we adopt the value ${\rm sin}^2\theta_W=0.2325$.
By following \cite{tony93b,Rampp02}, we evaluate $\eta_{NN}$ as
{\setlength\arraycolsep{2pt}
\begin{eqnarray}
\eta_{NN}&=&n_N\frac{\xi_N}{\sqrt{1+\xi_N^2}},\\
\xi_N&=&\frac{3}{2\beta E^F_N},\\
E^F_N&=&\frac{(\hbar c)^2}{2m_N c^2}(3\pi^2 n_N)^{2/3}.
\end{eqnarray}}

Next, for coherent scattering on heavy nuclei, zeroth and first order kernels become \citep{Bruenn85}
{\setlength\arraycolsep{2pt}
\begin{eqnarray}
\Phi^{\rm iso,0}_{(\varepsilon)}&=&\frac{2\pi}{(h c)^3}\frac{G_F^2}{4\pi h}n_A
\biggl(AC_V^0-\Bigl(\frac{1}{2}A-Z\Bigr)C_V^1\biggr)^2\biggl(\frac{2y-1+e^{-2y}}{y^2}\biggr),\\
\Phi^{\rm iso,1}_{(\varepsilon)}&=&\frac{2\pi}{(h c)^3}\frac{G_F^2}{4\pi h}n_A
\biggl(AC_V^0-\Bigl(\frac{1}{2}A-Z\Bigr)C_V^1\biggr)^2\biggl(\frac{2-3y+2y^2-(2+y)e^{-2y}}{y^3}\biggr),
\end{eqnarray}}
with $C_{V/A}^0=(h_{V/A}^p+h_{V/A}^n)/2$, $C_{V/A}^1=(h_{V/A}^p-h_{V/A}^n)$,
$y=4b\varepsilon^2$ and $b=4.8\times10^{-6}A^{2/3}$.
Here $n_A$, $A$ and $Z$ denote number density of heavy nuclei, the atomic number and the charge, respectively.

\subsection{Neutrino Electron Scattering}
\label{sec:Neutrino Electron Scattering}
Inelastic scattering of neutrinos off electrons plays important role to help neutrinos
to escape more freely from centre as a consequence of down-scattering and it thus enhances
deleptonization of central core.
The source term $S^{\rm nes,\mu}_{(\varepsilon)}$ is expressed in term of
the collision integral $B^{\rm nes}_{(\varepsilon,\Omega)}$ as \citep{Shibata11}
{\setlength\arraycolsep{2pt}
\begin{eqnarray}
\label{eq:Snes}
S^{\rm nes,\mu}_{(\varepsilon)}=\varepsilon^3 \int d\Omega B^{\rm nes}_{(\varepsilon,\Omega)} (u^\mu+l^\mu),
\end{eqnarray}}
where $l^\mu$ is a unit normal four vector orthogonal to $u^\mu$.
The collision integral $B^{\rm nes}_{(\varepsilon,\Omega)}$ along the propagation direction $\Omega$ is expressed as
{\setlength\arraycolsep{0pt}
\begin{eqnarray}
B^{\rm nes}_{(\varepsilon,\Omega)}=\int \varepsilon'^2d\varepsilon'd\Omega'&&\left[
f(\varepsilon',\Omega')\left\{1-f(\varepsilon,\Omega)\right\}R^{\rm nes,in}(\varepsilon,\varepsilon',\omega)\right. \nonumber \\
&&\left. -\left\{1-f(\varepsilon',\Omega')\right\}f(\varepsilon,\Omega)R^{\rm nes,out}(\varepsilon,\varepsilon',\omega)\right],
\end{eqnarray}}
where $R^{\rm nes,in/out}(\varepsilon,\varepsilon',\omega)$ is the angular dependent inward/outward scattering kernel
and $\omega$ is the cosine of scattering angle, i.e., angle between $\Omega$ and $\Omega'$.
With expanding the angular dependent inward/outward scattering kernel
$R^{\rm nes,in/out}(\varepsilon,\varepsilon',\omega)$ 
into a Legendre series with respect to $\omega$ and taking up to the first order as
\begin{eqnarray}
R^{\rm nes,in/out}(\varepsilon,\varepsilon',\omega)\approx\frac{1}{2} \Phi^{\rm nes,in/out,0}_{(\varepsilon,\varepsilon')}
+\frac{3}{2}\omega \Phi^{\rm nes,in/out,1}_{(\varepsilon,\varepsilon')},
\end{eqnarray}
and with decomposing the neutrino distribution function after scattering into isotropic and non-isotropic parts as 
\begin{eqnarray}
f(\varepsilon',\Omega')\approx f^0(\varepsilon')+f^{1,\mu}(\varepsilon')l_\mu',
\end{eqnarray}
final expression of the scattering integral is described as below
{\setlength\arraycolsep{2pt}
\begin{eqnarray}
\label{eq:Bnes_Bruenn85}
B^{\rm nes}_{(\varepsilon,\Omega)}=
f(\varepsilon,\Omega)\left(A^{0}_{(\varepsilon),\rm nes}+B^{0,\mu}_{(\varepsilon),\rm nes}l_\mu\right)
+C^{0}_{(\varepsilon),\rm nes}+C^{1,\mu}_{(\varepsilon),\rm nes}l_\mu.
\end{eqnarray}}
In the above, we used the same notations used in \cite{Bruenn85} (see their Eqs.(A34)-(A39)) for
$A^{0}_{(\varepsilon),\rm nes}$, $B^{0,\mu}_{(\varepsilon),\rm nes}$, $C^{0}_{(\varepsilon),\rm nes}$ and
$C^{1,\mu}_{(\varepsilon),\rm nes}$ with the exception of $B^{0,\mu}_{(\varepsilon),\rm nes}$ and
$C^{1,\mu}_{(\varepsilon),\rm nes}$ have three spatial components
(the zeroth component is determined from orthogonality $H^\mu u_\mu=0$) and of $B^{0,\mu}_{(\varepsilon),\rm nes}$
is a factor of 3 larger than that of Eq.(A35) in \cite{Bruenn85}.
They are explicitly written as
{\setlength\arraycolsep{2pt}
\begin{eqnarray}
\label{eq:Bruenn85A0_nes}
A^{0}_{(\varepsilon),\rm nes}&=&-\frac{2\pi}{(hc)^3}\int {\varepsilon'}^2 d\varepsilon' \left[
f^0(\varepsilon') \Phi^{\rm nes,in,0}_{(\varepsilon,\varepsilon')}
+(1-f^0(\varepsilon')) \Phi^{\rm nes,out,0}_{(\varepsilon,\varepsilon')}
\right], \\
\label{eq:Bruenn85B0_nes}
B^{0,\mu}_{(\varepsilon),\rm nes}&=&-\frac{2\pi}{(hc)^3}\int {\varepsilon'}^2 d\varepsilon' 
f^{1,\mu}(\varepsilon') \left(\Phi^{\rm nes,in,1}_{(\varepsilon,\varepsilon')}
-\Phi^{\rm nes,out,1}_{(\varepsilon,\varepsilon')}\right),\\
\label{eq:Bruenn85C0_nes}
C^{0}_{(\varepsilon),\rm nes}&=&\frac{2\pi}{(hc)^3}\int {\varepsilon'}^2 d\varepsilon' 
f^{0}(\varepsilon')\Phi^{\rm nes,in,0}_{(\varepsilon,\varepsilon')},\\
\label{eq:Bruenn85C1_nes}
C^{1,\mu}_{(\varepsilon),\rm nes}&=&\frac{2\pi}{(hc)^3}\int {\varepsilon'}^2 d\varepsilon' 
f^{1,\mu}(\varepsilon') \Phi^{\rm nes,in,1}_{(\varepsilon,\varepsilon')},
\end{eqnarray}}
where we employ
{\setlength\arraycolsep{2pt}
\begin{eqnarray}
f^0(\varepsilon)&=&\frac{J_{(\varepsilon)}}{4\pi\varepsilon^3},\\
f^{1,\mu}(\varepsilon)&=&\frac{3H^\mu_{(\varepsilon)}}{4\pi\varepsilon^3},\\
\end{eqnarray}}
for the isotropic and non-isotropic parts of the distribution function (see also \cite{Shibata11}).
For an explicit evaluation for $\Phi^{{\rm nes,in/out},0/1}_{(\varepsilon,\varepsilon')}$, we refer the reader to \cite{Yueh76,Bruenn85}.

Since coefficients (\ref{eq:Bruenn85A0_nes})-(\ref{eq:Bruenn85C1_nes}) do not depend on propagation direction $\Omega$,
the final integral Eq.(\ref{eq:Snes}) with imposing Eq.(\ref{eq:Bnes_Bruenn85}) becomes
{\setlength\arraycolsep{2pt}
\begin{eqnarray}
\label{eq:SnesFinal}
S^{\rm nes,\mu}_{(\varepsilon)}&=&\left(J_{(\varepsilon)}u^\mu+H_{(\varepsilon)}^\mu\right)A^{0}_{(\varepsilon),\rm nes}\nonumber \\
&&+\left(H_{\alpha(\varepsilon)}u^\mu+{{L_\alpha}^\mu}_{(\varepsilon)}\right)B^{0,\alpha}_{(\varepsilon),\rm nes}\nonumber \\
&&+4\pi\varepsilon^3u^\mu C^{0}_{(\varepsilon),\rm nes}\nonumber \\
&&+\frac{4\pi\varepsilon^3}{3}{h_\alpha}^\mu C^{1,\alpha}_{(\varepsilon),\rm nes},
\end{eqnarray}}
here $h_{\mu\nu}\equiv g_{\mu\nu}+u_\mu u_\nu$ is the projection operator.
Since, $\Phi^{\rm nes}$ has a unit [cm$^3$ s$^{-1}$],
$A^{0}_{(\varepsilon),\rm nes}$, $B^{0,\mu}_{(\varepsilon),\rm nes}$, $C^{0}_{(\varepsilon),\rm nes}$ and
$C^{1,\mu}_{(\varepsilon),\rm nes}$ thus have a unit [s$^{-1}$] which is required to let the dimension of
$S^{\rm nes,\mu}_{(\varepsilon)}$ to be [MeV$^3$ s$^{-1}$].


\subsection{Thermal Pair Production and Annihilation of Neutrinos}
\label{sec:Thermal Pair Production and Annihilation of Neutrinos}
As for the thermal pair production/annihilation process of neutrinos,
we take the same approach as neutrino electron scattering process.
The collision integral $B^{\rm tp}_{(\varepsilon,\Omega)}$ is expressed as
{\setlength\arraycolsep{0pt}
\begin{eqnarray}
B^{\rm nes}_{(\varepsilon,\Omega)}=\int \varepsilon'^2d\varepsilon'd\Omega'&&\left[
\left\{1-\bar f(\varepsilon',\Omega')\right\}\left\{1-f(\varepsilon,\Omega)\right\}R^{\rm tp,pro}(\varepsilon,\varepsilon',\omega)\right. \nonumber \\
&&\left. -\bar f(\varepsilon',\Omega')f(\varepsilon,\Omega)R^{\rm tp,ann}(\varepsilon,\varepsilon',\omega)\right],
\end{eqnarray}}
where $\bar f$ denotes anti-neutrino distribution function and
$R^{\rm tp,pro/ann}(\varepsilon,\varepsilon',\omega)$ is the angular dependent production/annihilation kernel.
We again expand the kernel
$R^{\rm tp,pro/ann}(\varepsilon,\varepsilon',\omega)$ up to the first order in $\omega$ as
\begin{eqnarray}
R^{\rm tp,pro/ann}(\varepsilon,\varepsilon',\omega)\approx\frac{1}{2} \Phi^{\rm tp,pro/ann,0}_{(\varepsilon,\varepsilon')}
+\frac{3}{2}\omega \Phi^{\rm tp,pro/ann,1}_{(\varepsilon,\varepsilon')},
\end{eqnarray}
and final expression of the scattering integral is thus described as below
{\setlength\arraycolsep{2pt}
\begin{eqnarray}
\label{eq:Btp_Bruenn85}
B^{\rm tp}_{(\varepsilon,\Omega)}=
f(\varepsilon,\Omega)\left(A^{0}_{(\varepsilon),\rm tp}+B^{0,\mu}_{(\varepsilon),\rm tp}l_\mu\right)
+C^{0}_{(\varepsilon),\rm tp}+C^{1,\mu}_{(\varepsilon),\rm tp}l_\mu,
\end{eqnarray}}
which has exactly the same expression as that of neutrino electron scattering (Eq.\ref{eq:Bnes_Bruenn85}).
Again coefficients have the same notations as in \cite{Bruenn85} (see their Eqs.(A43)-(A47)) and described as
{\setlength\arraycolsep{2pt}
\begin{eqnarray}
\label{eq:Bruenn85A0_tp}
A^{0}_{(\varepsilon),\rm tp}&=&-\frac{2\pi}{(hc)^3}\int {\varepsilon'}^2 d\varepsilon' \left[
\left\{1-\bar f^0(\varepsilon')\right\} \Phi^{\rm tp,pro,0}_{(\varepsilon,\varepsilon')}
+\bar f^0(\varepsilon') \Phi^{\rm tp,ann,0}_{(\varepsilon,\varepsilon')}
\right], \\
\label{eq:Bruenn85B0_tp}
B^{0,\mu}_{(\varepsilon),\rm tp}&=&\frac{2\pi}{(hc)^3}\int {\varepsilon'}^2 d\varepsilon' 
\bar f^{1,\mu}(\varepsilon') \left(\Phi^{\rm tp,pro,1}_{(\varepsilon,\varepsilon')}
-\Phi^{\rm tp,ann,1}_{(\varepsilon,\varepsilon')}\right),\\
\label{eq:Bruenn85C0_tp}
C^{0}_{(\varepsilon),\rm tp}&=&\frac{2\pi}{(hc)^3}\int {\varepsilon'}^2 d\varepsilon' 
\left\{1-\bar f^{0}(\varepsilon')\right\}\Phi^{\rm tp,pro,0}_{(\varepsilon,\varepsilon')},\\
\label{eq:Bruenn85C1_tp}
C^{1,\mu}_{(\varepsilon),\rm tp}&=&-\frac{2\pi}{(hc)^3}\int {\varepsilon'}^2 d\varepsilon' 
\bar f^{1,\mu}(\varepsilon') \Phi^{\rm tp,pro,1}_{(\varepsilon,\varepsilon')}.
\end{eqnarray}}
For an explicit evaluation for $\Phi^{{\rm tp,pro/ann},0/1}_{(\varepsilon,\varepsilon')}$, we refer the reader to \cite{Bruenn85}.
The final expression of the source term $S^{\rm tp,\mu}_{(\varepsilon)}$ is simply obtained by replacing
coefficients in Eq.(\ref{eq:SnesFinal}) with those of thermal process, i.e., with Eqs.(\ref{eq:Bruenn85A0_tp})-(\ref{eq:Bruenn85C1_tp}).

\subsection{Nucleon-nucleon Bremsstrahlung}
\label{sec:Nucleon-nucleon Bremsstrahlung}
The collision integral for Nucleon-nucleon Bremsstrahlung $B^{\rm br}_{(\varepsilon,\Omega)}$ has the same expression
as that of thermal pair production/annihilation of neutrinos and,
therefore, the same notations used in Sec. \ref{sec:Thermal Pair Production and Annihilation of Neutrinos} can be directly applicable.
$B^{\rm br}_{(\varepsilon,\Omega)}$ is written as
{\setlength\arraycolsep{0pt}
\begin{eqnarray}
B^{\rm br}_{(\varepsilon,\Omega)}=\int \varepsilon'^2d\varepsilon'd\Omega'&&\left[
\left\{1-\bar f(\varepsilon',\Omega')\right\}\left\{1-f(\varepsilon,\Omega)\right\}R^{\rm br,pro}(\varepsilon,\varepsilon',\omega)\right. \nonumber \\
&&\left. -\bar f(\varepsilon',\Omega')f(\varepsilon,\Omega)R^{\rm br,ann}(\varepsilon,\varepsilon',\omega)\right],
\end{eqnarray}}
here $R^{\rm br,pro/ann}(\varepsilon,\varepsilon',\omega)$ is the production/annihilation kernel.
As for the Bremsstrahlung process, we again expand the kernel
$R^{\rm br,pro/ann}(\varepsilon,\varepsilon',\omega)$ into a Legendre series, but take only the zeroth order term
in $\omega$ for simplicity. Then the kernel becomes
\begin{eqnarray}
R^{\rm br,pro/ann}(\varepsilon,\varepsilon',\omega)\approx\frac{1}{2} \Phi^{\rm br,pro/ann,0}_{(\varepsilon,\varepsilon')}.
\end{eqnarray}
Coefficients used in the final expression for the collision integral and the source term are simply evaluated by
replacing $\Phi^{\rm tp}$ in Eqs.(\ref{eq:Bruenn85A0_tp})-(\ref{eq:Bruenn85C1_tp}) with $\Phi^{\rm br}$.

Isotropic production kernel $\Phi^{\rm br,pro,0}_{(\varepsilon,\varepsilon')}$
can be evaluated by following manner.
{\setlength\arraycolsep{2pt}
\begin{eqnarray}
\Phi^{\rm br,pro,0}_{(\varepsilon,\varepsilon')}\equiv e^{-\beta(\varepsilon+\varepsilon')}\sum_{i\in{\rm nn,pp,np}}
{\rm min} \left(\phi_{\rm br}^{D,i}(\varepsilon,\varepsilon'),\ \phi_{\rm br}^{ND,i}(\varepsilon,\varepsilon')\right).
\end{eqnarray}}
In the above equation, indexes $D$ and $ND$ denote degenerate and non-degenerate limit
of free nucleons, respectively, and (nn, pp, np) represent bremsstrahlung due to
neutron-neutron, proton-proton and neutron-proton pair, respectively. 
$\phi_{\rm br}^{D,i}(\varepsilon,\varepsilon')$ and $\phi_{\rm br}^{ND,i}(\varepsilon,\varepsilon')$ are expressed as
{\setlength\arraycolsep{2pt}
\begin{eqnarray}
\phi_{\rm br}^{D,i}(\varepsilon,\varepsilon')&=&(\hbar c)^6\frac{G^2}{4}
\left(\frac{2\pi f}{m_\pi}\right)^4 \nonumber \\
&\times&\frac{m_i^4}{12\pi^9} \frac{1}{\beta^2(\varepsilon+\varepsilon')}
\frac{\left(4\pi^2+\beta^2(\varepsilon+\varepsilon')^2\right)}
{1-e^{-\beta(\varepsilon+\varepsilon')}}
\frac{\alpha_i}{S_i} \left[ \frac{2 \hbar c\left(3\pi^2 X_i n_b\right)^{\frac{1}{3}}}{(\hbar c)^6}\right], \\
\phi_{\rm br}^{ND,i}(\varepsilon,\varepsilon')&=&(\hbar c)^6\frac{G^2}{4}
\left(\frac{2\pi f}{m_\pi}\right)^4 \nonumber \\
&\times&\frac{2m_i^{3/2}}{\pi^{11/2}} \frac{1}{\beta^{1/2}(\varepsilon+\varepsilon')^2}
\sqrt{1+\beta(\varepsilon+\varepsilon')}
\frac{\alpha_i}{S_i} X_i^2 n_b^2.
\end{eqnarray}}
In the above equations, $\phi_{\rm br}^{D/ND,i}(\varepsilon,\varepsilon')$ have a dimension in [cm$^{3}$ s$^{-1}$],
$f=1$ and $G^2=c(\hbar c)^2G_F^2=1.55\times10^{-33}$ [MeV$^{-2}$ cm$^3$ s$^{-1}$].
We employed values for $\alpha_i$, $X_i$, $m_i$ and $S_i$ as
\begin{eqnarray}
\label{eq:coefficient_for_br}
\alpha_i=\left(
\begin{array}{c}
3g_A^2  \\
3g_A^2 \\
\frac{7}{2}g_A^2
 \end{array}
\right),\
X_i=\left(
\begin{array}{c}
X_n  \\
X_p \\
{\rm min}(X_n,X_p)
 \end{array}
\right),\
m_i=\left(
\begin{array}{c}
m_n  \\
m_p \\
\sqrt{m_nm_p}
 \end{array}
\right),\
S_i=\left(
\begin{array}{c}
4  \\
4 \\
1
 \end{array}
\right),\
{\rm for\ i=}\left(
\begin{array}{c}
  nn\\
pp \\
np
 \end{array}
\right). \nonumber \\
\end{eqnarray}

After deriving the production kernel, we can obtain the annihilation one by using a following relation
{\setlength\arraycolsep{0pt}
\begin{eqnarray}
R^{\rm br,ann}(\varepsilon,\varepsilon',\omega)=e^{\beta(\varepsilon+\varepsilon')}R^{\rm br,pro}(\varepsilon,\varepsilon',\omega).
\end{eqnarray}}

\subsection{Summary of the Source Term and Mean Free Paths}
\label{sec:Summary of the Source Term and Mean Free Paths}
By combining all the source terms mentioned above, the final expression becomes

{\setlength\arraycolsep{2pt}
\begin{eqnarray}
\label{eq:StotFinal}
S^{\mu}_{(\varepsilon)}&=&
\kappa^{\rm nae}_{(\varepsilon)}\left\{ \left(J^{\rm eq}_{(\varepsilon)}-J_{(\varepsilon)}\right) u^\mu-H^\mu_{(\varepsilon)}   \right\}
-\chi^{\rm iso}_{(\varepsilon)}H^\mu_{(\varepsilon)} \nonumber \\
&&+\sum_{X\in{\rm nes,tp,br}}\left\{\left(J_{(\varepsilon)}u^\mu+H_{(\varepsilon)}^\mu\right)A^{0}_{(\varepsilon),\rm X}
+4\pi\varepsilon^3u^\mu C^{0}_{(\varepsilon),\rm X}\right. \nonumber \\
&&\left. +\left(H_{\alpha(\varepsilon)}u^\mu+{{L_\alpha}^\mu}_{(\varepsilon)}\right)B^{0,\alpha}_{(\varepsilon),\rm X}
+\frac{4\pi\varepsilon^3}{3}{h_\alpha}^\mu C^{1,\alpha}_{(\varepsilon),\rm X}\right\},
\end{eqnarray}}
where $J^{\rm eq}_{(\varepsilon)}$ is written by
{\setlength\arraycolsep{2pt}
\begin{eqnarray}
\label{eq:Jeq}
J^{\rm eq}_{(\varepsilon)}=\frac{4\pi\varepsilon^3}{{\rm exp}\{\beta(\varepsilon-\mu_\nu)\}+1}.
\end{eqnarray}}

\begin{figure}[htpb]
\begin{center}
\includegraphics[width=150mm,angle=-90.]{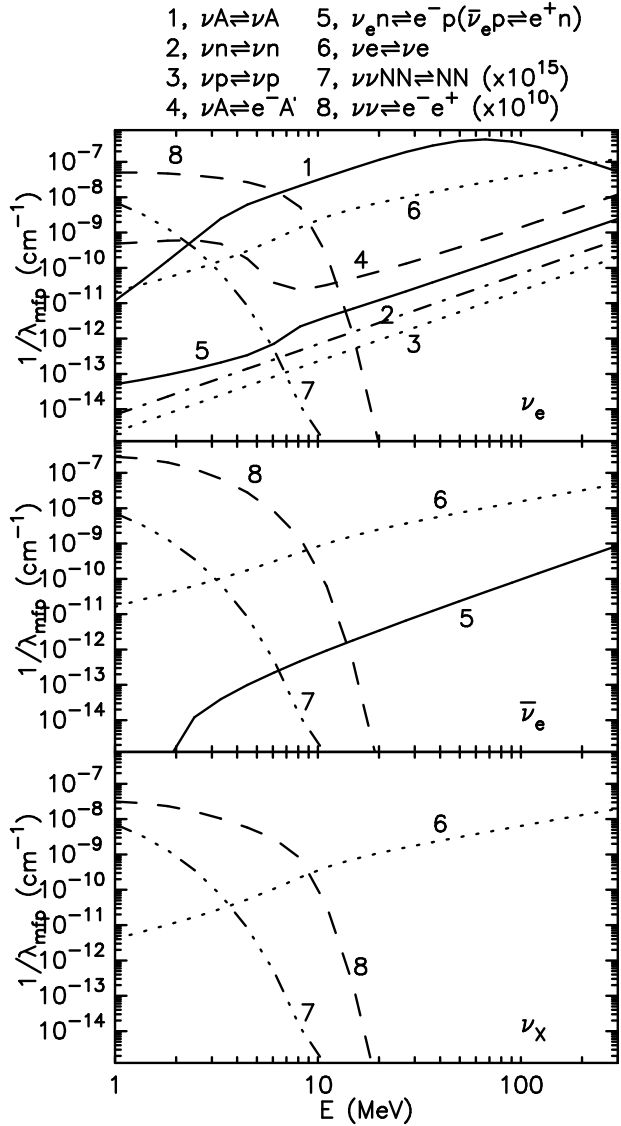}
\end{center}
  \caption{Energy dependence of inverse mean free paths for each process labeled at top part.
  Employed hydrodynamical background is $\rho=10^{10}$ g cm$^{-3}$, T=0.638 MeV and $Y_e=0.43$
  and we assumed the final state occupancy of the neutrinos is zero.
  With using the EOS of \cite{LSEOS} with compressibility parameter $K=220$ MeV,
  several representative thermodynamical quantities become
  $s=0.71$ $k_{\rm B}$ baryon$^{-1}$, $A=65.5$ and $Z=28.2$.}
\label{fig:Opacity10.eps}
\end{figure}

\begin{figure}[htpb]
\begin{center}
\includegraphics[width=150mm,angle=-90.]{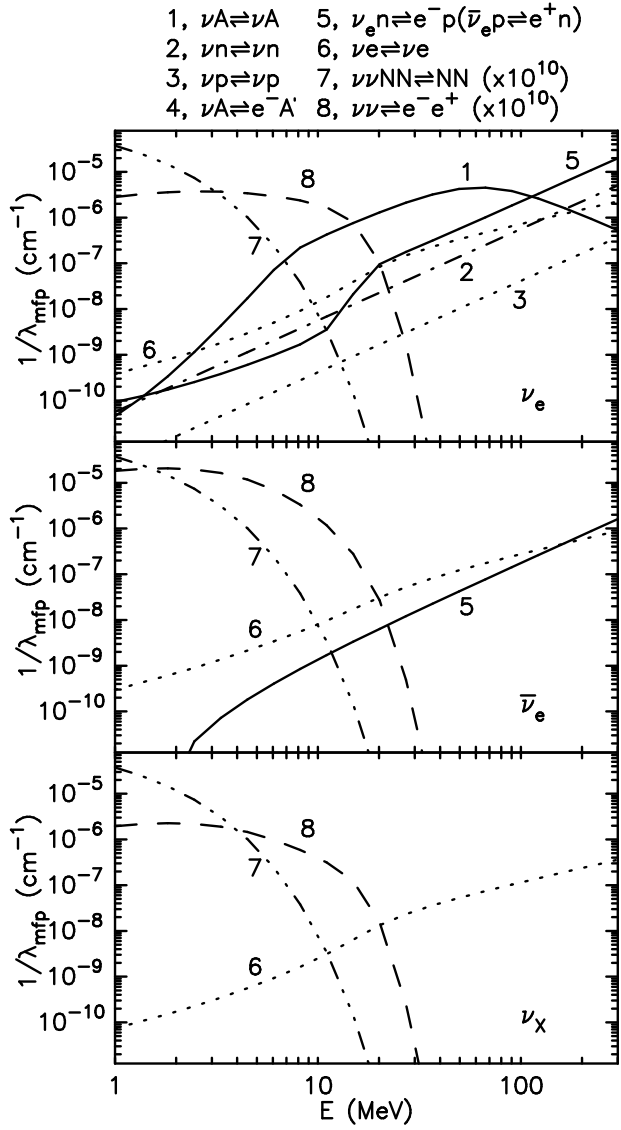}
\end{center}
 \caption{Same as Fig.\ref{fig:Opacity10.eps} but with $\rho=10^{11}$ g cm$^{-3}$, T=1.382 MeV and $Y_e=0.4$.
  This hydrodynamical profile returns $s=1.19$ $k_{\rm B}$ baryon$^{-1}$, $A=73.9$ and $Z=30.5$.
  Since $N=A-Z>40$, neutrino absorption on heavy nuclei is blocked with our current electron capture rate
  (Eq.\ref{eq:BlockHeavyNuclei}).}
\label{fig:OpacityBruenn85.eps}
\end{figure}

\begin{figure}[htpb]
\begin{center}
\includegraphics[width=150mm,angle=-90.]{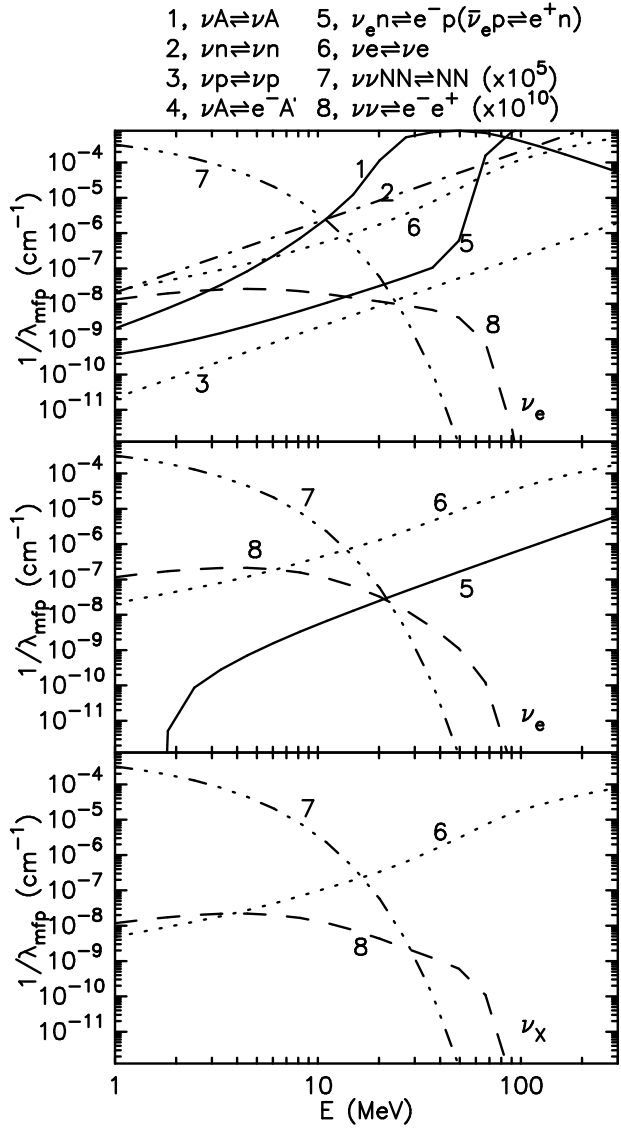}
\end{center}
  \caption{Same as Fig.\ref{fig:Opacity10.eps} but with $\rho=10^{13}$ g cm$^{-3}$, T=3 MeV and $Y_e=0.25$.
  This hydrodynamical profile returns $s=1.25$ $k_{\rm B}$ baryon$^{-1}$, $A=167.7$ and $Z=49.3$.
  Since $N=A-Z>40$, neutrino absorption on heavy nuclei is blocked with our current electron capture rate
  (Eq.\ref{eq:BlockHeavyNuclei}).}
\label{fig:Opacity13.eps}
\end{figure}

In the end, we plot inverse mean free paths of each process for reference.
Assumed hydrodynamical profiles are $\rho=10^{10}$ g cm$^{-3}$, T=0.638 MeV and $Y_e=0.43$
(Fig. \ref{fig:Opacity10.eps}), which corresponds to the central profile of ``s15s7b2'' in \cite{WW95},
$\rho=10^{11}$ g cm$^{-3}$, T=1.382 MeV and $Y_e=0.4$ (Fig. \ref{fig:OpacityBruenn85.eps}),
which is the same condition used in Fig. 36 in \cite{Bruenn85} and
$\rho=10^{13}$ g cm$^{-3}$, T=3 MeV and $Y_e=0.25$ (Fig. \ref{fig:Opacity13.eps}),
which is a typical profile after neutrino trapping during collapse phase.
As in \cite{Bruenn85}, we assumed the final state occupancy of the neutrinos is zero.
More explicitly, we plot $1/\lambda_{\rm mfp}$ [cm$^{-1}$] expressed by
\begin{eqnarray}
\label{eq:1/lambda}
\lambda_{\rm mfp}^{-1}=\frac{1}{c}\left\{
\begin{array}{cc}
\kappa^{\rm nae}_{(\nu,\varepsilon)}, &  ({\rm neutrino\ absorption\ and\ emission}) \\
\chi^{\rm iso}_{(\varepsilon)}, &  ({\rm isoenergy\ scattering}) \\ 
A^{0}_{(\varepsilon),\rm nes}, & ({\rm neutrino\ electron\ scattering})\\
A^{0}_{(\varepsilon),\rm tp}, & ({\rm thermal\ pair\ production\ and\ annihilation})\\ 
A^{0}_{(\varepsilon),\rm br}, & ({\rm nucleon-nucleon\ bremsstrahlung}).
 \end{array}
\right.
\end{eqnarray}
Since isoenergy scattering rate has the same value for all neutrino species, we only plotted for
electron type neutrino.
Note that, the inverse mean free path for bremsstrahlung is multiplied by $10^{15}$ (Fig. \ref{fig:Opacity10.eps}),
$10^{10}$ (Fig. \ref{fig:OpacityBruenn85.eps}) and $10^{5}$ (Fig. \ref{fig:Opacity13.eps})
and, for thermal process, it is multiplied by $10^{10}$
(Fig. \ref{fig:Opacity10.eps}-\ref{fig:Opacity13.eps}).

\newpage
\label{appB}
\section{Comparison in 1D Spherical Coordinate}
\label{sec:Comparison in 1D Spherical Coordinate}
In this appendix, we perform a core collapse simulation using 1D spherical symmetric M1 neutrino transport code.
The aim of this section is to test the local neutrino-matter interaction terms
implemented in our main 3D Cartesian grid based radiation-hydrodynamics code.
Since our 3D multi-energy neutrino radiation-hydrodynamics code with higher resolution
than the one used in Sec. \ref{sec:Core Collapse of a 15M star}
is still computationally demanding, it is not easy to check the spatial resolution dependence on the practical core collapse simulation.
Therefore, we can not assess whether the differences between our (pseudo-)1D results and previous 1D ones
come from the spatial advection terms or from the neutrino-matter interaction terms.
To expel the ambiguities coming from the spatial advection term,
we develop a new spherically symmetric M1 code in the Newtonian limit.
The basis of the new 1D code is the same as our main 3D Cartesian grid based code.
We solve Eq.(\ref{eq:ExpImpEq}) with the same neutrino opacities used in model ``s15Nso$\_$1d.b'' in \cite{Buras06a}.
The spatial advection term $\bf S_{\rm adv,s}$ in 1D spherical symmetry
is evaluated in a similar way as in \citet{Nakamura2D14,Horiuchi14}
and the advection terms in energy space $\bf S_{\rm adv,e}$ are also slightly modified for spherical polar coordinate.
To take the same approach as the recent M1 neutrino transport codes \citep{O'Connor14,just15},
we solve the advection terms in energy space $\bf S_{\rm adv,e}$ explicitly in time.

\begin{figure}[htpb] 
\begin{center}
\includegraphics[width=80mm]{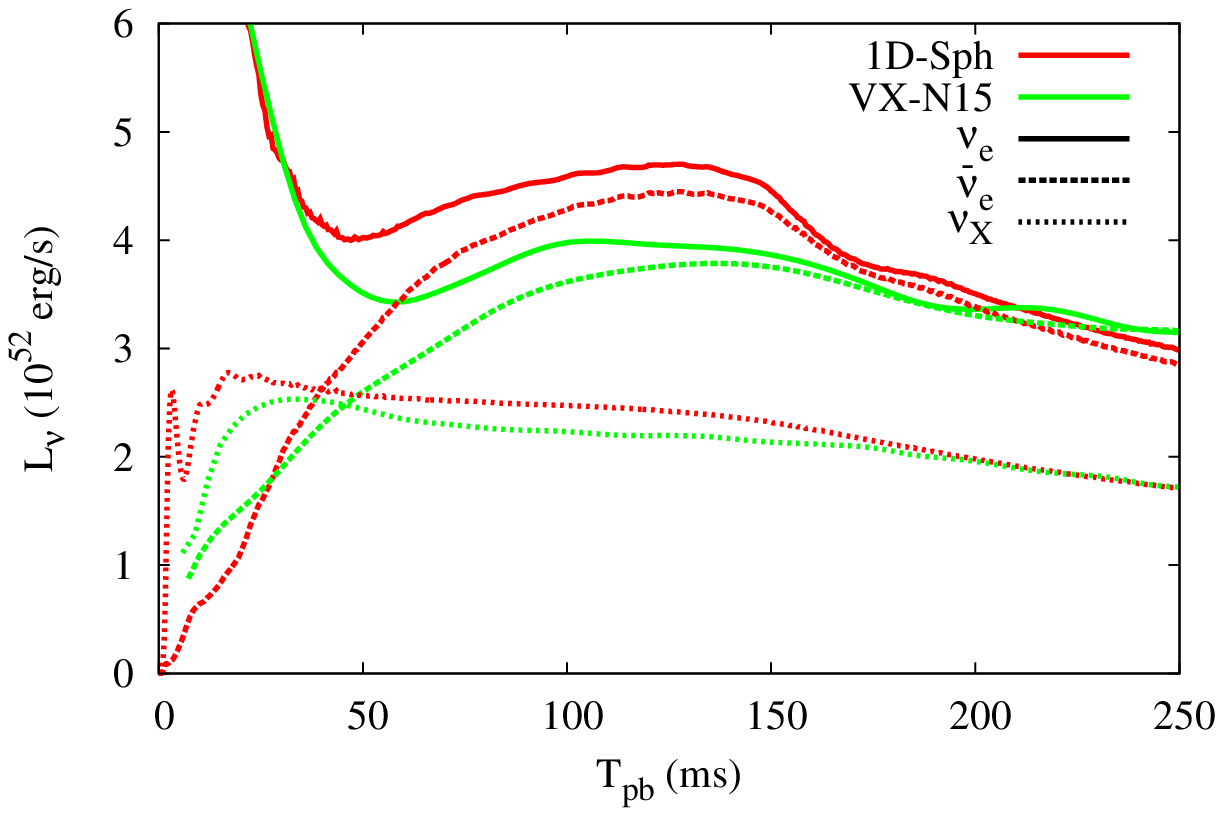}
\includegraphics[width=80mm]{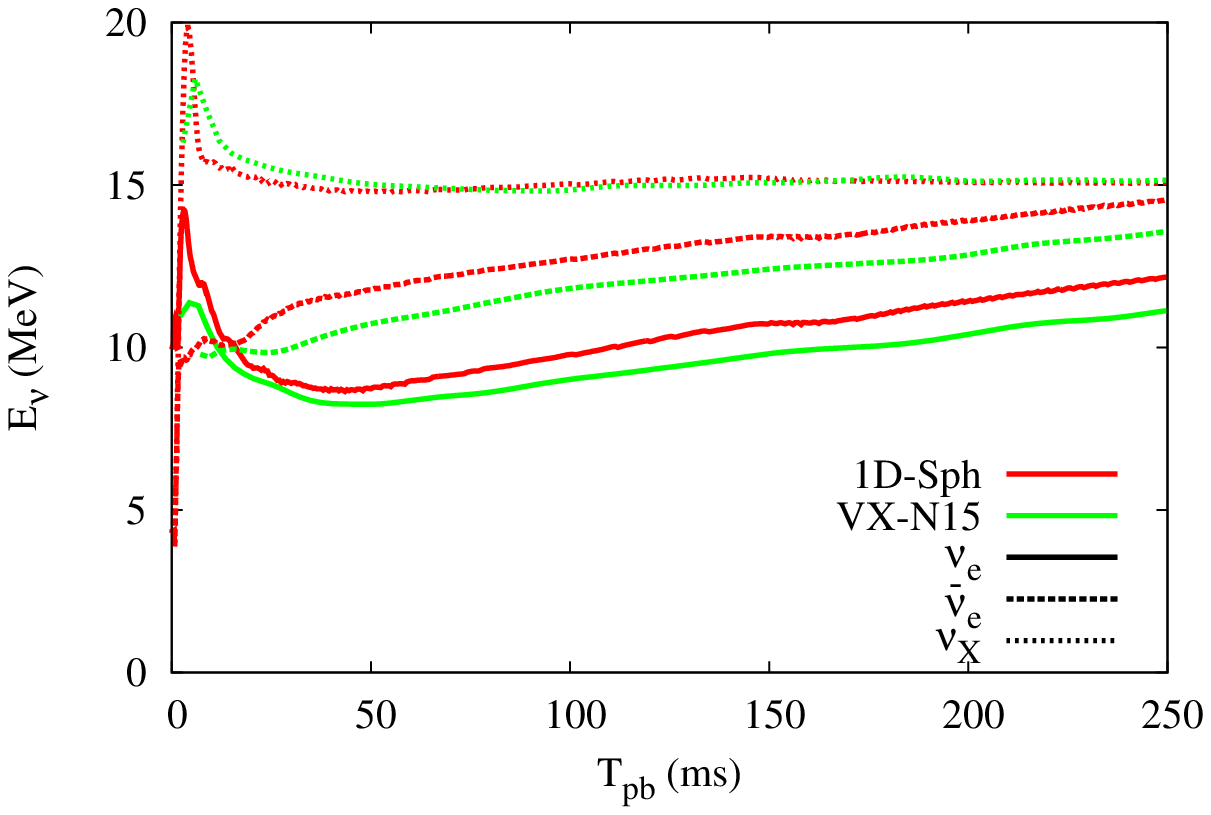}\\
\includegraphics[width=80mm]{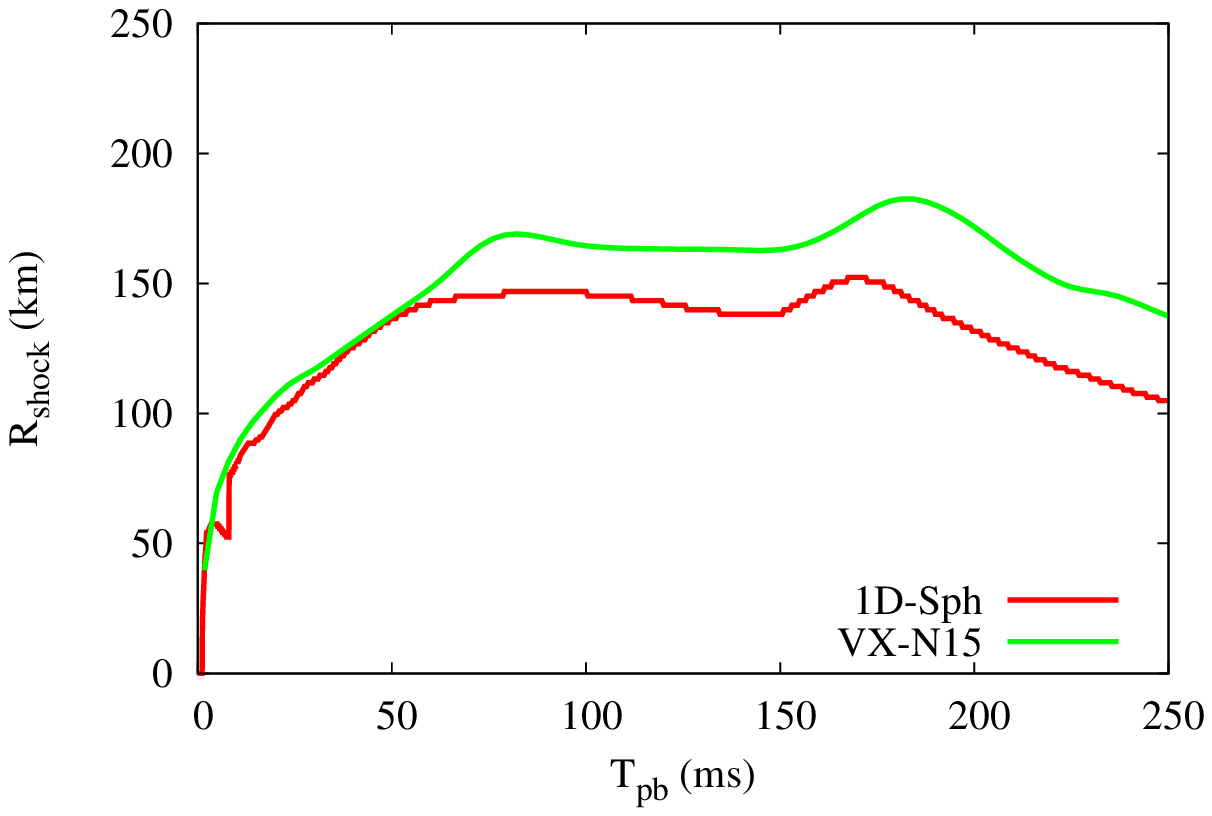}
\includegraphics[width=80mm]{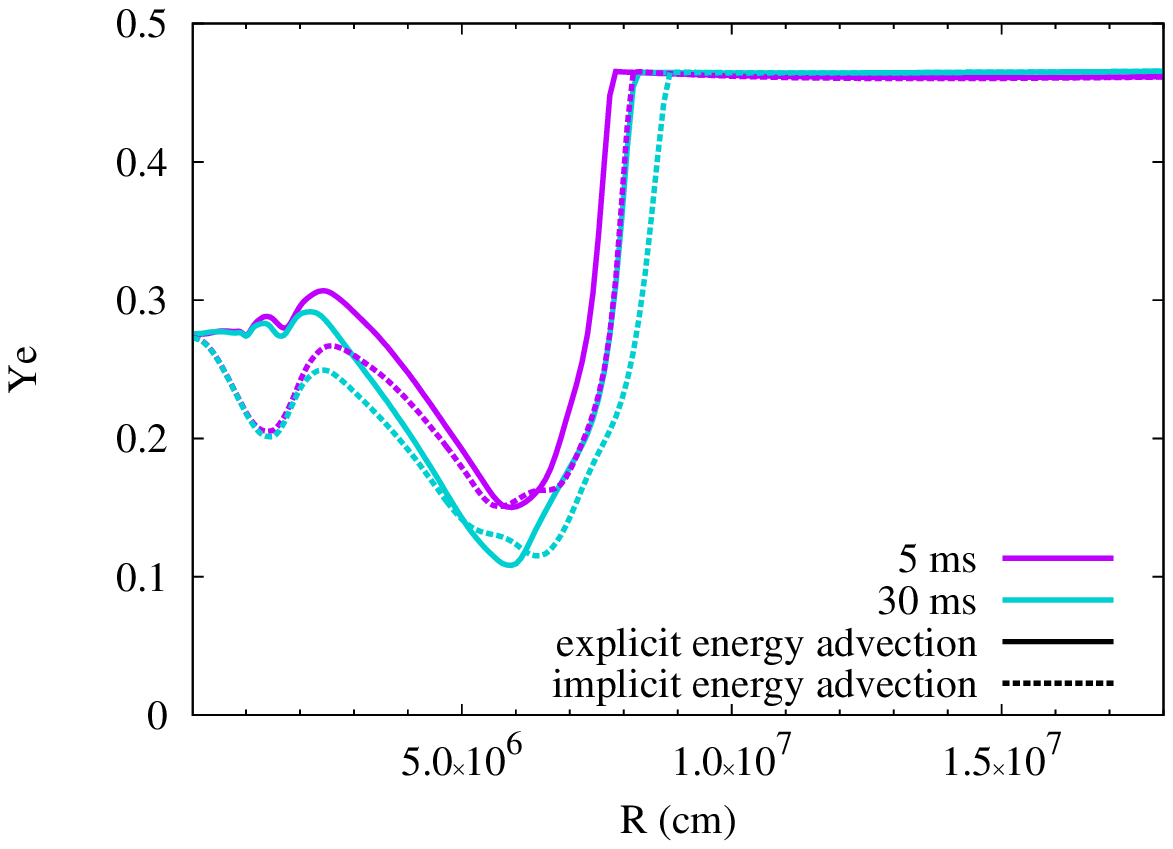}
\end{center}
  \caption{Top left panel: Time evolution of the neutrino luminosities measured at 400km. Results for our 1D-spherical M1 transport scheme (1D$-$Sph)
and for VERTEX-PROMETHEUS (VX$-$N15) are denoted by red and green lines, respectively.
Top right: Same as the top left one but for the mean neutrino energies.
Bottom left: Time evolution of the shock radius.
Bottom right: Radial profile of the electron fraction $Y_e$ at different time slices.
Solid and dotted lines are for explicit and implicit models, respectively.}
\label{fig:1D_Comparison}
\end{figure}

In Fig.\ref{fig:1D_Comparison}, we plot our results (1D$-$Sph) for the neutrino luminosity, mean energy, shock radius
and the $Y_e$ profile together with reference values of VERTEX-PROMETHEUS \cite[model ``s15Nso$\_$1d.b'' in][]{Buras06a} (VX$-$N15)
except the $Y_e$ profile.
In the $Y_e$ profile, we also plot results obtained from a model with solving $\bf S_{\rm adv,e}$ implicitly in time, i.e.,
as the same as our main 3D code.
Our results show a good agreement with VX$-$N15.
The maximum deviations are $\sim7\times10^{51}$ erg s$^{-1}$ in the luminosity and $\sim0.9$ MeV in the mean energy.
The mean energies, $\langle \varepsilon_{\nu_e}\rangle$ and $\langle \varepsilon_{\bar\nu_e}\rangle$,
show slightly higher values than the reference ones, they are, however, within the error bounds reported in
previous code comparisons \citep{Liebendorfer05,BMuller10}.
The neutrino luminosities show quite consistent behavior especially in late phase $T_{\rm pb}\ga150$ ms,
when the neutrino heating becomes more important.
Contrary to our main results reported in Sec. \ref{sec:Results}, the $Y_e$ trough observed at $R\sim10^6$ cm (see, Fig. \ref{f9})
disappears in the explicit model (solid lines),
while the $Y_e$ trough still exists in the implicit model (dotted).
These two models support the idea to evaluate both ${\bf S_{\rm adv,s}}$ and ${\bf S_{\rm adv,e}}$ at the same time slice.
We are going to examine this point in our main 3D Cartesian based code in the near future.
From this test, we confirm that the neutrino matter interaction terms are correctly implemented.

\newpage
\bibliographystyle{apj}
\bibliography{mybib}

\begin{thebibliography}{148}
\expandafter\ifx\csname natexlab\endcsname\relax\def\natexlab#1{#1}\fi

\bibitem[{{Ardeljan} {et~al.}(2000){Ardeljan}, {Bisnovatyi-Kogan}, \&
  {Moiseenko}}]{arde00}
{Ardeljan}, N.~V., {Bisnovatyi-Kogan}, G.~S., \& {Moiseenko}, S.~G. 2000,
  Astron. Astrophys., 355, 1181

\bibitem[{{Audit} {et~al.}(2002){Audit}, {Charrier}, {Chi{\`e}ze}, \&
  {Dubroca}}]{Audit02}
{Audit}, E., {Charrier}, P., {Chi{\`e}ze}, J.~., \& {Dubroca}, B. 2002, ArXiv
  Astrophysics e-prints

\bibitem[{{Bartl} {et~al.}(2014){Bartl}, {Pethick}, \& {Schwenk}}]{barti}
{Bartl}, A., {Pethick}, C.~J., \& {Schwenk}, A. 2014, Physical Review Letters,
  113, 081101

\bibitem[{{Baumgarte} \& {Shapiro}(1999)}]{Baumgarte99}
{Baumgarte}, T.~W. \& {Shapiro}, S.~L. 1999, \prd, 59, 024007

\bibitem[{{Bethe}(1990)}]{bethe}
{Bethe}, H.~A. 1990, Reviews of Modern Physics, 62, 801

\bibitem[{{Bethe} \& {Wilson}(1985)}]{Bethe85}
{Bethe}, H.~A. \& {Wilson}, J.~R. 1985, \apj, 295, 14

\bibitem[{{Blondin} \& {Mezzacappa}(2007)}]{Blondin07_nat}
{Blondin}, J.~M. \& {Mezzacappa}, A. 2007, \nat, 445, 58

\bibitem[{{Blondin} {et~al.}(2003){Blondin}, {Mezzacappa}, \&
  {DeMarino}}]{Blondin03}
{Blondin}, J.~M., {Mezzacappa}, A., \& {DeMarino}, C. 2003, \apj, 584, 971

\bibitem[{{Brandt} {et~al.}(2011){Brandt}, {Burrows}, {Ott}, \&
  {Livne}}]{Brandt}
{Brandt}, T.~D., {Burrows}, A., {Ott}, C.~D., \& {Livne}, E. 2011, \apj, 728, 8

\bibitem[{{Bruenn}(1985)}]{Bruenn85}
{Bruenn}, S.~W. 1985, \apjs, 58, 771

\bibitem[{{Bruenn} {et~al.}(2001){Bruenn}, {De Nisco}, \&
  {Mezzacappa}}]{Bruenn01}
{Bruenn}, S.~W., {De Nisco}, K.~R., \& {Mezzacappa}, A. 2001, \apj, 560, 326

\bibitem[{{Bruenn} \& {Haxton}(1991)}]{Bruenn91}
{Bruenn}, S.~W. \& {Haxton}, W.~C. 1991, \apj, 376, 678

\bibitem[{{Bruenn} \& {Mezzacappa}(1997)}]{bruenn97}
{Bruenn}, S.~W. \& {Mezzacappa}, A. 1997, \prd, 56, 7529

\bibitem[{{Bruenn} {et~al.}(2009){Bruenn}, {Mezzacappa}, {Hix}, {Blondin},
  {Marronetti}, {Messer}, {Dirk}, \& {Yoshida}}]{Bruenn09}
{Bruenn}, S.~W., {Mezzacappa}, A., {Hix}, W.~R., {Blondin}, J.~M.,
  {Marronetti}, P., {Messer}, O.~E.~B., {Dirk}, C.~J., \& {Yoshida}, S. 2009,
  Journal of Physics Conference Series, 180, 012018

\bibitem[{{Bruenn} {et~al.}(2013){Bruenn}, {Mezzacappa}, {Hix}, {Lentz},
  {Bronson Messer}, {Lingerfelt}, {Blondin}, {Endeve}, {Marronetti}, \&
  {Yakunin}}]{Bruenn13}
{Bruenn}, S.~W., {Mezzacappa}, A., {Hix}, W.~R., {Lentz}, E.~J., {Bronson
  Messer}, O.~E., {Lingerfelt}, E.~J., {Blondin}, J.~M., {Endeve}, E.,
  {Marronetti}, P., \& {Yakunin}, K.~N. 2013, \apjl, 767, L6

\bibitem[{{Buras} {et~al.}(2006{\natexlab{a}}){Buras}, {Janka}, {Rampp}, \&
  {Kifonidis}}]{Buras06b}
{Buras}, R., {Janka}, H.-T., {Rampp}, M., \& {Kifonidis}, K.
  2006{\natexlab{a}}, \aap, 457, 281

\bibitem[{{Buras} {et~al.}(2006{\natexlab{b}}){Buras}, {Rampp}, {Janka}, \&
  {Kifonidis}}]{Buras06a}
{Buras}, R., {Rampp}, M., {Janka}, H.-T., \& {Kifonidis}, K.
  2006{\natexlab{b}}, \aap, 447, 1049

\bibitem[{{Burrows}(2013)}]{Burrows13}
{Burrows}, A. 2013, Reviews of Modern Physics, 85, 245

\bibitem[{{Burrows} {et~al.}(2007){Burrows}, {Dessart}, {Livne}, {Ott}, \&
  {Murphy}}]{burr07}
{Burrows}, A., {Dessart}, L., {Livne}, E., {Ott}, C.~D., \& {Murphy}, J. 2007,
  Astrophys. J., 664, 416

\bibitem[{{Burrows} {et~al.}(1995){Burrows}, {Hayes}, \& {Fryxell}}]{burr95}
{Burrows}, A., {Hayes}, J., \& {Fryxell}, B.~A. 1995, \apj, 450, 830

\bibitem[{{Burrows} {et~al.}(2006){Burrows}, {Reddy}, \&
  {Thompson}}]{Burrows06}
{Burrows}, A., {Reddy}, S., \& {Thompson}, T.~A. 2006, Nuclear Physics A, 777,
  356

\bibitem[{{Cardall} {et~al.}(2013{\natexlab{a}}){Cardall}, {Endeve}, \&
  {Mezzacappa}}]{cardall13}
{Cardall}, C.~Y., {Endeve}, E., \& {Mezzacappa}, A. 2013{\natexlab{a}}, \prd,
  88, 023011

\bibitem[{{Cardall} {et~al.}(2013{\natexlab{b}}){Cardall}, {Endeve}, \&
  {Mezzacappa}}]{cardall12}
---. 2013{\natexlab{b}}, \prd, 87, 103004

\bibitem[{{Cernohorsky} \& {Bludman}(1994)}]{Cernohorsky94}
{Cernohorsky}, J. \& {Bludman}, S.~A. 1994, \apj, 433, 250

\bibitem[{{Couch}(2013)}]{couch13a}
{Couch}, S.~M. 2013, \apj, 775, 35

\bibitem[{{Couch} \& {Ott}(2013)}]{couch_ott}
{Couch}, S.~M. \& {Ott}, C.~D. 2013, \apjl, 778, L7

\bibitem[{{Dimmelmeier} {et~al.}(2002){Dimmelmeier}, {Font}, \&
  {M{\"u}ller}}]{Dimmelmeier02}
{Dimmelmeier}, H., {Font}, J.~A., \& {M{\"u}ller}, E. 2002, \aap, 388, 917

\bibitem[{{Fern{\'a}ndez} {et~al.}(2014){Fern{\'a}ndez}, {M{\"u}ller},
  {Foglizzo}, \& {Janka}}]{Fernandez14}
{Fern{\'a}ndez}, R., {M{\"u}ller}, B., {Foglizzo}, T., \& {Janka}, H.-T. 2014,
  \mnras, 440, 2763

\bibitem[{{Fern{\'a}ndez} \& {Thompson}(2009)}]{rodrigo09}
{Fern{\'a}ndez}, R. \& {Thompson}, C. 2009, \apj, 697, 1827

\bibitem[{{Fischer} {et~al.}(2013){Fischer}, {Langanke}, \&
  {Mart{\'{\i}}nez-Pinedo}}]{tobias13}
{Fischer}, T., {Langanke}, K., \& {Mart{\'{\i}}nez-Pinedo}, G. 2013, \prc, 88,
  065804

\bibitem[{{Foglizzo} {et~al.}(2007){Foglizzo}, {Galletti}, {Scheck}, \&
  {Janka}}]{Foglizzo07}
{Foglizzo}, T., {Galletti}, P., {Scheck}, L., \& {Janka}, H.-T. 2007, \apj,
  654, 1006

\bibitem[{{Foglizzo} {et~al.}(2015){Foglizzo}, {Kazeroni}, {Guilet}, {Masset},
  {Gonz{\'a}lez}, {Krueger}, {Novak}, {Oertel}, {Margueron}, {Faure}, {Martin},
  {Blottiau}, {Peres}, \& {Durand}}]{thierry15}
{Foglizzo}, T., {Kazeroni}, R., {Guilet}, J., {Masset}, F., {Gonz{\'a}lez}, M.,
  {Krueger}, B.~K., {Novak}, J., {Oertel}, M., {Margueron}, J., {Faure}, J.,
  {Martin}, N., {Blottiau}, P., {Peres}, B., \& {Durand}, G. 2015, ArXiv
  e-prints

\bibitem[{{Foglizzo} {et~al.}(2012){Foglizzo}, {Masset}, {Guilet}, \&
  {Durand}}]{thierry12}
{Foglizzo}, T., {Masset}, F., {Guilet}, J., \& {Durand}, G. 2012, Physical
  Review Letters, 108, 051103

\bibitem[{{Foglizzo} {et~al.}(2006){Foglizzo}, {Scheck}, \&
  {Janka}}]{Foglizzo06}
{Foglizzo}, T., {Scheck}, L., \& {Janka}, H.-T. 2006, \apj, 652, 1436

\bibitem[{{Fryer} {et~al.}(1999){Fryer}, {Benz}, {Herant}, \&
  {Colgate}}]{fryer1999}
{Fryer}, C., {Benz}, W., {Herant}, M., \& {Colgate}, S.~A. 1999, \apj, 516, 892

\bibitem[{{Guilet} \& {Foglizzo}(2012)}]{jerome12}
{Guilet}, J. \& {Foglizzo}, T. 2012, \mnras, 421, 546

\bibitem[{{Hanke} {et~al.}(2012){Hanke}, {Marek}, {M{\"u}ller}, \&
  {Janka}}]{Hanke12}
{Hanke}, F., {Marek}, A., {M{\"u}ller}, B., \& {Janka}, H.-T. 2012, \apj, 755,
  138

\bibitem[{{Hanke} {et~al.}(2013){Hanke}, {M{\"u}ller}, {Wongwathanarat},
  {Marek}, \& {Janka}}]{Hanke13}
{Hanke}, F., {M{\"u}ller}, B., {Wongwathanarat}, A., {Marek}, A., \& {Janka},
  H.-T. 2013, \apj, 770, 66

\bibitem[{{Hannestad} \& {Raffelt}(1998)}]{Hannestad98}
{Hannestad}, S. \& {Raffelt}, G. 1998, \apj, 507, 339

\bibitem[{Harten {et~al.}(1983)Harten, Lax, \& Leer}]{Harten83}
Harten, A., Lax, P.~D., \& Leer, B.~v. 1983, SIAM review, 25, 35

\bibitem[{{Herant} {et~al.}(1994){Herant}, {Benz}, {Hix}, {Fryer}, \&
  {Colgate}}]{herant}
{Herant}, M., {Benz}, W., {Hix}, W.~R., {Fryer}, C.~L., \& {Colgate}, S.~A.
  1994, \apj, 435, 339

\bibitem[{{Horiuchi} {et~al.}(2014){Horiuchi}, {Nakamura}, {Takiwaki},
  {Kotake}, \& {Tanaka}}]{Horiuchi14}
{Horiuchi}, S., {Nakamura}, K., {Takiwaki}, T., {Kotake}, K., \& {Tanaka}, M.
  2014, \mnras, 445, L99

\bibitem[{{Horowitz}(2002)}]{horowitz02}
{Horowitz}, C.~J. 2002, \prd, 65, 043001

\bibitem[{{Iwakami} {et~al.}(2008){Iwakami}, {Kotake}, {Ohnishi}, {Yamada}, \&
  {Sawada}}]{Iwakami08}
{Iwakami}, W., {Kotake}, K., {Ohnishi}, N., {Yamada}, S., \& {Sawada}, K. 2008,
  \apj, 678, 1207

\bibitem[{{Iwakami} {et~al.}(2009){Iwakami}, {Kotake}, {Ohnishi}, {Yamada}, \&
  {Sawada}}]{Iwakami09}
---. 2009, \apj, 700, 232

\bibitem[{{Iwakami} {et~al.}(2014){Iwakami}, {Nagakura}, \&
  {Yamada}}]{iwakami14}
{Iwakami}, W., {Nagakura}, H., \& {Yamada}, S. 2014, \apj, 793, 5

\bibitem[{{Janka}(1992)}]{janka92}
{Janka}, H.-T. 1992, \aap, 256, 452

\bibitem[{{Janka}(2012)}]{Janka12}
---. 2012, Annual Review of Nuclear and Particle Science, 62, 407

\bibitem[{{Janka} \& {M\"uller}(1996)}]{Janka96}
{Janka}, H.-T. \& {M\"uller}, E. 1996, \aap, 306, 167

\bibitem[{{Juodagalvis} {et~al.}(2010){Juodagalvis}, {Langanke}, {Hix},
  {Mart{\'{\i}}nez-Pinedo}, \& {Sampaio}}]{juoda}
{Juodagalvis}, A., {Langanke}, K., {Hix}, W.~R., {Mart{\'{\i}}nez-Pinedo}, G.,
  \& {Sampaio}, J.~M. 2010, Nuclear Physics A, 848, 454

\bibitem[{{Just} {et~al.}(2015){Just}, {Obergaulinger}, \& {Janka}}]{just15}
{Just}, O., {Obergaulinger}, M., \& {Janka}, H.-T. 2015, ArXiv e-prints

\bibitem[{{Kanno} {et~al.}(2013){Kanno}, {Harada}, \& {Hanawa}}]{Kanno13}
{Kanno}, Y., {Harada}, T., \& {Hanawa}, T. 2013, \pasj, 65, 72

\bibitem[{{Kitaura} {et~al.}(2006){Kitaura}, {Janka}, \&
  {Hillebrandt}}]{kitaura}
{Kitaura}, F.~S., {Janka}, H.-T., \& {Hillebrandt}, W. 2006, \aap, 450, 345

\bibitem[{{Kotake}(2013)}]{Kotake13}
{Kotake}, K. 2013, Comptes Rendus Physique, 14, 318

\bibitem[{{Kotake} {et~al.}(2006{\natexlab{a}}){Kotake}, {Ohnishi}, {Yamada},
  \& {Sato}}]{kotake06fld}
{Kotake}, K., {Ohnishi}, N., {Yamada}, S., \& {Sato}, K. 2006{\natexlab{a}},
  Journal of Physics Conference Series, 31, 95

\bibitem[{{Kotake} {et~al.}(2006{\natexlab{b}}){Kotake}, {Sato}, \&
  {Takahashi}}]{kotake06}
{Kotake}, K., {Sato}, K., \& {Takahashi}, K. 2006{\natexlab{b}}, Reports on
  Progress in Physics, 69, 971

\bibitem[{{Kotake} {et~al.}(2004){Kotake}, {Sawai}, {Yamada}, \&
  {Sato}}]{kota04b}
{Kotake}, K., {Sawai}, H., {Yamada}, S., \& {Sato}, K. 2004, Astrophys. J.,
  608, 391

\bibitem[{{Kotake} {et~al.}(2012{\natexlab{a}}){Kotake}, {Sumiyoshi}, {Yamada},
  {Takiwaki}, {Kuroda}, {Suwa}, \& {Nagakura}}]{Kotake12_ptep}
{Kotake}, K., {Sumiyoshi}, K., {Yamada}, S., {Takiwaki}, T., {Kuroda}, T.,
  {Suwa}, Y., \& {Nagakura}, H. 2012{\natexlab{a}}, Progress of Theoretical and
  Experimental Physics, 2012, 010000

\bibitem[{{Kotake} {et~al.}(2012{\natexlab{b}}){Kotake}, {Takiwaki}, {Suwa},
  {Iwakami Nakano}, {Kawagoe}, {Masada}, \& {Fujimoto}}]{Kotake12}
{Kotake}, K., {Takiwaki}, T., {Suwa}, Y., {Iwakami Nakano}, W., {Kawagoe}, S.,
  {Masada}, Y., \& {Fujimoto}, S.-i. 2012{\natexlab{b}}, Advances in Astronomy,
  2012

\bibitem[{{Kotake} {et~al.}(2003){Kotake}, {Yamada}, \& {Sato}}]{Kotake03}
{Kotake}, K., {Yamada}, S., \& {Sato}, K. 2003, \prd, 68, 044023

\bibitem[{{Kuroda} {et~al.}(2012){Kuroda}, {Kotake}, \& {Takiwaki}}]{KurodaT12}
{Kuroda}, T., {Kotake}, K., \& {Takiwaki}, T. 2012, \apj, 755, 11

\bibitem[{{Kuroda} {et~al.}(2014){Kuroda}, {Takiwaki}, \& {Kotake}}]{KurodaT14}
{Kuroda}, T., {Takiwaki}, T., \& {Kotake}, K. 2014, \prd, 89, 044011

\bibitem[{{Kuroda} \& {Umeda}(2010)}]{KurodaT10}
{Kuroda}, T. \& {Umeda}, H. 2010, \apjs, 191, 439

\bibitem[{{Langanke} \& {Mart{\'{\i}}nez-Pinedo}(2000)}]{Langanke00}
{Langanke}, K. \& {Mart{\'{\i}}nez-Pinedo}, G. 2000, Nuclear Physics A, 673,
  481

\bibitem[{{Langanke} {et~al.}(2003){Langanke}, {Mart{\'{\i}}nez-Pinedo},
  {Sampaio}, {Dean}, {Hix}, {Messer}, {Mezzacappa}, {Liebend{\"o}rfer},
  {Janka}, \& {Rampp}}]{langanke03}
{Langanke}, K., {Mart{\'{\i}}nez-Pinedo}, G., {Sampaio}, J.~M., {Dean}, D.~J.,
  {Hix}, W.~R., {Messer}, O.~E., {Mezzacappa}, A., {Liebend{\"o}rfer}, M.,
  {Janka}, H.-T., \& {Rampp}, M. 2003, Physical Review Letters, 90, 241102

\bibitem[{{Lattimer} \& {Douglas Swesty}(1991)}]{LSEOS}
{Lattimer}, J.~M. \& {Douglas Swesty}, F. 1991, Nuclear Physics A, 535, 331

\bibitem[{{Lentz} {et~al.}(2012{\natexlab{a}}){Lentz}, {Mezzacappa}, {Bronson
  Messer}, {Hix}, \& {Bruenn}}]{lentz12a}
{Lentz}, E.~J., {Mezzacappa}, A., {Bronson Messer}, O.~E., {Hix}, W.~R., \&
  {Bruenn}, S.~W. 2012{\natexlab{a}}, \apj, 760, 94

\bibitem[{{Lentz} {et~al.}(2012{\natexlab{b}}){Lentz}, {Mezzacappa}, {Bronson
  Messer}, {Liebend{\"o}rfer}, {Hix}, \& {Bruenn}}]{lentz12b}
{Lentz}, E.~J., {Mezzacappa}, A., {Bronson Messer}, O.~E., {Liebend{\"o}rfer},
  M., {Hix}, W.~R., \& {Bruenn}, S.~W. 2012{\natexlab{b}}, \apj, 747, 73

\bibitem[{{Levermore}(1984)}]{levermore84}
{Levermore}, C.~D. 1984, \jqsrt, 31, 149

\bibitem[{{Liebend{\"o}rfer} {et~al.}(2004){Liebend{\"o}rfer}, {Messer},
  {Mezzacappa}, {Bruenn}, {Cardall}, \& {Thielemann}}]{Liebendorfer04}
{Liebend{\"o}rfer}, M., {Messer}, O.~E.~B., {Mezzacappa}, A., {Bruenn}, S.~W.,
  {Cardall}, C.~Y., \& {Thielemann}, F.-K. 2004, \apjs, 150, 263

\bibitem[{{Liebend{\"o}rfer} {et~al.}(2001){Liebend{\"o}rfer}, {Mezzacappa},
  {Thielemann}, {Messer}, {Hix}, \& {Bruenn}}]{Liebendorfer01}
{Liebend{\"o}rfer}, M., {Mezzacappa}, A., {Thielemann}, F.-K., {Messer}, O.~E.,
  {Hix}, W.~R., \& {Bruenn}, S.~W. 2001, \prd, 63, 103004

\bibitem[{{Liebend{\"o}rfer} {et~al.}(2005){Liebend{\"o}rfer}, {Rampp},
  {Janka}, \& {Mezzacappa}}]{Liebendorfer05}
{Liebend{\"o}rfer}, M., {Rampp}, M., {Janka}, H.-T., \& {Mezzacappa}, A. 2005,
  \apj, 620, 840

\bibitem[{{Liebend{\"o}rfer} {et~al.}(2009){Liebend{\"o}rfer}, {Whitehouse}, \&
  {Fischer}}]{Liebendorfer09}
{Liebend{\"o}rfer}, M., {Whitehouse}, S.~C., \& {Fischer}, T. 2009, \apj, 698,
  1174

\bibitem[{{Livne} {et~al.}(2004){Livne}, {Burrows}, {Walder}, {Lichtenstadt},
  \& {Thompson}}]{Livne04}
{Livne}, E., {Burrows}, A., {Walder}, R., {Lichtenstadt}, I., \& {Thompson},
  T.~A. 2004, \apj, 609, 277

\bibitem[{{Marek} {et~al.}(2006){Marek}, {Dimmelmeier}, {Janka}, {M{\"u}ller},
  \& {Buras}}]{Marek06}
{Marek}, A., {Dimmelmeier}, H., {Janka}, H.-T., {M{\"u}ller}, E., \& {Buras},
  R. 2006, \aap, 445, 273

\bibitem[{{Marek} \& {Janka}(2009)}]{Marek09}
{Marek}, A. \& {Janka}, H.-T. 2009, \apj, 694, 664

\bibitem[{{Mart{\'{\i}}nez-Pinedo} {et~al.}(2012){Mart{\'{\i}}nez-Pinedo},
  {Fischer}, {Lohs}, \& {Huther}}]{gabriel12}
{Mart{\'{\i}}nez-Pinedo}, G., {Fischer}, T., {Lohs}, A., \& {Huther}, L. 2012,
  Physical Review Letters, 109, 251104

\bibitem[{{Masada} {et~al.}(2014){Masada}, {Takiwaki}, \& {Kotake}}]{masada14}
{Masada}, Y., {Takiwaki}, T., \& {Kotake}, K. 2014, ArXiv e-prints

\bibitem[{{Masada} {et~al.}(2012){Masada}, {Takiwaki}, {Kotake}, \&
  {Sano}}]{masada12}
{Masada}, Y., {Takiwaki}, T., {Kotake}, K., \& {Sano}, T. 2012, \apj, 759, 110

\bibitem[{{May} \& {White}(1966)}]{May}
{May}, M.~M. \& {White}, R.~H. 1966, Physical Review, 141, 1232

\bibitem[{{Meakin} {et~al.}(2011){Meakin}, {Sukhbold}, \& {Arnett}}]{meakin11}
{Meakin}, C.~A., {Sukhbold}, T., \& {Arnett}, W.~D. 2011, \apss, 336, 123

\bibitem[{{Melson} {et~al.}(2015){Melson}, {Janka}, \& {Marek}}]{melson15}
{Melson}, T., {Janka}, H.-T., \& {Marek}, A. 2015, ArXiv e-prints

\bibitem[{{Mezzacappa} \& {Bruenn}(1993{\natexlab{a}})}]{tony93c}
{Mezzacappa}, A. \& {Bruenn}, S.~W. 1993{\natexlab{a}}, \apj, 405, 669

\bibitem[{{Mezzacappa} \& {Bruenn}(1993{\natexlab{b}})}]{tony93a}
---. 1993{\natexlab{b}}, \apj, 410, 740

\bibitem[{{Mezzacappa} \& {Bruenn}(1993{\natexlab{c}})}]{tony93b}
---. 1993{\natexlab{c}}, \apj, 405, 637

\bibitem[{{Mezzacappa} {et~al.}(2014){Mezzacappa}, {Bruenn}, {Lentz}, {Hix},
  {Messer}, {Harris}, {Lingerfelt}, {Endeve}, {Yakunin}, {Blondin}, \&
  {Marronetti}}]{Mezzacappa2014}
{Mezzacappa}, A., {Bruenn}, S.~W., {Lentz}, E.~J., {Hix}, W.~R., {Messer},
  O.~E.~B., {Harris}, J.~A., {Lingerfelt}, E.~J., {Endeve}, E., {Yakunin},
  K.~N., {Blondin}, J.~M., \& {Marronetti}, P. 2014, in Astronomical Society of
  the Pacific Conference Series, Vol. 488, 8th International Conference of
  Numerical Modeling of Space Plasma Flows (ASTRONUM 2013), ed. N.~V.
  {Pogorelov}, E.~{Audit}, \& G.~P. {Zank}, 102

\bibitem[{{Mezzacappa} \& {Matzner}(1989)}]{Mezzacappa89}
{Mezzacappa}, A. \& {Matzner}, R.~A. 1989, \apj, 343, 853

\bibitem[{{Minerbo}(1978)}]{minerbo78}
{Minerbo}, G.~N. 1978, \jqsrt, 20, 541

\bibitem[{{M{\"o}sta} {et~al.}(2014){M{\"o}sta}, {Richers}, {Ott}, {Haas},
  {Piro}, {Boydstun}, {Abdikamalov}, {Reisswig}, \& {Schnetter}}]{moesta14}
{M{\"o}sta}, P., {Richers}, S., {Ott}, C.~D., {Haas}, R., {Piro}, A.~L.,
  {Boydstun}, K., {Abdikamalov}, E., {Reisswig}, C., \& {Schnetter}, E. 2014,
  \apjl, 785, L29

\bibitem[{{M{\"u}ller} \& {Janka}(2014)}]{BMuller14}
{M{\"u}ller}, B. \& {Janka}, H.-T. 2014, \apj, 788, 82

\bibitem[{{M\"uller} \& {Janka}(2014)}]{bernhard14}
{M\"uller}, B. \& {Janka}, H.-T. 2014, ArXiv e-prints

\bibitem[{{M{\"u}ller} {et~al.}(2010){M{\"u}ller}, {Janka}, \&
  {Dimmelmeier}}]{BMuller10}
{M{\"u}ller}, B., {Janka}, H.-T., \& {Dimmelmeier}, H. 2010, \apjs, 189, 104

\bibitem[{{M{\"u}ller} {et~al.}(2012{\natexlab{a}}){M{\"u}ller}, {Janka}, \&
  {Heger}}]{BMuller12b}
{M{\"u}ller}, B., {Janka}, H.-T., \& {Heger}, A. 2012{\natexlab{a}}, \apj, 761,
  72

\bibitem[{{M{\"u}ller} {et~al.}(2012{\natexlab{b}}){M{\"u}ller}, {Janka}, \&
  {Marek}}]{BMuller12a}
{M{\"u}ller}, B., {Janka}, H.-T., \& {Marek}, A. 2012{\natexlab{b}}, \apj, 756,
  84

\bibitem[{{M\"uller} \& {Janka}(1997)}]{mujan}
{M\"uller}, E. \& {Janka}, H.-T. 1997, \aap, 317, 140

\bibitem[{{Murphy} \& {Burrows}(2008)}]{Murphy08}
{Murphy}, J.~W. \& {Burrows}, A. 2008, \apj, 688, 1159

\bibitem[{{Nagakura} {et~al.}(2014){Nagakura}, {Sumiyoshi}, \&
  {Yamada}}]{Nagakura14}
{Nagakura}, H., {Sumiyoshi}, K., \& {Yamada}, S. 2014, \apjs, 214, 16

\bibitem[{{Nakamura} {et~al.}(2014{\natexlab{a}}){Nakamura}, {Kuroda},
  {Takiwaki}, \& {Kotake}}]{Nakamura3D14}
{Nakamura}, K., {Kuroda}, T., {Takiwaki}, T., \& {Kotake}, K.
  2014{\natexlab{a}}, \apj, 793, 45

\bibitem[{{Nakamura} {et~al.}(2014{\natexlab{b}}){Nakamura}, {Takiwaki},
  {Kuroda}, \& {Kotake}}]{Nakamura2D14}
{Nakamura}, K., {Takiwaki}, T., {Kuroda}, T., \& {Kotake}, K.
  2014{\natexlab{b}}, ArXiv e-prints

\bibitem[{{Obergaulinger} {et~al.}(2006){Obergaulinger}, {Aloy}, {Dimmelmeier},
  \& {M{\"u}ller}}]{ober06a}
{Obergaulinger}, M., {Aloy}, M.~A., {Dimmelmeier}, H., \& {M{\"u}ller}, E.
  2006, Astron. Astrophys., 457, 209

\bibitem[{{Obergaulinger} \& {Janka}(2011)}]{Obergaulinger11}
{Obergaulinger}, M. \& {Janka}, H.-T. 2011, ArXiv e-prints

\bibitem[{{Obergaulinger} {et~al.}(2014){Obergaulinger}, {Janka}, \&
  {Aloy}}]{martin14}
{Obergaulinger}, M., {Janka}, H.-T., \& {Aloy}, M.~A. 2014, \mnras, 445, 3169

\bibitem[{{O'Connor}(2014)}]{O'Connor14}
{O'Connor}, E. 2014, ArXiv e-prints

\bibitem[{{O'Connor} \& {Ott}(2011)}]{oconnor}
{O'Connor}, E. \& {Ott}, C.~D. 2011, \apj, 730, 70

\bibitem[{{O'Connor} \& {Ott}(2013)}]{O'Connor13}
---. 2013, \apj, 762, 126

\bibitem[{{Ohnishi} {et~al.}(2006){Ohnishi}, {Kotake}, \& {Yamada}}]{Ohnishi06}
{Ohnishi}, N., {Kotake}, K., \& {Yamada}, S. 2006, \apj, 641, 1018

\bibitem[{{Ott} {et~al.}(2012){Ott}, {Abdikamalov}, {O'Connor}, {Reisswig},
  {Haas}, {Kalmus}, {Drasco}, {Burrows}, \& {Schnetter}}]{Ott12a}
{Ott}, C.~D., {Abdikamalov}, E., {O'Connor}, E., {Reisswig}, C., {Haas}, R.,
  {Kalmus}, P., {Drasco}, S., {Burrows}, A., \& {Schnetter}, E. 2012, \prd, 86,
  024026

\bibitem[{{Ott} {et~al.}(2008){Ott}, {Burrows}, {Dessart}, \& {Livne}}]{Ott08}
{Ott}, C.~D., {Burrows}, A., {Dessart}, L., \& {Livne}, E. 2008, \apj, 685,
  1069

\bibitem[{{Perego} {et~al.}(2014){Perego}, {Gafton}, {Cabez{\'o}n}, {Rosswog},
  \& {Liebend{\"o}rfer}}]{perego}
{Perego}, A., {Gafton}, E., {Cabez{\'o}n}, R., {Rosswog}, S., \&
  {Liebend{\"o}rfer}, M. 2014, \aap, 568, A11

\bibitem[{{Peres} {et~al.}(2014){Peres}, {Penner}, {Novak}, \&
  {Bonazzola}}]{peres}
{Peres}, B., {Penner}, A.~J., {Novak}, J., \& {Bonazzola}, S. 2014, Classical
  and Quantum Gravity, 31, 045012

\bibitem[{{Pomraning}(1981)}]{pomraning81}
{Pomraning}, G.~C. 1981, \jqsrt, 26, 385

\bibitem[{{Pons} {et~al.}(2000){Pons}, {Ib{\'a}{\~n}ez}, \&
  {Miralles}}]{Pons00}
{Pons}, J.~A., {Ib{\'a}{\~n}ez}, J.~M., \& {Miralles}, J.~A. 2000, \mnras, 317,
  550

\bibitem[{{Rampp} \& {Janka}(2002)}]{Rampp02}
{Rampp}, M. \& {Janka}, H.-T. 2002, \aap, 396, 361

\bibitem[{{Rosswog} \& {Liebend{\"o}rfer}(2003)}]{ross03}
{Rosswog}, S. \& {Liebend{\"o}rfer}, M. 2003, \mnras, 342, 673

\bibitem[{{Ruffert} {et~al.}(1996){Ruffert}, {Janka}, \&
  {Schaefer}}]{ruffert96}
{Ruffert}, M., {Janka}, H.-T., \& {Schaefer}, G. 1996, \aap, 311, 532

\bibitem[{{Sanchis-Gual} {et~al.}(2014){Sanchis-Gual}, {Montero}, {Font},
  {M{\"u}ller}, \& {Baumgarte}}]{gual}
{Sanchis-Gual}, N., {Montero}, P.~J., {Font}, J.~A., {M{\"u}ller}, E., \&
  {Baumgarte}, T.~W. 2014, \prd, 89, 104033

\bibitem[{{Sato}(1975)}]{Sato75}
{Sato}, K. 1975, Progress of Theoretical Physics, 54, 1325

\bibitem[{{Sawai} {et~al.}(2013){Sawai}, {Yamada}, \& {Suzuki}}]{sawai13}
{Sawai}, H., {Yamada}, S., \& {Suzuki}, H. 2013, \apjl, 770, L19

\bibitem[{{Scheck} {et~al.}(2006){Scheck}, {Kifonidis}, {Janka}, \&
  {M{\"u}ller}}]{scheck06}
{Scheck}, L., {Kifonidis}, K., {Janka}, H., \& {M{\"u}ller}, E. 2006, \aap,
  457, 963

\bibitem[{{Scheidegger} {et~al.}(2010){Scheidegger}, {K{\"a}ppeli},
  {Whitehouse}, {Fischer}, \& {Liebend{\"o}rfer}}]{Scheidegger10}
{Scheidegger}, S., {K{\"a}ppeli}, R., {Whitehouse}, S.~C., {Fischer}, T., \&
  {Liebend{\"o}rfer}, M. 2010, \aap, 514, A51

\bibitem[{{Schwartz}(1967)}]{schwartz}
{Schwartz}, R.~A. 1967, Annals of Physics, 43, 42

\bibitem[{{Sekiguchi}(2010)}]{Sekiguchi10}
{Sekiguchi}, Y. 2010, Progress of Theoretical Physics, 124, 331

\bibitem[{{Shibata} {et~al.}(2011){Shibata}, {Kiuchi}, {Sekiguchi}, \&
  {Suwa}}]{Shibata11}
{Shibata}, M., {Kiuchi}, K., {Sekiguchi}, Y., \& {Suwa}, Y. 2011, Progress of
  Theoretical Physics, 125, 1255

\bibitem[{{Shibata} {et~al.}(2014){Shibata}, {Nagakura}, {Sekiguchi}, \&
  {Yamada}}]{shibata14}
{Shibata}, M., {Nagakura}, H., {Sekiguchi}, Y., \& {Yamada}, S. 2014, \prd, 89,
  084073

\bibitem[{{Shibata} \& {Nakamura}(1995)}]{Shibata95}
{Shibata}, M. \& {Nakamura}, T. 1995, \prd, 52, 5428

\bibitem[{{Shibata} \& {Sekiguchi}(2004)}]{shibaseki}
{Shibata}, M. \& {Sekiguchi}, Y.-I. 2004, \prd, 69, 084024

\bibitem[{{Shibata} \& {Sekiguchi}(2005)}]{shibata05}
---. 2005, \prd, 71, 024014

\bibitem[{{Skinner} \& {Ostriker}(2013)}]{Skinner13}
{Skinner}, M.~A. \& {Ostriker}, E.~C. 2013, \apjs, 206, 21

\bibitem[{{Sumiyoshi} \& {R{\"o}pke}(2008)}]{sumi08}
{Sumiyoshi}, K. \& {R{\"o}pke}, G. 2008, \prc, 77, 055804

\bibitem[{{Sumiyoshi} {et~al.}(2014){Sumiyoshi}, {Takiwaki}, {Matsufuru}, \&
  {Yamada}}]{sumi14}
{Sumiyoshi}, K., {Takiwaki}, T., {Matsufuru}, H., \& {Yamada}, S. 2014, ArXiv
  e-prints

\bibitem[{{Sumiyoshi} \& {Yamada}(2012)}]{Sumiyoshi12}
{Sumiyoshi}, K. \& {Yamada}, S. 2012, \apjs, 199, 17

\bibitem[{{Sumiyoshi} {et~al.}(2005){Sumiyoshi}, {Yamada}, {Suzuki}, {Shen},
  {Chiba}, \& {Toki}}]{Sumiyoshi05}
{Sumiyoshi}, K., {Yamada}, S., {Suzuki}, H., {Shen}, H., {Chiba}, S., \&
  {Toki}, H. 2005, \apj, 629, 922

\bibitem[{{Suwa} {et~al.}(2011){Suwa}, {Kotake}, {Takiwaki},
  {Liebend{\"o}rfer}, \& {Sato}}]{Suwa11}
{Suwa}, Y., {Kotake}, K., {Takiwaki}, T., {Liebend{\"o}rfer}, M., \& {Sato}, K.
  2011, \apj, 738, 165

\bibitem[{{Suwa} {et~al.}(2010){Suwa}, {Kotake}, {Takiwaki}, {Whitehouse},
  {Liebend{\"o}rfer}, \& {Sato}}]{Suwa10}
{Suwa}, Y., {Kotake}, K., {Takiwaki}, T., {Whitehouse}, S.~C.,
  {Liebend{\"o}rfer}, M., \& {Sato}, K. 2010, \pasj, 62, L49

\bibitem[{{Suwa} {et~al.}(2013){Suwa}, {Takiwaki}, {Kotake}, {Fischer},
  {Liebend{\"o}rfer}, \& {Sato}}]{suwa13}
{Suwa}, Y., {Takiwaki}, T., {Kotake}, K., {Fischer}, T., {Liebend{\"o}rfer},
  M., \& {Sato}, K. 2013, \apj, 764, 99

\bibitem[{{Suwa} {et~al.}(2014){Suwa}, {Yamada}, {Takiwaki}, \&
  {Kotake}}]{suwa14}
{Suwa}, Y., {Yamada}, S., {Takiwaki}, T., \& {Kotake}, K. 2014, ArXiv e-prints

\bibitem[{{Swesty} \& {Myra}(2009)}]{Swesty09}
{Swesty}, F.~D. \& {Myra}, E.~S. 2009, \apjs, 181, 1

\bibitem[{{Takiwaki} {et~al.}(2012){Takiwaki}, {Kotake}, \&
  {Suwa}}]{Takiwaki12}
{Takiwaki}, T., {Kotake}, K., \& {Suwa}, Y. 2012, \apj, 749, 98

\bibitem[{{Takiwaki} {et~al.}(2014){Takiwaki}, {Kotake}, \&
  {Suwa}}]{Takiwaki14}
---. 2014, \apj, 786, 83

\bibitem[{{Thompson} {et~al.}(2003){Thompson}, {Burrows}, \&
  {Pinto}}]{Thompson03}
{Thompson}, T.~A., {Burrows}, A., \& {Pinto}, P.~A. 2003, \apj, 592, 434

\bibitem[{{Thorne}(1981)}]{Thorne81}
{Thorne}, K.~S. 1981, \mnras, 194, 439

\bibitem[{{Wilson}(1985)}]{Wilson85}
{Wilson}, J.~R. 1985, in Numerical Astrophysics, ed. J.~M. {Centrella}, J.~M.
  {Leblanc}, \& R.~L. {Bowers}, 422

\bibitem[{{Woosley} \& {Bloom}(2006)}]{Woosley06}
{Woosley}, S.~E. \& {Bloom}, J.~S. 2006, \araa, 44, 507

\bibitem[{{Woosley} \& {Weaver}(1995)}]{WW95}
{Woosley}, S.~E. \& {Weaver}, T.~A. 1995, \apjs, 101, 181

\bibitem[{{Yamada}(1997)}]{Yamada97}
{Yamada}, S. 1997, \apj, 475, 720

\bibitem[{{Yamada} {et~al.}(1999){Yamada}, {Janka}, \& {Suzuki}}]{Yamada99}
{Yamada}, S., {Janka}, H.-T., \& {Suzuki}, H. 1999, \aap, 344, 533

\bibitem[{{Yueh} \& {Buchler}(1976)}]{Yueh76}
{Yueh}, W.~R. \& {Buchler}, J.~R. 1976, \apss, 41, 221

\bibitem[{{Zhang} {et~al.}(2013){Zhang}, {Howell}, {Almgren}, {Burrows},
  {Dolence}, \& {Bell}}]{zhang13}
{Zhang}, W., {Howell}, L., {Almgren}, A., {Burrows}, A., {Dolence}, J., \&
  {Bell}, J. 2013, \apjs, 204, 7

\end{thebibliography}
\end{document}